\documentclass[fleqn]{2022AMSE}

\usepackage{color}
\usepackage{adjustbox}
\usepackage{hyperref}
\usepackage{graphicx}
\usepackage{upgreek}
\usepackage{float}
\usepackage{booktabs} 
\usepackage{tabularx} 
\usepackage{amsmath} 
\usepackage{siunitx} 
\usepackage{bm}
\usepackage{xcolor}
\usepackage{bm}

\begin{document}
\setlength{\mathindent}{0cm}

\ensubject{Fluid Dynamics}
\ArticleType{RESEARCH PAPER}%??Article
%\SpecialTopic{SPECIAL TOPIC: }%???????
\Year{}
\Vol{}
\DOI{}
\ArtNo{}
\ReceiveDate{???}
\AcceptDate{???}
\OnlineDate{???}

\title{Physics-Informed Transformer operator for the prediction of three-dimensional turbulence}

\author[1,2,3]{Zhihong Guo}{}
\author[1,2,3]{Sunan Zhao}{}
\author[1,2,3]{Huiyu Yang}{}
\author[1,2,3]{Yunpeng Wang}{}
\author[1,2,3]{Jianchun Wang}{wangjc@sustech.edu.cn}

\AuthorMark{Zhihong Guo}%Y. Ji%\authorcr????????  Z. Wang

\AuthorCitation{Zhihong Guo, Sunan Zhao, Huiyu Yang, Yunpeng Wang, and Jianchun Wang}

\address[1]{Department of Mechanics and Aerospace Engineering, Southern University of Science and Technology, Shenzhen 518055, China}
\address[2]{Guangdong Provincial Key Laboratory of Turbulence Research and Applications, Southern University of Science and Technology, Shenzhen 518055, China}
\address[3]{Shenzhen Key Laboratory of Complex Aerospace Flows, Southern University of Science and Technology, Shenzhen 518055, China}

\contributions{Executive Editor: ???}

\abstract
		{Data-driven turbulence prediction methods often face challenges related to data dependency and lack of physical interpretability. In this paper, we propose a physics-informed Transformer operator (PITO) and its implicit variant (PIITO) for predicting three-dimensional (3D) turbulence, which aRre developed based on the vision Transformer (ViT) architecture with an appropriate patch size. Given the current flow field, the Transformer operator computes its prediction for the next time step. By embedding the large-eddy simulation (LES) equations into the loss function, PITO and PIITO can learn solution operators without using labeled data. Furthermore, PITO can automatically learn the subgrid scale (SGS) coefficient using a single set of flow data during training. Both PITO and PIITO exhibit excellent stability and accuracy on the predictions of various statistical properties and flow structures for the situation of long-term extrapolation exceeding 25 times the training horizon in decaying homogeneous isotropic turbulence (HIT), and outperform the physics-informed Fourier neural operator (PIFNO). Furthermore, PITO exhibits a remarkable accuracy on the predictions of forced HIT where PIFNO fails. Notably, PITO and PIITO reduce graphics processing unit (GPU) memory consumption by 79.5\% and 91.3\% while requiring only 31.5\% and 3.1\% of the parameters, respectively, compared to PIFNO. Moreover, both PITO and PIITO models are much faster compared to traditional LES method.}


\setlength{\textheight}{23.6cm}
\thispagestyle{empty}

\maketitle
\setlength{\parindent}{1em}

% \setlength{\textheight}{23.6cm}
% \Authorfootnote
% \tableofcontents%?????
\vspace{-1mm}
\begin{multicols}{2}
\section{Introduction}
Turbulence is ubiquitous in both natural phenomena and engineering applications \cite{pope}. Due to the complex multiscale nature of turbulence, traditional computational fluid dynamics (CFD) methods face trade-offs between computational cost and accuracy \cite{moin1998direct,sagaut2006large,smagorinsky1963general}. Various machine learning (ML) methods have been developed to improve or replace CFD simulations of turbulence \cite{brunton2020machine,karniadakis2021physics}. Most early machine learning methods were primarily used to enhance the accuracy of turbulence closure models for Reynolds-averaged Navier-Stokes simulation (RANS) and large-eddy simulation (LES) \cite{beck2019deep,xie2020modeling,kurz2023deep,maulik2018data,zhu2019machine,ling2016reynolds,wang2018investigations,zhou2019subgrid,novati2021automating,yang2019predictive}.\Authorfootnote 

In recent years, deep neural network methods have been developed for directly predicting flow fields \cite{srinivasan2019predictions,ruhling2023dyffusion,nakamura2021convolutional,bukka2021assessment,raissi2017physics,cai2021physics,sun2026reconstruction,zhu2026physics}. Wu et al. combined a physics-embedded super-resolution generative adversarial neural network (SRGAN) with down-sampling modules and proposed the SRGAN-based
energy cascade reconstruction (EC-SRGAN) framework for the high-fidelity reconstruction of wall turbulence, which can accurately reproduce small-scale velocity fields and energy spectra, and achieve zero-shot generalization to turbulent boundary layer flows \cite{Wu2024high}. Wu et al. further proposed a scale-oriented zonal generative adversarial network (SoZoGAN) framework which integrates a zonal decomposition strategy with a SRGAN library to achieve robust zero-shot generation across diverse flows \cite{Wu2025_general}. 
Du et al. proposed the conditional neural field latent diffusion (CoNFiLD), which enables the generation of diverse three-dimensional (3D) turbulent flows by integrating latent diffusion processes with conditional neural fields \cite{du2024conditional}. Moreover, Jiang et al. proposed the DiAFNO model which integrates the implicit adaptive Fourier neural operator (IAFNO) with diffusion model for the prediction of 3D turbulence, achieving long-term stable and highly accurate predictions in channel flows \cite{jiang2025integrating}. Oommen et al. proposed an adversarially trained neural operator (adv-NO), which leverages generative adversarial network (GAN) mechanisms to resolve spectral bias for high-fidelity flow forecasting, while employing a conditional diffusion model for accurate 3D reconstruction from sparse observations \cite{oommen2026learning}.

Raissi et al. proposed physics-informed neural networks (PINNs), a method that embeds physical laws directly into the loss function to address forward and inverse partial differential equations (PDE) problems \cite{raissi2019physics}. Wang et al. proposed the turbulent flow net (TF-Net), a physics-informed architecture designed to capture the dynamics of two-dimensional (2D) Rayleigh-B\'enard convection \cite{wang2020towards}. Jin et al. developed NSFnets, a PINN-based framework that predicts flow fields given their initial and boundary conditions \cite{jin2021nsfnets}. Although PINNs have demonstrated the capability to achieve satisfactory results even with sparse or no labeled data by embedding physical laws into the loss function, their application in complex scenarios is a great challenge \cite{zhang2026physics,wang2022and}. Wang et al. proposed an adaptive weighting algorithm to mitigate gradient pathologies in PINNs by dynamically balancing the gradient magnitudes of different loss terms \cite{wang2021understanding}. Jagtap et al. proposed adaptive activation functions to accelerate the convergence and improve the accuracy of PINNs by dynamically adjusting the slopes of neurons during training \cite{jagtap2020adaptive}. Yang et al. introduced Bayesian physics-informed neural networks (B-PINNs), which combine Bayesian inference with PINNs to enable stable predictions and uncertainty quantification of physical fields using noisy experimental data \cite{yang2021b}. Wang et al. achieved the simulations of 3D turbulent flows by improving PINN training strategies \cite{wang2025simulating}. Wang et al. introduced ShampoO with Adam in the Preconditioner's eigenbasis (SOAP), a quasi-Newton method that resolves gradient conflicts through second-order optimization, achieving breakthrough accuracy in turbulence simulations at high Reynolds numbers \cite{wang2025gradient}. Zhang et al. used coordinate transformation to mitigate large gradients, enabling accurate PINN simulations of high-Reynolds-number flat plate and NACA0012 airfoil flows \cite{zhang2025physics}, while Cao et al. utilized mesh transformation to resolve the sharp leading-edge transitions in subsonic airfoil flows \cite{cao2024solver}. Moreover, Cao et al. stabilized PINN training by introducing a Jacobian-based gain-tuning strategy, enabling the robust simulation of complex 3D flows \cite{cao2025analysis}. To improve the temporal modeling of PINNs, Zhao et al. proposed Pinnsformer, which leverages Transformer blocks to better capture temporal dependencies than conventional MLPs \cite{zhao2023pinnsformer}.

Previous studies primarily focused on neural networks for learning finite-dimensional mappings, often struggling to generalize to other flow types or boundary conditions. To address this limitation, the framework of neural operators (NO) was proposed to learn operators between infinite-dimensional function spaces \cite{lu2021learning,li2023geometry,li2024Transformer,luo2024fourier,peng2023linear,hao2023gnot}. Among these, Li et al. proposed the Fourier neural operator (FNO), which approximates the solution operator by parameterizing the integral kernel in the Fourier domain. In 2D Navier-Stokes turbulence, FNO outperformed convolutional networks in terms of accuracy and efficiency \cite{li2020fourier}. Consequently, numerous extensions of FNO were proposed. You et al. proposed the implicit Fourier neural operator (IFNO), which significantly reduces memory requirements and the number of trainable parameters \cite{you2022learning}. Tran et al. proposed the factorized Fourier neural operator (F-FNO), which achieves more efficient operator learning by performing low-rank factorization on the frequency-domain weight tensor \cite{tran2021factorized}. Additionally, Peng et al. embedded an attention mechanism into the FNO model, enhancing its predictive accuracy for 2D turbulence \cite{peng2022attention}. While early FNO-based studies primarily focused on one-dimensional (1D) and 2D problems, Li et al. extended the framework to 3D space and applied it to the predictions of 3D homogeneous isotropic turbulence \cite{li2022fourier}. To further address the long-term instability issue in FNO, Li et al. subsequently developed an implicit U-Net enhanced FNO (IUFNO), achieving stable and accurate long-term predictions of 3D turbulent flows \cite{li2023long}. Wang et al. extended IUFNO to 3D turbulent channel flows, achieving fast, accurate and stable long-term predictions \cite{wang2024prediction}.

Although learning parameters in the frequency domain enables FNO to predict turbulence effectively and accurately, it should be noted that FNO relies on the Fourier transform and is therefore primarily applicable to periodic boundaries \cite{li2023fourier}. Recently, Transformer models have demonstrated outstanding performance in computer vision and natural language processing \cite{vaswani2017attention,khan2022Transformers,liu2021swin,dosovitskiy2020image}. Similar to FNO, Transformer also has the ability to capture global information and is increasingly being employed to solve partial differential equations \cite{wu2024transolver,hao2024dpot,chen2024omniarch} and perform super-resolution flow reconstruction \cite{wang2026reconstruction,wang2025parametric,zheng2024aerodit,fan2025cascaded,lei2025efficient}. Cao et al. established a theoretical connection between self-attention and Petrov-Galerkin projection, and proposed Galerkin Transformer to enhance approximation efficiency in PDE problems \cite{cao2021choose}. Patil et al. proposed a convolutional Transformer model, achieving accurate long-term predictions for 2D wake flows \cite{patil2023autoregressive}. Xu et al. proposed a super-resolution Transformer (SRTT) to recover fine-scale structures of turbulent flows from low-resolution data \cite{xu2023super}.

Despite these successes, the high computational cost of standard attention remains a challenge for 3D simulations. To address this issue, Niu et al. proposed a 3D spatiotemporal model incorporating a channel attention mechanism to predict unsteady airflow over complex three-dimensional aircraft structures \cite{niu2025spatiotemporal}. Li et al. proposed the  factorized Transformer (FactFormer) for 3D PDE simulations via axial factorization \cite{li2023scalable}. Yang et al. achieved stable long-term predictions of 3D turbulence by proposing the implicit factorized Transformer (IFactFormer) \cite{yang2024implicit}. Yang et al. subsequently improved the model by replacing serial attention with a parallel mechanism, which achieves more accurate long-term predictions of 3D turbulent channel flows \cite{yang2026implicit}.

Another effective approach is the vision Transformer (ViT), proposed by Dosovitskiy et al., which reduces computational costs while achieving excellent results in computer vision by partitioning images into patches \cite{dosovitskiy2020image}. Recently, the ViT has been widely applied to fluid mechanics and aerodynamics problems \cite{miotto2023flow,jinhua2025general,sun2024improving,cui2024Transformer,liu2025efficient,zeng2025swin,huo2025aero,wu2024fast}. Deng et al. proposed a modified ViT-based network for the rapid and accurate predictions of transonic flow fields over supercritical airfoils \cite{deng2023prediction}. Ovadia et al. proposed the vision Transformer neural operator (ViTO), which exhibits an excellent predictive accuracy in 2D partial differential equations \cite{ovadia2024vito}. Wang et al. proposed the continuous vision Transformer (CViT) for 2D PDEs \cite{wang2024cvit}.

However, purely data-driven neural operators often suffer from reliance on high-fidelity training data and a lack of physical interpretability. Recently, the physics-informed method has been introduced into neural operators to solve PDEs, including physics-informed DeepONets \cite{wang2021learning} and physics-informed Fourier neural operator \cite{li2024physics}. It has been shown that physics-informed neural operator (PINO) models perform with high accuracy across various differential equations \cite{goswami2023physics,lin2023operator,jiao2024solving,chen2024physics,wang2024beyond,ehlers2025bridging,VINO,zhang2025omnifluids}. By leveraging equation constraints, PINO can learn partial differential equations with minimal or even no data. Zhao et al. proposed the large-eddy simulation nets model (LESnets), which performs excellently in predictions of 3D free shear turbulence and isotropic decaying turbulence without using any labeled data \cite{zhao2025lesnets}. Recently, Lorsung et al. proposed the physics informed token Transformer (PITT) to solve 1D and 2D PDEs \cite{lorsung2024physics}. Due to the increase in sequence length, applying Transformer methods to 3D problems requires significant computational costs. Consequently, we consider the ViT method as an effective solution to handle the computational complexity of 3D problems. 

In this study, we propose the ViT-based physics-informed Transformer operator (PITO) and its implicit variant (PIITO) for predicting 3D turbulence. By partitioning the fluid domain into patches, PITO and PIITO significantly reduce the computational overhead in 3D turbulent flows. Compared to the physics-informed Fourier neural operator (PIFNO), PITO and PIITO demonstrate improved performance in both decaying and forced homogeneous isotropic turbulence.

The organization of the rest of this paper is as follows. Section 2 introduces the governing equations for large-eddy simulation, the architectures of ViTO, PITO, and their implicit variants (IViTO and PIITO). Section 3 presents the numerical results of PITO and PIITO in decaying and forced HIT, compared to PIFNO. Section 4 discusses the impact of key hyperparameters on model performance. Section 5 introduces the automatic optimization strategy for the coefficient of subgrid-scale model. Finally, the conclusions are summarized in Section 6.

\section{Methodology}
\subsection{Governing equations of large eddy simulation}
The Navier-Stokes equations governing incompressible fluid motion are given by \cite{pope}:
\begin{equation}
    \frac{\partial u_i}{\partial x_i} = 0,
\end{equation}
\begin{equation}
\frac{\partial u_i}{\partial t} + \frac{\partial u_i u_j}{\partial x_j} = - \frac{\partial p}{\partial x_i} + \nu \frac{\partial^2 u_i}{\partial x_j \partial x_j} + \mathcal{F}_i.
\end{equation}
Here, $u_i$ denotes the velocity component in the $i$-th direction. $\nu$ is the kinematic viscosity, $p$ is the ratio of pressure to density, and $\mathcal{F}_i$ is the external force component in the $i$-th direction. In this paper, we adopt Einstein's summation convention to handle repeated indices.

Due to the multi-scale nature of turbulence at high Reynolds number, directly resolving all spatial scales requires extremely fine grids, resulting in computational costs that are expensive for practical engineering applications. The LES method applies a spatial filter to the Navier-Stokes equations, enabling the use of coarser grids to resolve large-scale turbulent structures while modeling the unresolved small-scale motions. The filtered governing equations can be written as \cite{pope}:
\begin{equation}
\frac{\partial \bar{u}_i}{\partial x_i} = 0 ,
\label{ns1}
\end{equation}
\begin{equation}
\frac{\partial \bar{u}_i}{\partial t} + \frac{\partial \bar{u}_i \bar{u}_j}{\partial x_j} = - \frac{\partial \bar{p}}{\partial x_i} + \nu \frac{\partial^2 \bar{u}_i}{\partial x_j \partial x_j} - \frac{\partial \tau_{ij}}{\partial x_i} + \bar{\mathcal{F}}_i ,
\label{ns2}
\end{equation}
where $\bar{u}_i$ represents the filtered velocity. $\tau_{ij}$ denotes the unclosed sub-grid scale (SGS) stress defined by $\tau_{ij} = \overline{u_iu_j}-\bar{u}_i\bar{u}_j$. 

A widely used approach to close the SGS stress is the Smagorinsky (SM) model, which is a representative functional model based on the eddy-viscosity hypothesis \cite{smagorinsky1963general,germano1991dynamic}. The SM model assumes that the SGS stress is proportional to the resolved strain rate tensor  $\bar S_{ij}$, formulated as:
\begin{equation}\tau_{ij} - \frac{1}{3}\tau_{kk}\delta_{ij} = -2(C_\text{smag} \Delta)^2 |\bar{S}| \bar{S}_{ij},\end{equation}
where $C_\text{smag}$ is the Smagorinsky coefficient and $\Delta$ is the filter width. $\bar{S}_{ij} = \frac{1}{2} \left( \frac{\partial \bar{u}_i}{\partial x_j} + \frac{\partial \bar{u}_j}{\partial x_i} \right)$ denotes the filtered rate-of-strain tensor. $ |\bar{S}| = \sqrt{2 \bar{S}_{ij} \bar{S}_{ij}} $ represents the characteristic strain rate. 

The integral length scale and the large-eddy turnover time are defined as \cite{pope}:
\begin{equation}
    L_I = \frac{3\pi}{2 (u^\text{rms})^2} \int_{0}^{\infty} \frac{E(k)}{k} dk, \quad \tau = \frac{L_I}{u^\text{rms}},
\end{equation}
respectively. Here, $E(k)$ denotes the energy spectrum. $u^\text{rms}$ is the root-mean-square velocity magnitude. Moreover, the Kolmogorov length scale $\eta$, Taylor length scale $\lambda$, and the Taylor Reynolds number $Re_\lambda$ are respectively defined by: 
\begin{equation}
    \eta = \left( \frac{\nu^3}{\varepsilon} \right)^{1/4}, \quad 
    \lambda = \sqrt{\frac{5\nu}{\varepsilon}} u^\text{rms}, \quad 
    Re_{\lambda} = \frac{u^\text{rms} \lambda}{\sqrt{3}\nu}.
    \label{eq:scales}
\end{equation}
Here, $\epsilon = 2\nu \langle S_{ij} S_{ij} \rangle$ represents the average dissipation rate.

\subsection{Vision Transformer operator}
In recent years, Transformer \cite{vaswani2017attention} based on self-attention has become the preferred models in  natural language processing (NLP). However, it is computationally prohibitive to apply self-attention directly to images due to the quadratic cost associated with the number of pixels. To address this issue, Dosovitskiy et al. proposed the vision Transformer (ViT) \cite{dosovitskiy2020image}. ViT reduces the sequence length by partitioning images into patches, thereby reducing the computational overhead of self-attention. After extensive data pre-training, it achieved superior results compared to convolutional neural networks (CNNs).

\begin{figure*}
    \centering
    \includegraphics[width=0.7\textwidth]{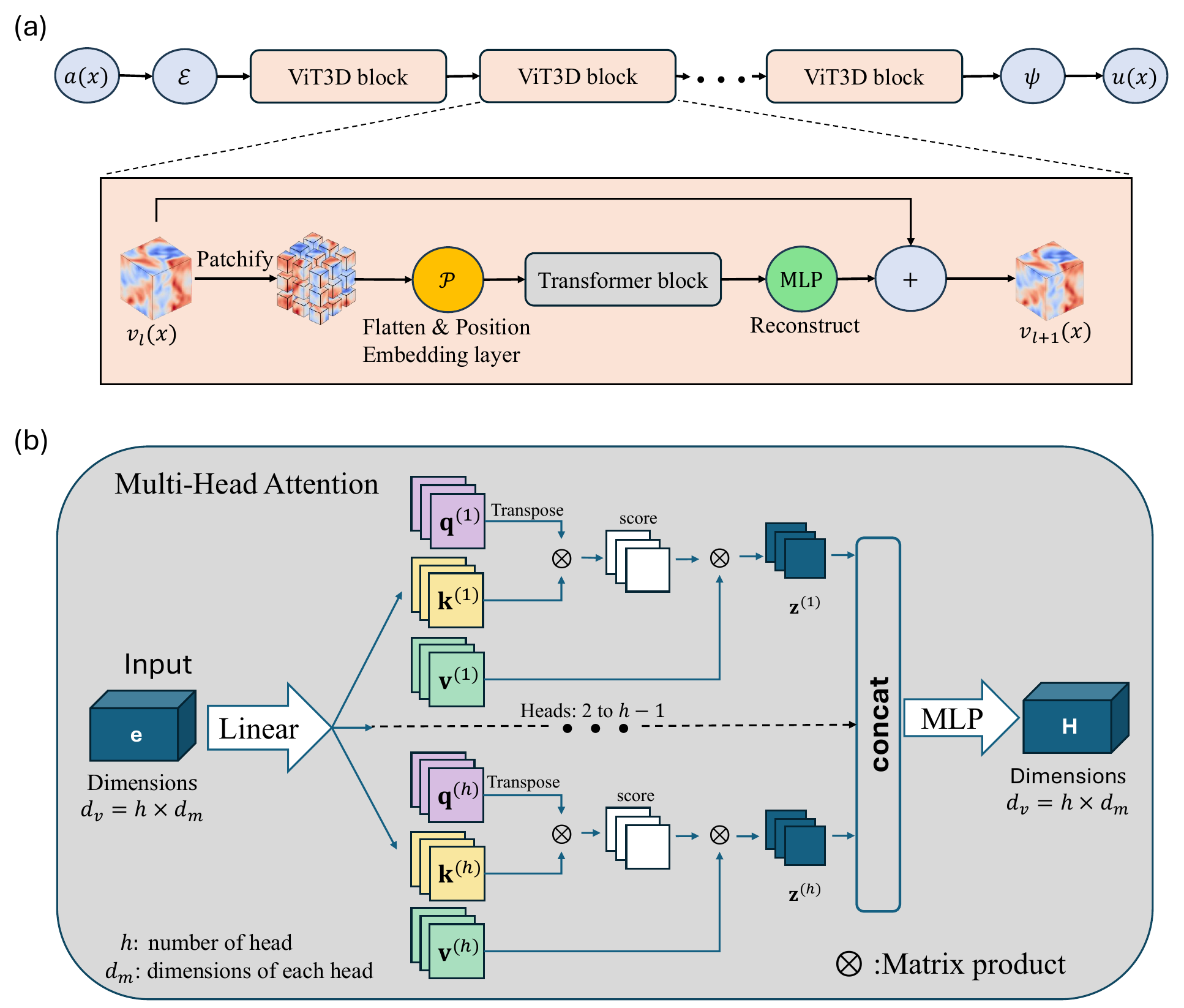} % 调整宽度比例为合适大小
    \caption{(a) The architecture of the ViTO; (b) The Multi-Head attention mechanism in the Transformer block.}
    \label{ViTO}
\end{figure*}

Inspired by the success of ViT, we propose the three-dimensional vision Transformer operator (ViTO) to learn the dynamics of turbulence. To handle the massive computational demands, we introduced a 3D patch partitioning strategy. Specifically, for a three-dimensional flow field with the dimensions $H \times W \times D$, we divide the entire  domain into a series of non-overlapping cubic patches, each with dimensions $P \times P \times P$, where $P$ denotes the patch size. These cubic blocks form the token sequence of the Transformer, significantly reducing the sequence length and addressing the issue of high computational complexity. The model architecture of ViTO is illustrated in Fig.~\ref{ViTO}(a). The input $a(x)$ is first mapped to a high-dimensional latent space via a linear lifting layer $\mathcal{E}$. These feature fields are then iteratively updated through a sequence of ViT3D blocks, where they are partitioned into patches and encoded to capture global dependencies through self-attention mechanisms. Finally, the features from the last block pass through a projection layer $\psi$ to map the high-dimensional features back to the target domain, yielding the predicted flow field $u(x)$.

The key component of ViTO is the self-attention mechanism \cite{dosovitskiy2020image}. Self-attention computes spatial correlations between vectors, and assigns attention weights accordingly, dynamically aggregating physical information from the entire fluid domain. The input 3D fluid field is first partitioned into non-overlapping patches. These patches are subsequently flattened and position embeddings are added to retain the spatial information of the original 3D patches. Then the vectors are linearly projected into a latent vector space of dimension $d_v$ to form the input vectors. Since turbulence is a multiscale complex system, we decompose flow information into $h$ distinct feature components. We set $d_v = d_m/h$, where $h$ is the number of heads. Each head in the $d_m$-dimensional subspace focuses on specific flow feature, which are subsequently integrated to update the global state.

The detailed computation process of multi-head attention is illustrated in Fig.~\ref{ViTO}(b). In the self-attention layer, the input is a sequence of vectors $\mathbf{e} = \{\mathbf{e}_1, \mathbf{e}_2, \dots, \mathbf{e}_N\}$, where $\mathbf{e}_i \in \mathbb{R}^{d_m}$ represents the feature embedding of the $i$-th fluid patch. The input vector $\mathbf{e}_i$ undergoes a linear mapping to yield three feature vectors-Query ($\mathbf{q}_i$), Key ($\mathbf{k}_i$), and Value ($\mathbf{v}_i$)-via learnable projection matrices $W^Q$, $W^K$, and $W^V$.
The transformation is formulated as: 
\begin{equation} 
\mathbf{q}_i = \mathbf{e}_i W^Q, \quad \mathbf{k}_i = \mathbf{e}_i W^K, \quad \mathbf{v}_i = \mathbf{e}_i W^V,
\end{equation}
where $W^Q$, $W^K$ and $W^V \in \mathbb{R}^{d_m \times d_v}$ are the learnable weight matrices of the models. 

The model calculates the relationships between all fluid patches based on the Query and Key vectors. The correlation between the $i$-th fluid patch and the $j$-th fluid patch is computed as follows:
\begin{equation}
\hat{A}_{ij} = \mathbf{q}_i \mathbf{k}_j^T.
\end{equation}

To ensure numerical stability and prevent gradient saturation, the relevance score will be scaled by a factor of $1/\sqrt{d_k}$, where $d_k$ is the dimension of the query and key vectors. These scores are then normalized using the softmax function to obtain the final attention weights.
\begin{equation}
A_{ij} = \operatorname{softmax}_j \left( \frac{\hat{A}_{ij}}{\sqrt{d_k}} \right) = \frac{\exp\left( \hat{A}_{ij} / \sqrt{d_k} \right)}{\sum_{m=1}^{N} \exp\left( \hat{A}_{im} / \sqrt{d_k} \right)} .
\end{equation}

Subsequently, the layer integrates the global context into the local representation. The output $\mathbf{z}_i$ is derived by taking the expectation of the value vectors $\mathbf{v}_j$ with respect to the learned probability distribution $A_{ij}$.
\begin{equation}
\mathbf{z}_i = \sum_{j=1}^{N} A_{ij} \mathbf{v}_j.
\label{eq:weighted_sum}
\end{equation}

The proposed ViTO architecture functions as a neural operator approximating the solution to the governing PDEs. Let $D \in \mathbb{R}^d$ denote a bounded open set; $\mathcal{A} = \mathcal{A}(D;\mathbb{R}^{d_a})$ and $\mathcal{U}=\mathcal{U}(D;\mathbb{R}^{d_u})$ are the separable Banach spaces of functions. The goal is to learn the global solution operator $\mathcal{G}: \mathcal{A} \to \mathcal{U}$ that models the evolution of the turbulent velocity field. In this study, given the current flow field, the ViTO computes its prediction for the next time step.

To approximate the nonlinear solutions operator $\mathcal{G}$, the network is constructed as an iterative architecture. Each layer updates the feature field $v_l$ via a kernel integral operator $\mathcal{K}$, defined generally as \cite{zhao2025lesnets,li2023long,li2020fourier}:
\begin{equation}
(\mathcal{K}v_l)(x) = \int_{D} \kappa\Big(x, y, v_l(x), v_l(y); \theta\Big) v_l(y) dy, \quad \forall x \in D.
\end{equation}
Here, $v_l(x)$ and $v_l(y)$ represent the feature state at the target position $x$ and source position $y$, respectively. $\kappa$ measures the influence exerted by position state at $y$ on position state at $x$. $\theta$ represents learnable parameters in each layer.

In ViTO, the attention mechanism serves as a direct parameterization of the integral operator $\mathcal{K}$. The continuous integral operator is discretized and parameterized via the attention mechanism as follows \cite{yang2026implicit}:
\begin{equation}
\begin{split}
\mathbf{z}_i &= \sum_{j=1}^{N} A_{ij} \mathbf{v}_j \\
&\approx \int_{D} \kappa\Big(x, y, v_l(x), v_l(y); \theta\Big) v_l(y) dy, \quad \forall x \in D.
\end{split}
\label{eq:dengxiao op}
\end{equation}

Specifically, the integral over D is approximated by the sum of vectors representing discrete patches:$\int_D (\cdot) dy \approx \sum_{j} (\cdot)$. The term $v_l(y)$ corresponds to the Value vector $\mathbf{v}_j$ of the source patch. Most importantly, the kernel $\kappa\Big(x, y, v_l(x), v_l(y)\Big)$ is dynamically computed as the attention weight $A_{ij}$. The attention mechanism of ViTO achieves global function mapping by calculating the correlation between Query ($\mathbf{q}_i$) and Key ($\mathbf{k}_i$). 

Finally, to recover the flow fields from the latent feature space, we employ a simple linear reconstruction layer. Subsequently, the feature vectors are reshaped back into the three-dimensional physical space.

Based on this operator definition, ViTO architecture adopts an explicit iterative strategy. The model consists of a stack of $L$ Transformer layers, where each layer $l$ possesses a distinct set of learnable parameters $\theta_l$. The update rule for the feature field from layer $l$ to $l+1$ is given by \cite{yang2026implicit}:
\begin{equation}
v_{l+1} = {v}_l + \sigma\big(\mathcal{K}({v}_l; \theta_l)\big).
\end{equation}
Here, $\sigma$ represents nonlinear activation and $\theta_l$ denotes the layer-specific learnable parameters for the $l$-th Transformer block. Stacking these layers with independent parameters enhances the expressive power.

\begin{figure*}
    \centering
    \includegraphics[width=0.85\textwidth]{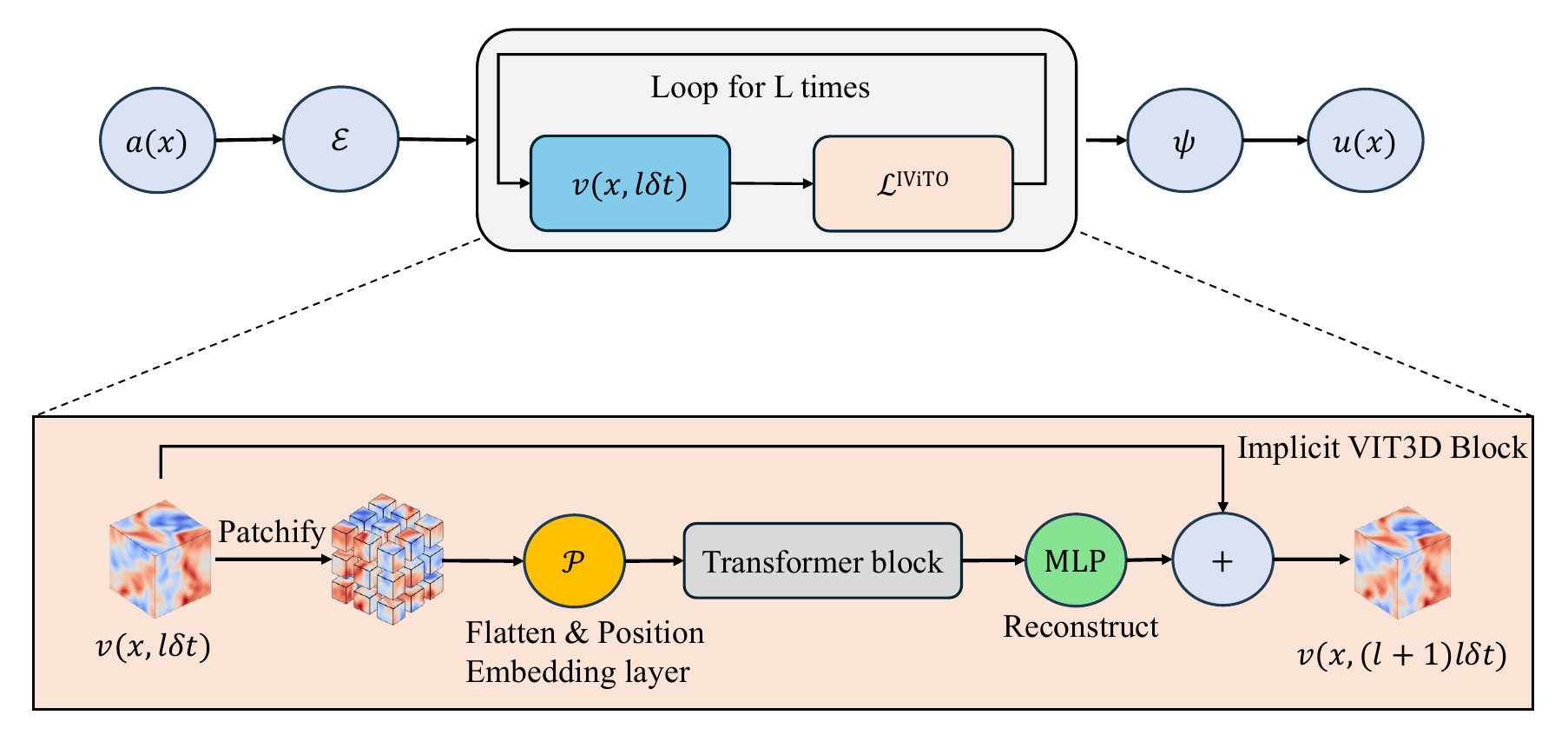} % 调整宽度比例为合适大小
    \caption{The architecture of the IViTO}
    \label{IViTO}
\end{figure*}

Although stacking independent Transformer modules enhances the model's depth and expressive power, it inevitably introduces parameter redundancy. To further enhance parameter efficiency, we introduced an implicit variant of the model, termed IViTO. The architecture of the IViTO is illustrated in Fig.~\ref{IViTO}. Unlike stacking $L$ standard blocks with independent parameters as shown in Fig.~\ref{ViTO}(a), IViTO employs a weight-sharing mechanism that iteratively applies the same vision Transformer operator \cite{el2021implicit,bai2019deep}. Its iteration process can be written as \cite{yang2026implicit}:
\begin{equation}
v(x,(l+1)\delta t) = {v}(x,l\delta t) + \delta t \cdot \sigma\big( \mathcal{K}({v}(x,l\delta t); \theta) \big).
\label{eq:ivito_iteration}
\end{equation}
In this implicit iteration, $\delta t= 1/L$ denotes the time step size for each iterations, while $\theta$ denotes the shared learnable parameters employed across the $L$ iterations.

\subsection{Physics-Informed Transformer operator}
Although neural operators have powerful approximation capabilities, relying solely on data-driven approaches requires substantial amounts of high-quality datasets and cannot guarantee that predicted flow fields comply with fundamental physical laws. To address this issue, we propose the physics-informed Transformer operator (PITO) and its implicit variant (PIITO), which directly embed governing partial differential equations (PDEs) as the loss function. To further reduce reliance on data, the loss function does not incorporate data loss. Here, we introduce the governing equations of LES into the PDE loss, with the loss function taking the following form in decaying HIT:
\begin{equation}
\begin{split}
    &\mathcal{L}_1 = \| \nabla \cdot \bar{\mathbf{u}} \|_{L^2(T;D)}^2, \\
    &\mathcal{L}_2 = \| \partial_t \bar{\mathbf{u}} + \bar{\mathbf{u}} \cdot \nabla \bar{\mathbf{u}} + \nabla \bar{p} - \nu \nabla^2 \bar{\mathbf{u}} - \nabla \cdot \boldsymbol\tau \|_{L^2(T;D)}^2, \\
    &\mathcal{L}_{PDE} = \lambda_1\mathcal{L}_1 + \lambda_2\mathcal{L}_2,
\end{split}
\label{eq:physics_loss}
\end{equation}
where $\bar{\mathbf{u}}$ denotes the filtered velocity and the forcing term $\bar{\mathcal{F}} = 0$. Given an initial field $\bar{\mathbf{u}}_1$, the model predicts a sequence of time steps $\bar{\mathbf{u}}_{2}, \dots, \bar{\mathbf{u}}_{T}$, and the PDE loss $\mathcal{L}_{PDE}$ is then computed over the entire sequence $\{\bar{\mathbf{u}}_1, \dots, \bar{\mathbf{u}}_T\}$. Only the initial field needs to be provided, which can be generated randomly or from numerical simulations. Significant disparities in the numerical magnitudes of the loss terms can lead to gradient competition, where the optimization is dominated by the larger term. In this study, we set the ratio of the weights $\lambda_1$ to $\lambda_2$ to 1:1. The time derivative term is obtained by a finite difference method:
\begin{equation}
\frac{\partial \bar{\mathbf{u}}}{\partial t} \bigg|_{t=t_n} \approx 
\begin{cases} 
\displaystyle \frac{\bar{\mathbf{u}}_{n+1} - \bar{\mathbf{u}}_n}{\Delta t}, & n = 1, \\[12pt]
\displaystyle \frac{\bar{\mathbf{u}}_{n+1} - \bar{\mathbf{u}}_{n-1}}{2\Delta t}, & 2 \le n \le T-1 ,\\[12pt]
\displaystyle \frac{\bar{\mathbf{u}}_{n} - \bar{\mathbf{u}}_{n-1}}{\Delta t}, & n = T.
\end{cases}
\end{equation}
Here, $n$ denotes the index of time step, $\Delta t$ is the time step size, $t_n = n \cdot \Delta t$ represents the time at the $n$-th step and $T$ is the total number of time steps.

In numerical simulations of statistically steady-state turbulence, the external forcing term $\mathcal{F}$ is typically implemented through spectral space manipulations, scaling velocity modes in the low-wavenumber region to preserve the target energy spectrum \cite{eswaran1988examination}. Let $\hat{\mathbf{u}}$ denote the velocity field in spectral space. We define rescaling operator $\mathcal{R}_f$ formulated as \cite{zhao2025lesnets}:
\begin{equation}
\begin{split}
    & \hat{\mathbf{u}}^f(k) = \alpha(k) \hat{\mathbf{u}}(k), \\ % & 在最前面表示左对齐
    & \text{where } \alpha(k) = 
    \begin{cases} 
        \sqrt{E_0(1)/E(1)}, & 0.5 \le k \le 1.5, \\ 
        \sqrt{E_0(2)/E(2)}, & 1.5 \le k \le 2.5, \\ 
        1, & \text{otherwise}.
    \end{cases}
\end{split}
\label{eq:forcing}
\end{equation}
Here, $\hat{\mathbf{u}}^f$ denotes the forced velocity field in spectral space, and $k$ represents the wavenumber magnitude. $\mathcal{R}_f$ performs a forward Fourier transform, applies the spectral scaling $\alpha(k)$ to the low-wavenumber modes, and transforms the field back to physical space.

This forcing mechanism is essentially a global constraint on the energy shell, rather than a local additive source term in the physical space governing equations, and can be rewritten as follows:

\begin{equation} 
\bar{\mathbf{u}}_{t}^f = \mathcal{R}_{f} \left[ \bar{\mathbf{u}}_{t-1} + \Delta t \left( {-\bar{\mathbf{u}}_{t}} \cdot \nabla \bar{\mathbf{u}}_{t}  -\nabla \bar{p} + \nu \nabla^2 \bar{\mathbf{u}}_{t} + \nabla \cdot \boldsymbol\tau \right) \right].
\end{equation}

Therefore, the forced PDE loss function for stationary HIT $\mathcal{L}_{fPDE}$ is given by:
\begin{equation} 
\mathcal{L}_{PDE}^{f} = \left\| \bar{\mathbf{u}} - \bar{\mathbf{u}}^f \right\|_{L^2(T; D)}^2.
\end{equation}
Since $\mathcal{R}_f$ explicitly projects the velocity field onto a divergence-free space, the mass conservation constraint is inherently satisfied, eliminating the need for an additional continuity loss term.
\begin{figure*}
    \centering
    \includegraphics[width=\textwidth]{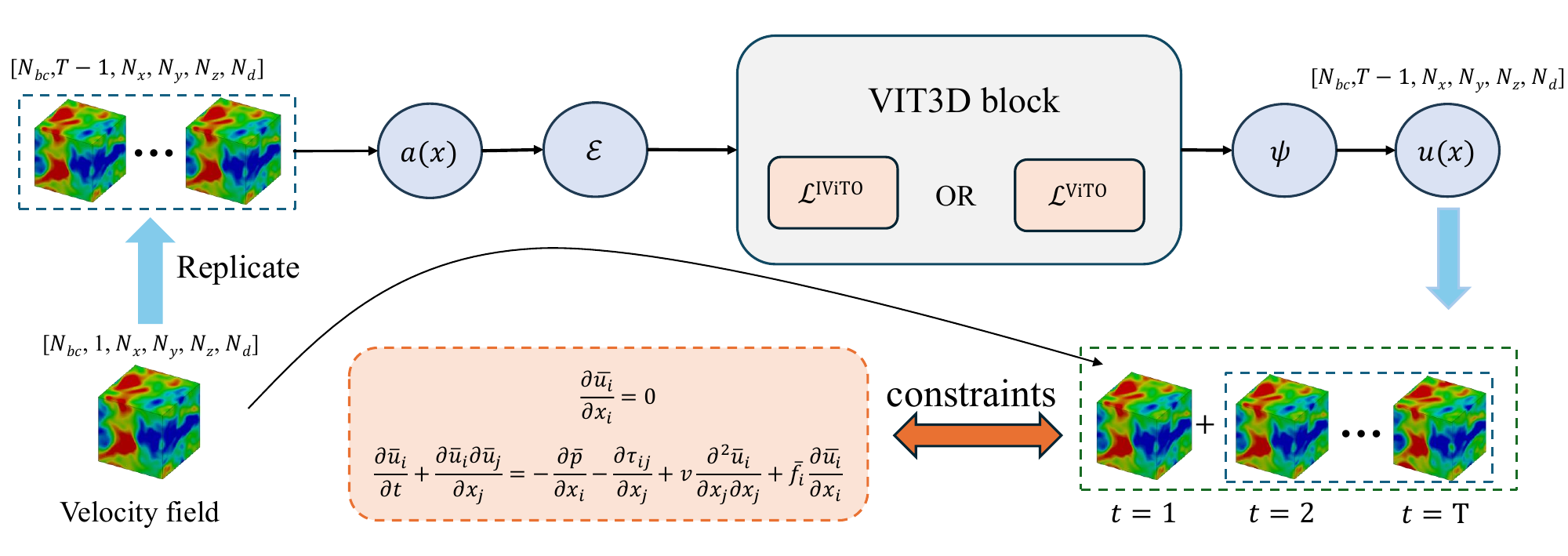} % 调整宽度比例为合适大小
    \caption{The architecture of the PITO and PIITO frameworks constrained by LES equations.}
    \label{model}
\end{figure*}

The architecture of the PITO and PIITO constrained by LES is shown in the Fig.~\ref{model}. The neural operator learns a mapping from an instantaneous flow field $\bar{\mathbf{u}}_t$ to a sequence of future states $\bar{\mathbf{u}}_{t+1}, \dots, \bar{\mathbf{u}}_{t+T}$. To achieve continuous long-term predictions, the final predicted state $\bar{\mathbf{u}}_{t+T}$ is recursively fed back into the model as the new initial condition for the subsequent iteration. During the training process, the model takes only the instantaneous velocity field $\bar{\mathbf{u}}_t$ as input, with positional information concatenated during the data preprocessing stage. Finally, the input ${\bar{\mathbf{u}}}_t$ and the predicted sequence are combined to compute the PDE loss.

While physics-informed Transformer models have rarely been applied to three-dimensional problems, the physics-informed Fourier neural operator (PIFNO) has demonstrated a effective performance for 3D turbulence in our previous study \cite{zhao2025lesnets}. The implementation details of the surrogate model FNO are provided in Appendix A.

\section{Numerical results}
In this section, we conduct a series of numerical simulations to assess the performance of the PITO, PIITO, and PIFNO models in the predictions of three-dimensional decaying and forced homogeneous isotropic turbulence. The performance of these physics-informed models is directly compared to the traditional LES with Smagorinsky model and the filtered DNS (fDNS) data.
\subsection{Decaying homogeneous isotropic turbulence at statistically stationary initial condition}
The direct numerical simulation of decaying homogeneous isotropic turbulence (HIT) is conducted in a $(2\pi)^3$ cubic domain under periodic boundary conditions, employing a uniform grid resolution of $128^3$ \cite{zhao2025lesnets}. The pseudo-spectral method is applied for spatial discretization of the governing equations, coupled with a second-order Adams-Bashforth scheme for time integration \cite{canuto2007spectral,peyret2002spectral}. In addition, high-wavenumber Fourier modes are truncated in accordance with the two-thirds rule to suppress aliasing. 

The statistically stationary initial condition for the decay HIT is obtained from a fully developed turbulent flow sustained by external forcing. In order to maintain a stationary turbulent velocity field, large-scale forcing is applied by holding the velocity spectrum in the two lowest wavenumber shells fixed at the values $E_0(1) = 1.242477$ and $E_0(2) = 0.391356$. The kinematic viscosity is set to $\nu = 0.015625$ and the corresponding Taylor Reynolds number is $Re_\lambda \approx 60$. The DNS time step $\Delta t$ is set to 0.001. The simulation is initialized with a random velocity field following 
a Gaussian distribution in spectral space. We maintain 10,000 time steps using the forcing term until the flow becomes statistically steady. Thereafter, the forcing term is removed, allowing the turbulence to decay over 5000 time steps, enabling the development of isotropic decay turbulence \cite{zhao2025lesnets}.

The DNS data are filtered to obtain large-scale flow fields with the sharp spectral filter. The filtering is performed on a $32^3$ grid with a cutoff wavenumber of $k_c=10$ \cite{zhao2025lesnets}. Here, we generated 5,000 stationary initial turbulent velocity fields for training and 20 samples for testing. We save the numerical solution for each time node at every 20 DNS time steps. To evaluate the performance and generalization ability of the model, the \textit{a posteriori} analysis employs five independent benchmark cases with different initial conditions, where each simulates a complete 5000-step decay period of homogeneous isotropic turbulence. We use the high-fidelity fDNS data for both training and testing. The data are formatted as tensors of size [$N_t$,32,32,32,3], where $N_t$ is the number of input time steps. The physics-informed models are trained on a dataset of 5000 samples, each containing one ($N_t=1$) velocity field at the time $t=10000\Delta t$. The test set consists of 20 additional samples. Each sample is a sequence of length $N_t=11$, containing velocity fields from $t=10,000\Delta t$ to $t=10,200\Delta t$. The detailed simulation parameters are provided in Table 1 \cite{zhao2025lesnets}.

\begin{table*}
    \centering
    \caption{Parameters for DNS and fDNS of decaying HIT at stationary initial condition.}
    \label{tab:parameters}
    \begin{tabular}{ c c c c c c c c }
        \toprule
        {Res.(DNS)} & {Res.(fDNS)} & {Domain} & ${Re_\lambda}$ & ${\nu}$ & ${\Delta t}$ & ${k_c}$ & ${\tau}$ \\
        \midrule
        $128^3$ & $32^3$ & $(2\pi)^3$ & 60 & 0.015625 & 0.001 & 10 & 1.00 \\
        \bottomrule
    \end{tabular}
\end{table*}

In this study, the physics-informed models are constrained by the LES equations \eqref{ns1} and \eqref{ns2} closed with the Smagorinsky model (Smagorinsky coefficient $C_\text{smag}= 0.1$). Both Transformer neural operators use the same patch size of 4. The model employs the lifting layer $\mathcal{E}$ with $d_v=128$ and $L=10$ vision Transformer layers. Moreover, in the attention layer of the Transformer, the number of attention heads is set to $h=17$, where each head has a dimension of $d_m = 60$. For the PIFNO, we utilize the parameters that performed best in numerical simulations: the Fourier modes are $k_{\max}$ = 12 and the number of Fourier layers and  the channel dimension are set to 6 and 80, respectively. The configurations of physics-informed models correspond to the best performance achieved by each model.

\begin{table*}
    \centering
    \caption{The training hyperparameters of decaying HIT at stationary initial condition.}

    \begin{tabular}{c c c c c c}
        \toprule
        {Epochs} & {Batch size} & {Learning rate} & {Decay period} & {Decay factor}\\
        \midrule
        30000 & 1 & 0.001 &  (4000,10000,20000) & 0.1 \\
        \bottomrule
    \end{tabular}
    \label{para}
\end{table*}

The training hyperparameters for all data-driven models are summarized in Table \ref{para}. The decay period values (4000, 10000, 20000) represent a learning rate schedule \cite{zhao2025lesnets}. The learning rate is multiplied by the decay factor at these designated epochs to refine the optimization process. Here, we utilize the Adam optimizer for training and the GELU function for activation \cite{kingma2014adam,hendrycks2016gaussian}. Physics-Informed models minimize the PDE loss defined by \eqref{ns1} and \eqref{ns2}.

\begin{figure*}[!htbt]
    \centering
    % (a) FNO
    \includegraphics[width=.38\textwidth]{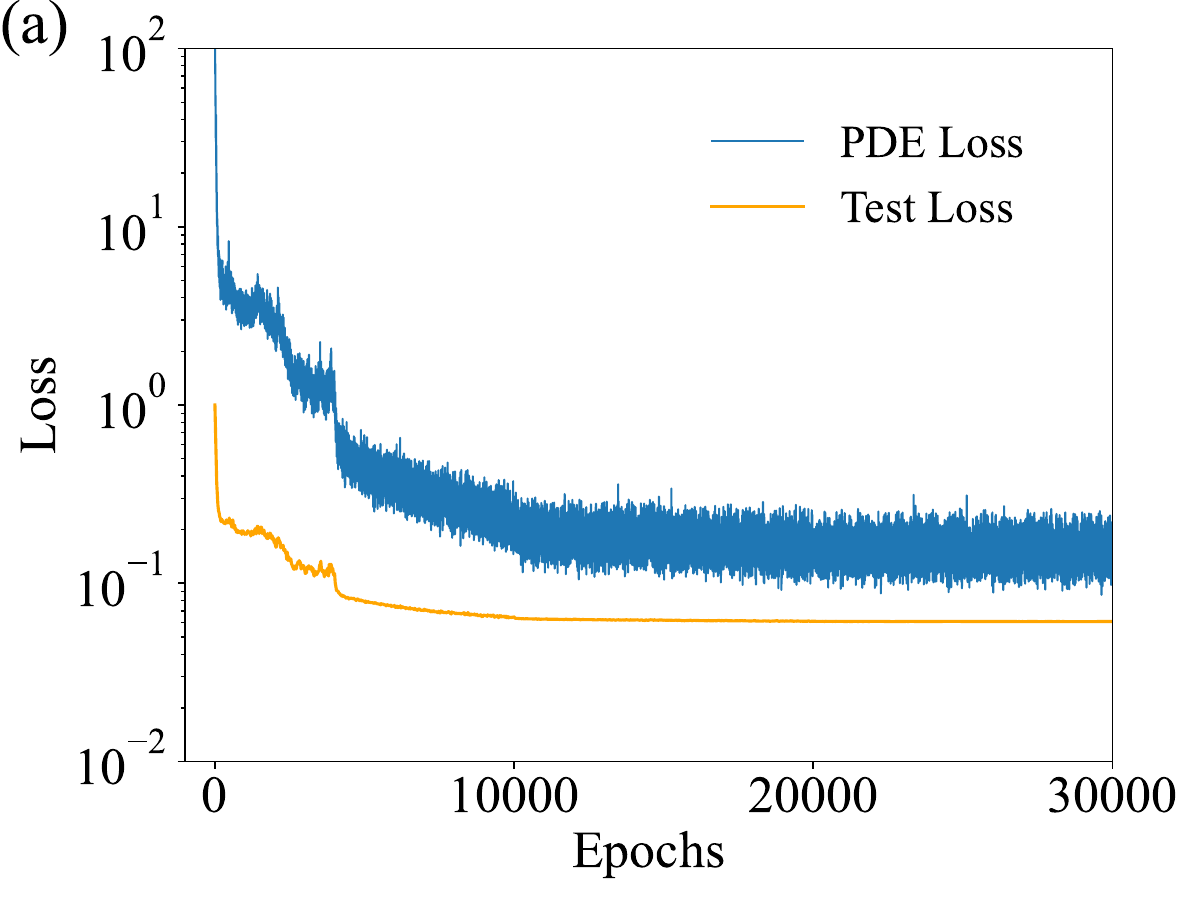}
    \includegraphics[width=.38\textwidth]{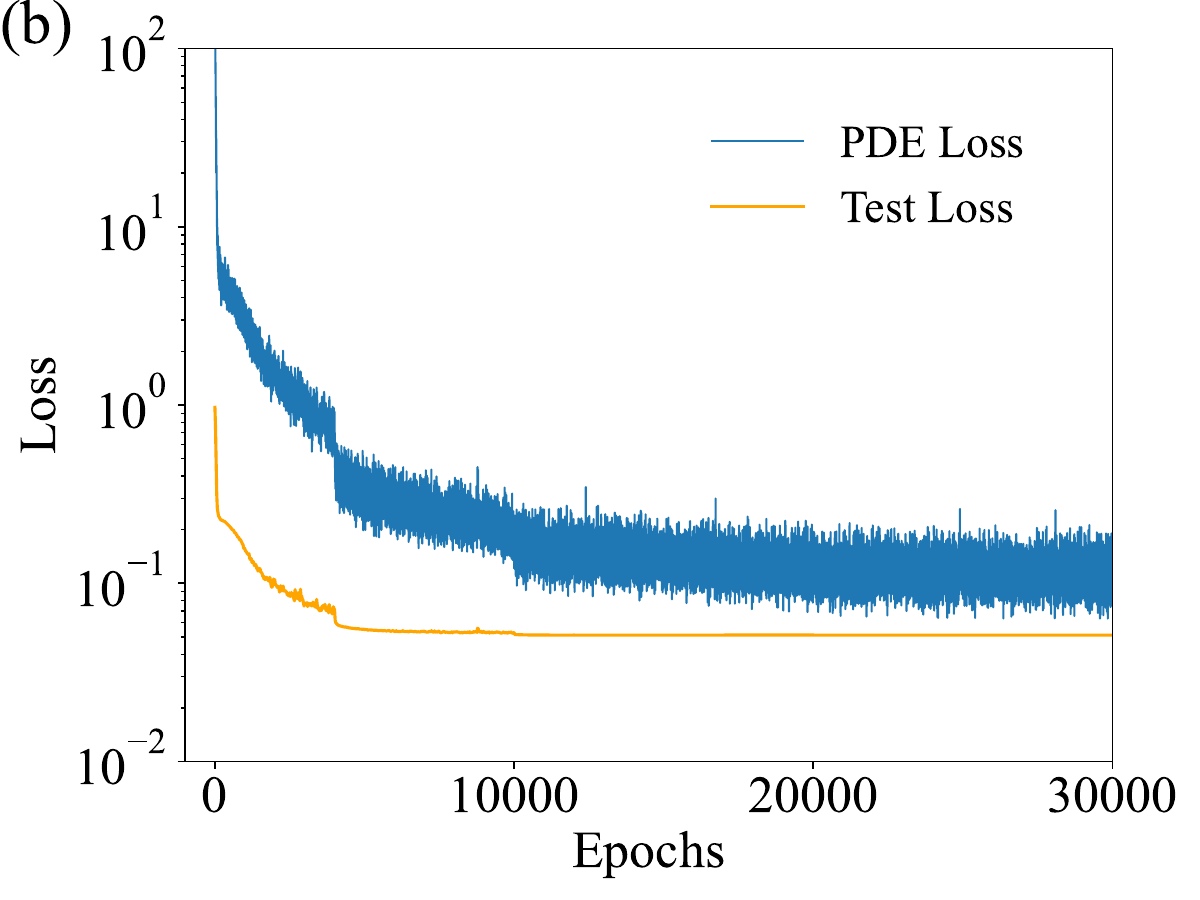}
    \includegraphics[width=.38\textwidth]{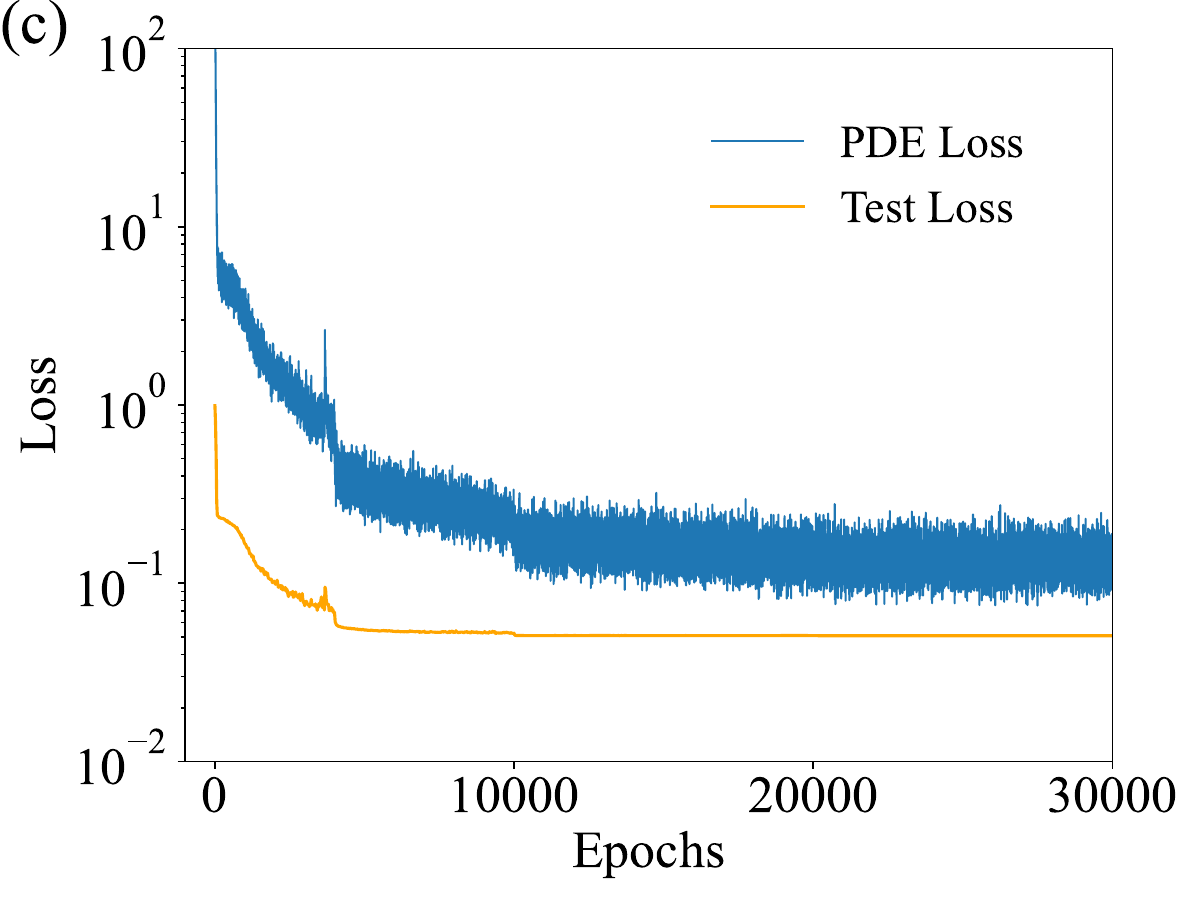}
    \includegraphics[width=.38\textwidth]{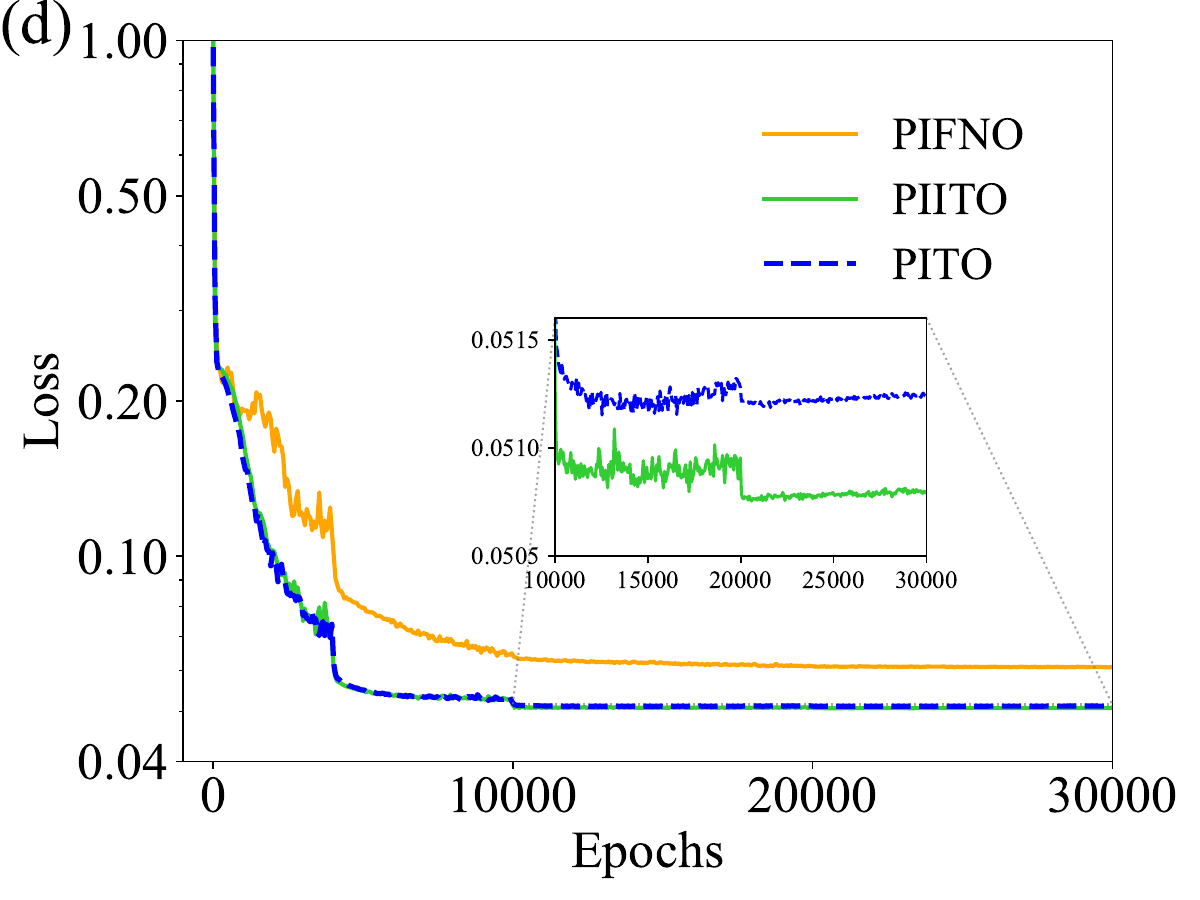}
    \caption{Evolutions of loss curves in decaying HIT at the stationary initial condition: (a) PIFNO; (b) PITO; (c) PIITO and (d) comparison of test loss during training.}
    \label{fig:dhit_loss_evolutions}
\end{figure*}

Fig.~\ref{fig:dhit_loss_evolutions} compares the training and testing loss curves. It can be seen that PDE loss starts at a significantly higher value (near $10^2$), while the test loss initializes around $10^0$ and exhibits a rapid decaying. PITO and PIITO share similar trends with PIFNO, and both models achieve lower test losses as compared to PIFNO.

\begin{table*}
    \centering
    \caption{Training efficiency and minimum losses of different models in decaying HIT at stationary initial condition.}
    \label{tab:loss_comparisonDHIT}
    \begin{tabular}{cccc} % 列格式：左对齐，四个列居中
        \toprule
        {Model} & {Training (s/epoch)} & {PDE Loss} & {Test Loss} \\
        \midrule
        PIFNO     & 0.39 & 0.08624      & 0.06090 \\
        PITO       & 0.18 & 0.06342      & 0.05115 \\
        PIITO       & 0.16  & 0.07494 & 0.05075 \\
        \bottomrule
    \end{tabular}
\end{table*}

Table \ref{tab:loss_comparisonDHIT} summarizes the minimum training, minimum testing loss and the training time per epoch achieved by the physics-informed models. Both PITO and PIITO achieve a twofold speedup in training compared to PIFNO; specifically, PIITO exhibits the highest training efficiency. Moreover, the PITO and PIITO achieve lower PDE and testing losses compared to PIFNO. These results demonstrate that the architectures based on Transformer offer superior representation capabilities compared to the Fourier-based approach.

In the \textit{a posteriori} test, each model employs the same five statistically steady turbulent velocity fields as the benchmark fDNS. Through iterative inference, we obtained the velocity fields extrapolated over 250 time steps (i.e. $N_t=250$), while the training of the physics-informed models is restricted to the initial $N_t = 11$ time steps. The statistical results in our study are obtained by ensemble averaging predictions from five different initializations. The learning time step of the model is $\Delta t_n = t_{n+1} - t_n = 20 \Delta t$, while predicting 10 steps forward, covering a duration of $0.2\tau$.

\begin{figure*}
    \centering
    % (a) FNO
    \includegraphics[width=.38\textwidth]{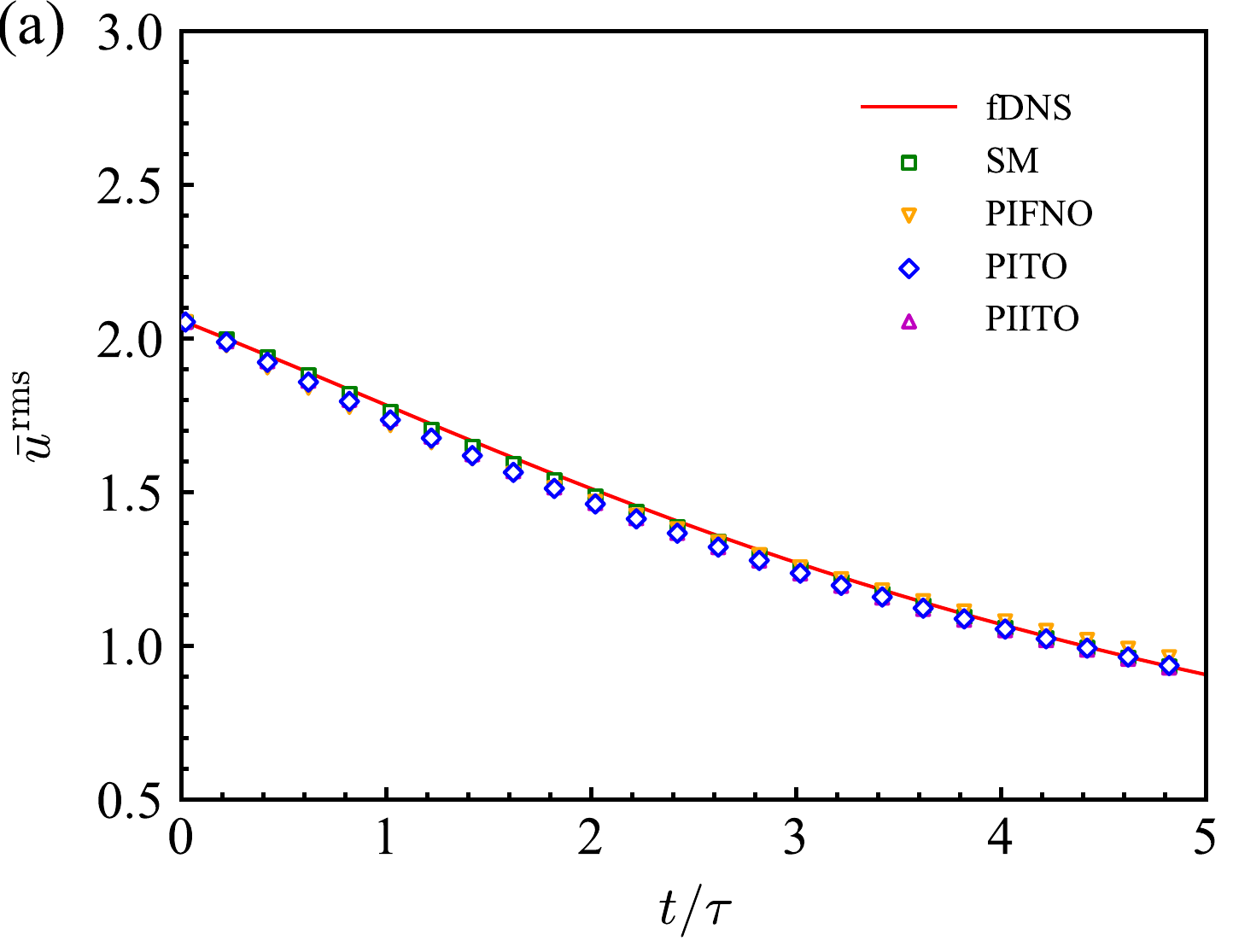} 
    \includegraphics[width=.38\textwidth]{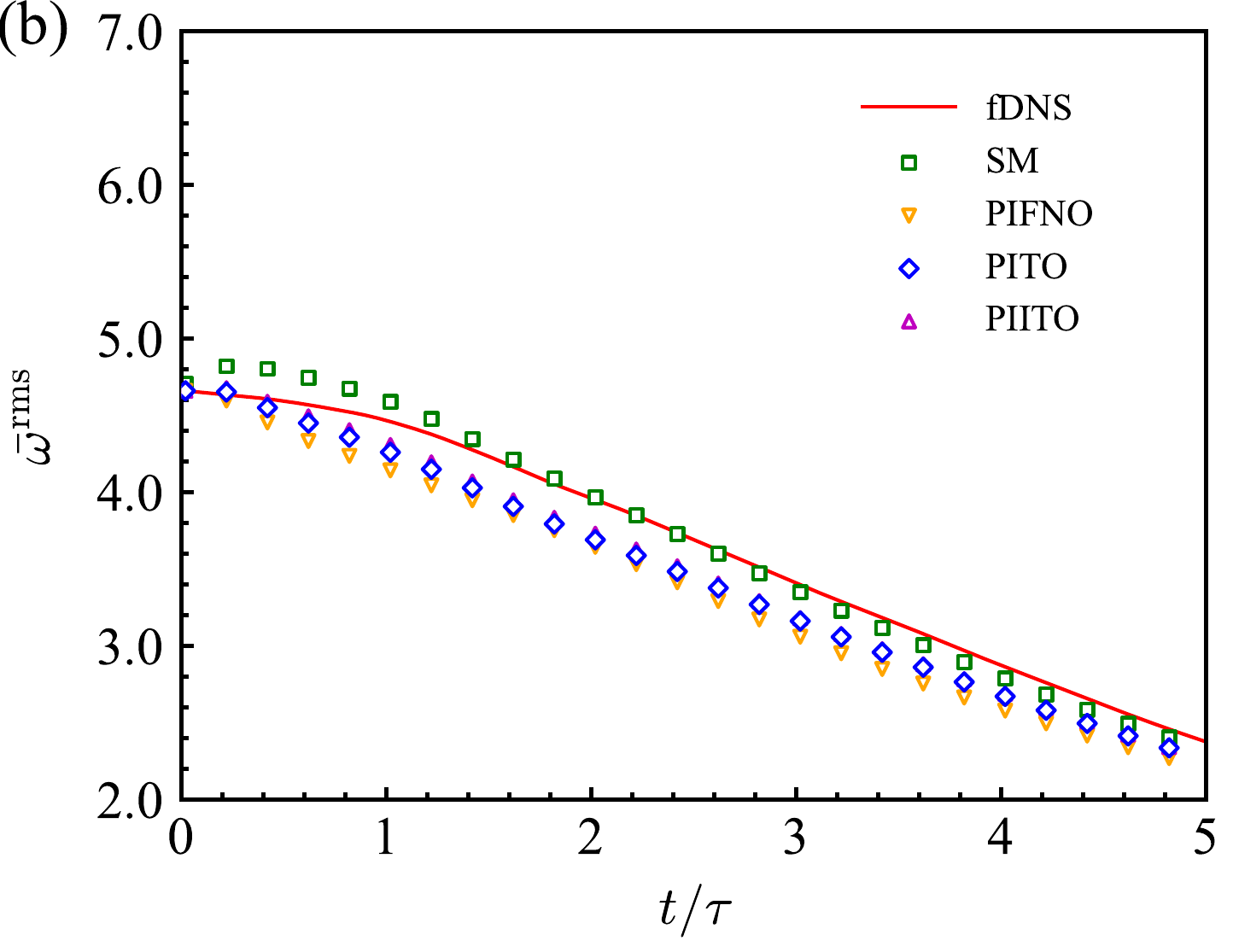}
    \caption{Temporal evolutions of (a) the root-mean-square (rms) velocity and (b) the rms of vorticity for different models in decaying HIT at the stationary initial condition.}
    \label{DHIT u w}
\end{figure*}

Fig.~\ref{DHIT u w} shows the temporal evolutions of the root mean square (rms) values of velocity and vorticity. All physics-informed models accurately capture the decay trend of velocity. Specifically, compared to the PIFNO, the results of PITO and PIITO are generally closer to fDNS and SM benchmarks when predicting the decay of vorticity and velocity. All models exhibit a stable long-term performance over the extended period up to $N_t$ = 250.

\begin{figure*}
    \centering
    \includegraphics[width=.38\textwidth]{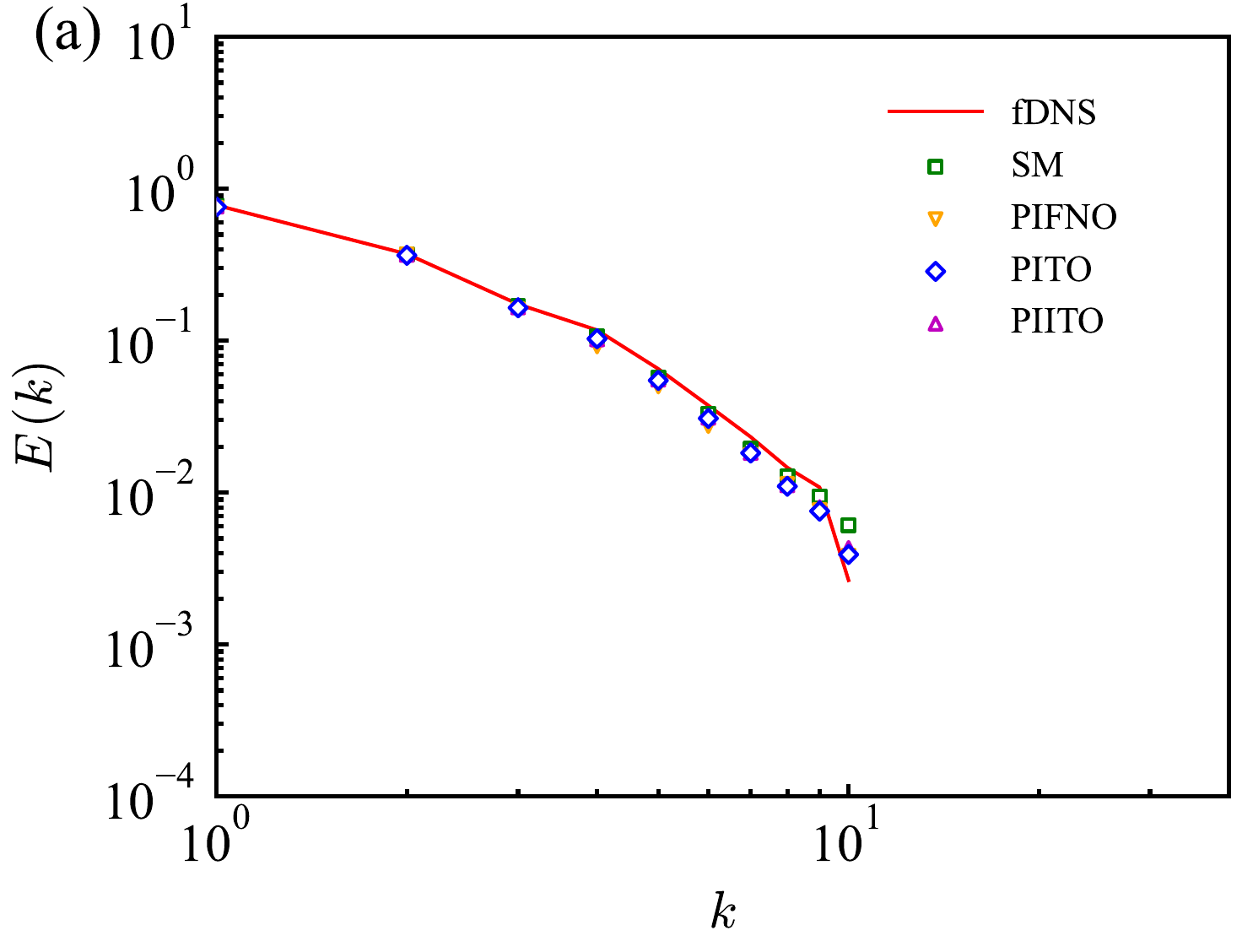}
    \includegraphics[width=.38\textwidth]{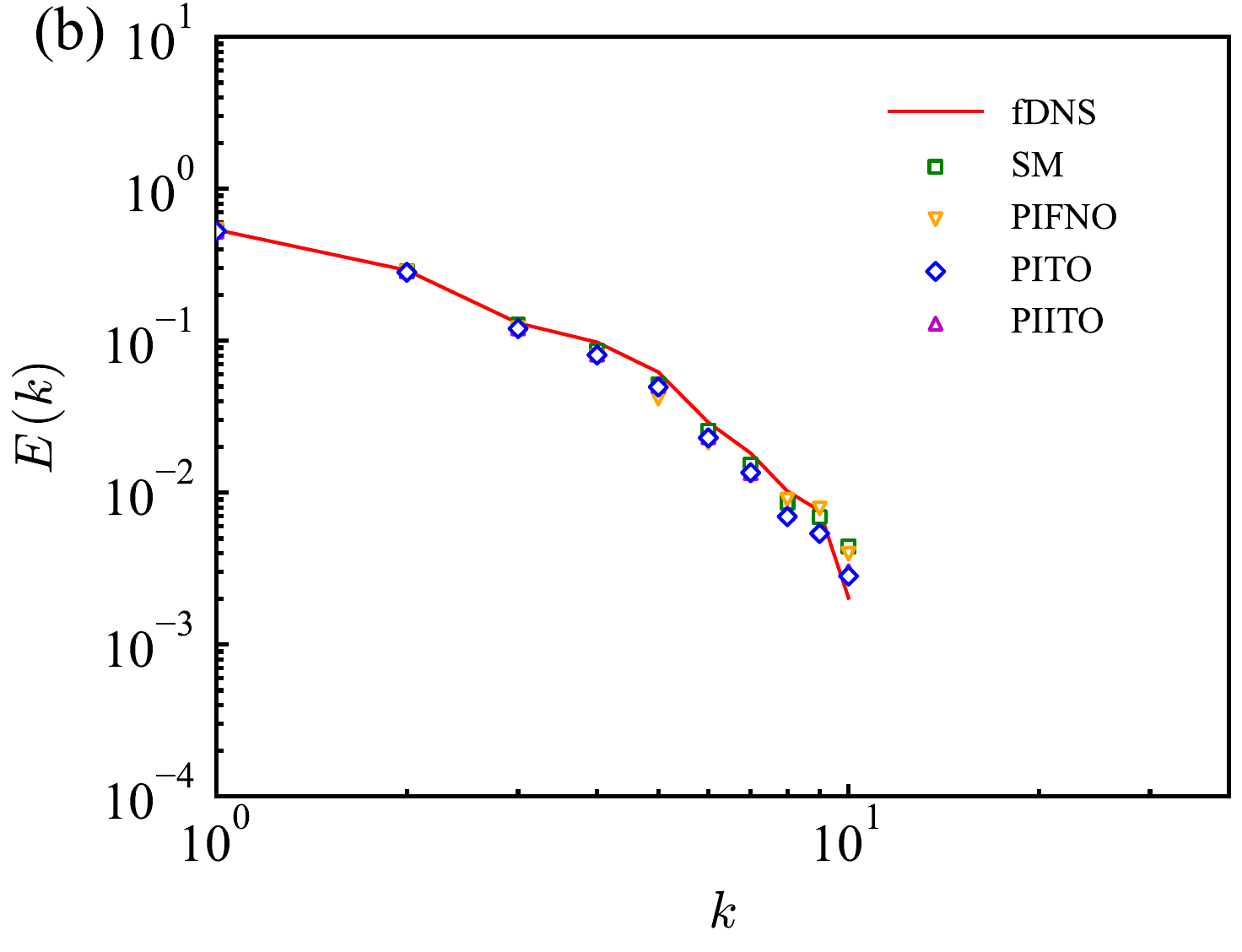}
    \includegraphics[width=.38\textwidth]{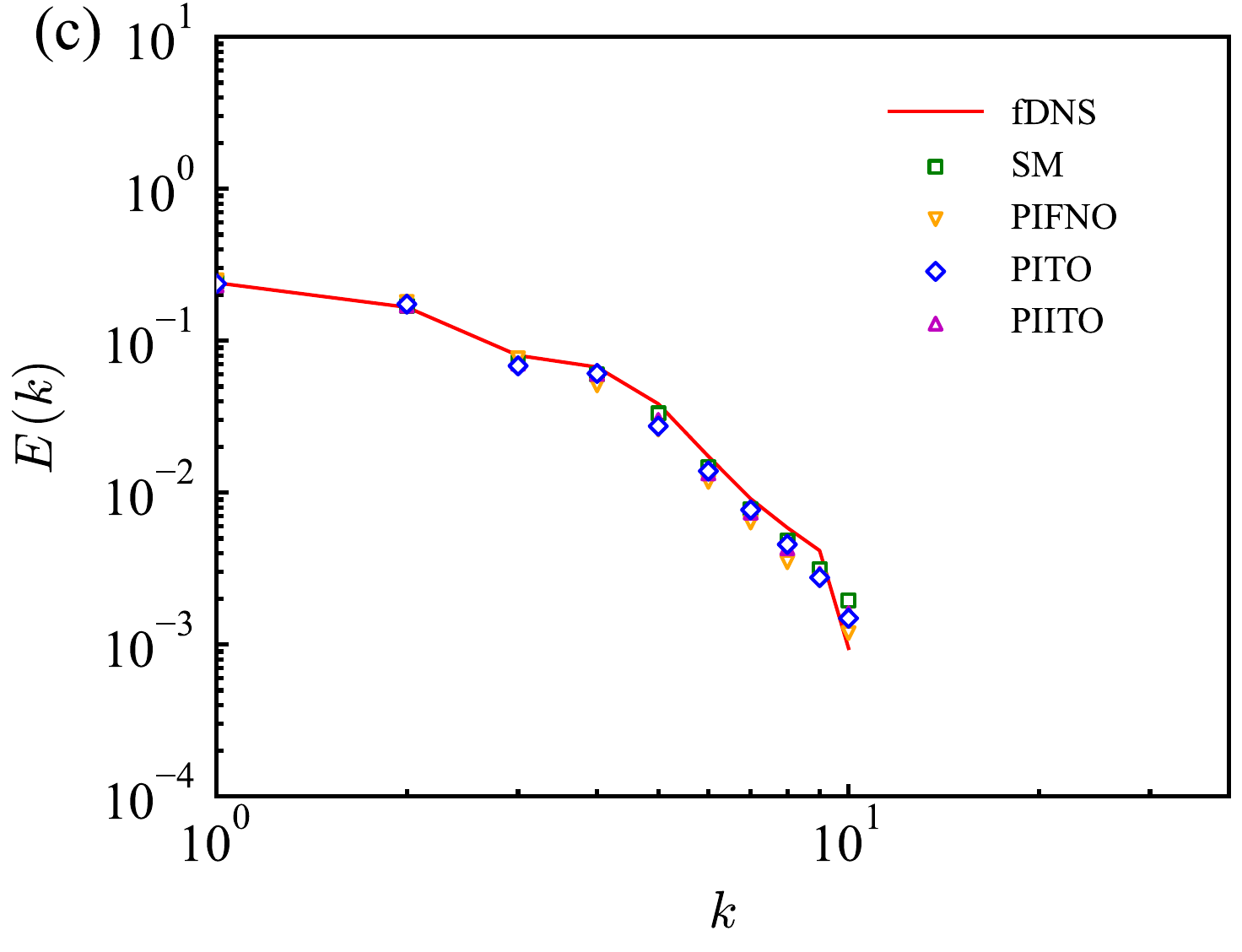}
    \includegraphics[width=.38\textwidth]{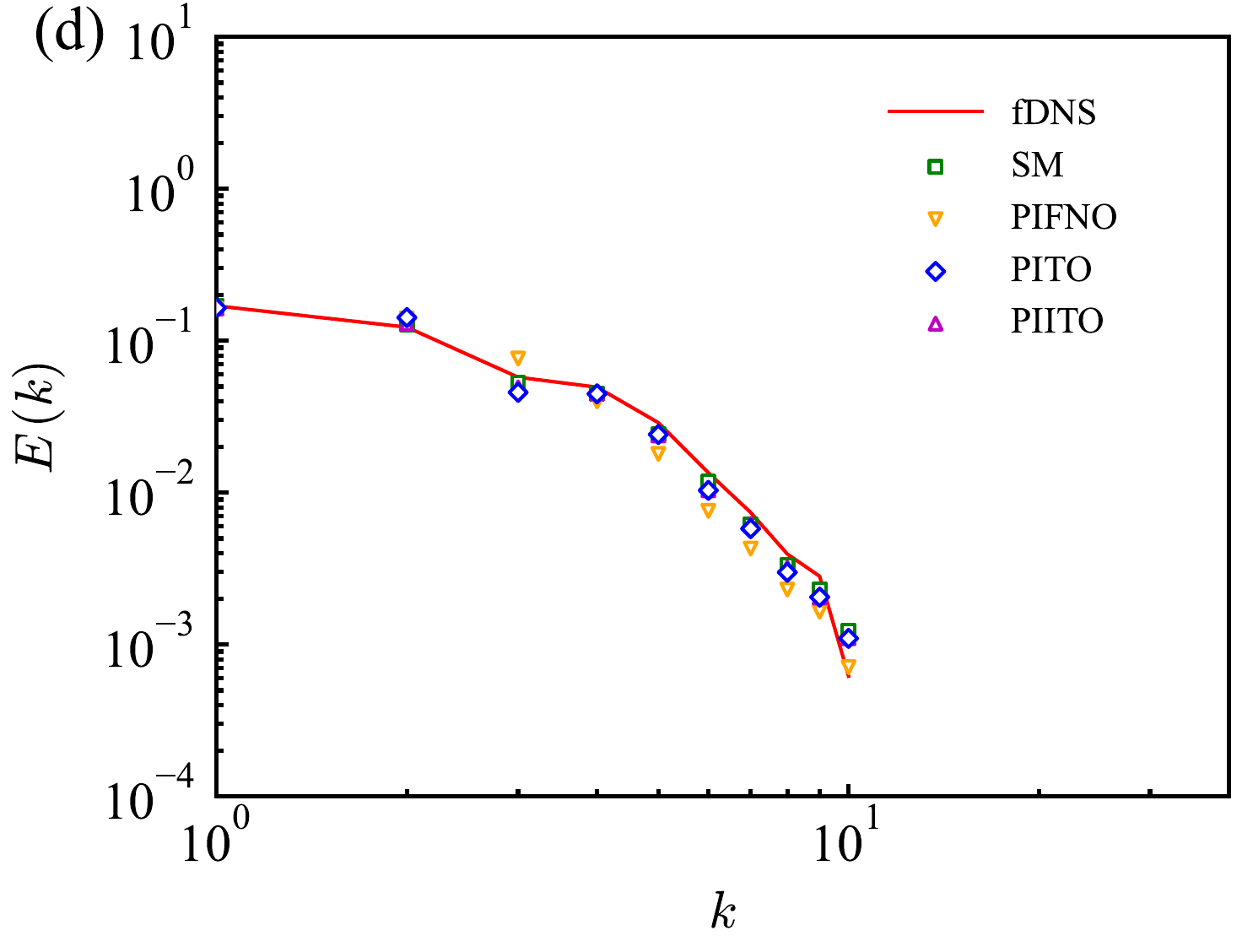} % 图片内容缩放 0.75 倍
    \caption{
Comparison of turbulent kinetic energy spectra predicted by fDNS, SM, and physics-informed models starting from stationary initial condition at (a) $t \approx \tau$ (b) $t \approx2\tau$ (c) $t \approx4\tau$ (d) $t \approx5\tau$.}
    \label{DHIT.energy}
\end{figure*}

The velocity spectra predicted by the physics-informed models, and SM at different instants are shown in Fig.~\ref{DHIT.energy}. At the first two large-scale wavenumbers, the predictions from all models closely match those from fDNS and SM. However, for the wavenumbers $k > 2$, the energy spectrum predicted by PIFNO exhibits significant errors at $t\approx 5\tau$, whereas the PITO and PIITO maintain a good consistency with fDNS and SM.

\begin{figure*}
    \centering
    \includegraphics[width=0.8\textwidth]{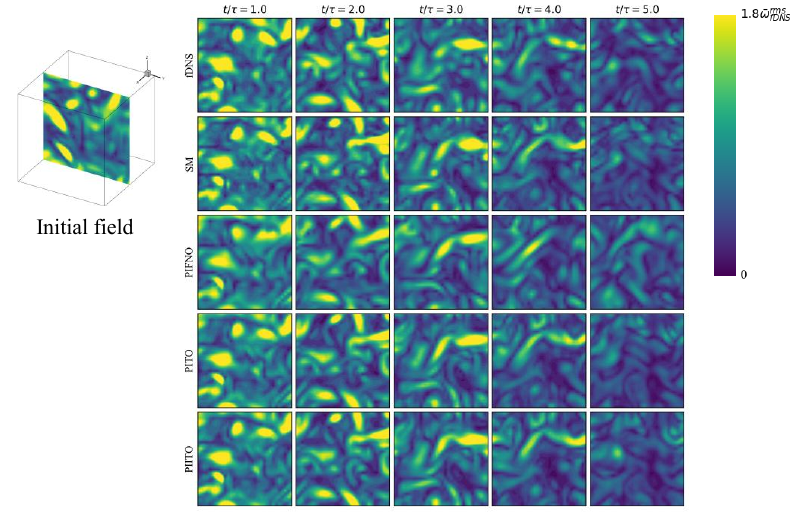} % 调整宽度比例为合适大小
    \caption{Comparison of contour lines for vorticity magnitude across a two-dimensional cross-section during multiple time steps of predictions by different models in the decaying HIT at stationary initial condition.}
    \label{2D qiemian}
\end{figure*}

\begin{figure*}
    \centering
    \includegraphics[width=0.6\textwidth]{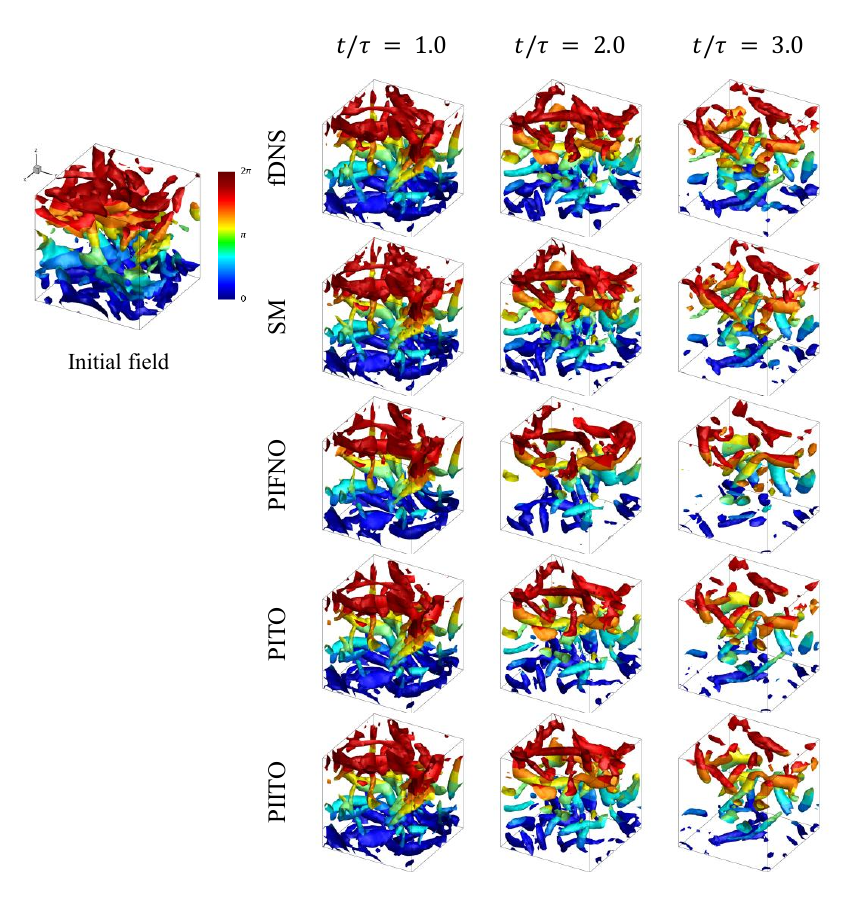} % 调整宽度比例为合适大小
    \caption{Comparison of instantaneous iso-surfaces of vorticity magnitude predicted by different models in decaying HIT at stationary initial condition.}
    \label{3D DHIT}
\end{figure*}

Fig.~\ref{2D qiemian} compares the vorticity magnitude field predicted by different models. At five distinct time instants, we select a YZ-plane slice to examine the vorticity distribution. We present the fDNS results as primary benchmark in the first row, and the subsequent five rows present the predictions results for SM, PIFNO, PITO, and PIITO respectively. The traditional SM model closely approximates the vorticity distribution of fDNS. PITO and PIITO keep consistent with the ground truth, while PIFNO exhibits minor errors, particularly at $t\approx4\tau$ and $t\approx5\tau$. We also demonstrated the three-dimensional vorticity structures of different models in Fig.~\ref{3D DHIT}. Compared to  PIFNO, both PITO and PIITO predict vortex structures more accurately.

%-------修改两个tabel
%
%
%
%

Machine learning approaches for numerical simulations are distinguished by their lower computational costs. Table \ref{compution perfrom} compares the computational efficiency of the traditional model SM with three physics-informed models. Here, the physics-informed models are trained and tested on an NVIDIA A100 PCIe 40GB graphics processing unit (GPU) paired with an Intel 6248R central processing unit (CPU) @3.0GHz, while the SM model is simulated on an Intel Xeon Gold 6148 CPU (16 cores @ 2.40 GHz).

Table \ref{compution perfrom} shows that PITO and PIITO reduce GPU memory consumption by 79.5\% and 91.3\%, respectively, compared to PIFNO. Moreover, PITO and PIITO require fewer parameters, representing only 31.5\% and 3.1\% of the PIFNO parameter count. These results indicate that PITO and PIITO effectively capture complex fluid features through a more streamlined architecture. Inference results show that all physics-informed models provide a 40-fold acceleration over the traditional SM method, with PIITO achieving the lowest inference time.
\begin{table*}
    \centering
    \caption{Comparison of computational efficiency among different models in decaying HIT at stationary initial condition.}
    \label{compution perfrom}
    \begin{tabular}{cccc} % 列格式：左对齐，四个列居中
        \toprule
        {Model}  & {parameters ($\times10^6$)} & {GPU Memory (GB)}  &{inference (s)}\\
        \midrule
        SM          & N/A & N/A & 66.49\\
        PIFNO          & 1062 & 38.83 & 1.561\\
        PITO             & 334.2 & 7.977& 1.629\\
        PIITO         & 33.49 & 3.395& 1.556\\
        \bottomrule
    \end{tabular}
\end{table*}

\subsection{Decaying homogeneous isotropic turbulence at random initial condition}
The physics-informed models, constrained by LES equations, can evolve from a random initial condition to achieve the same turbulent state as the traditional LES method. Therefore, we evaluate the performance of the proposed physics-informed models on a more challenging task for decaying isotropic turbulence in the situation of random initial condition.

\begin{figure*}
    \centering
    \includegraphics[width=0.5\textwidth]{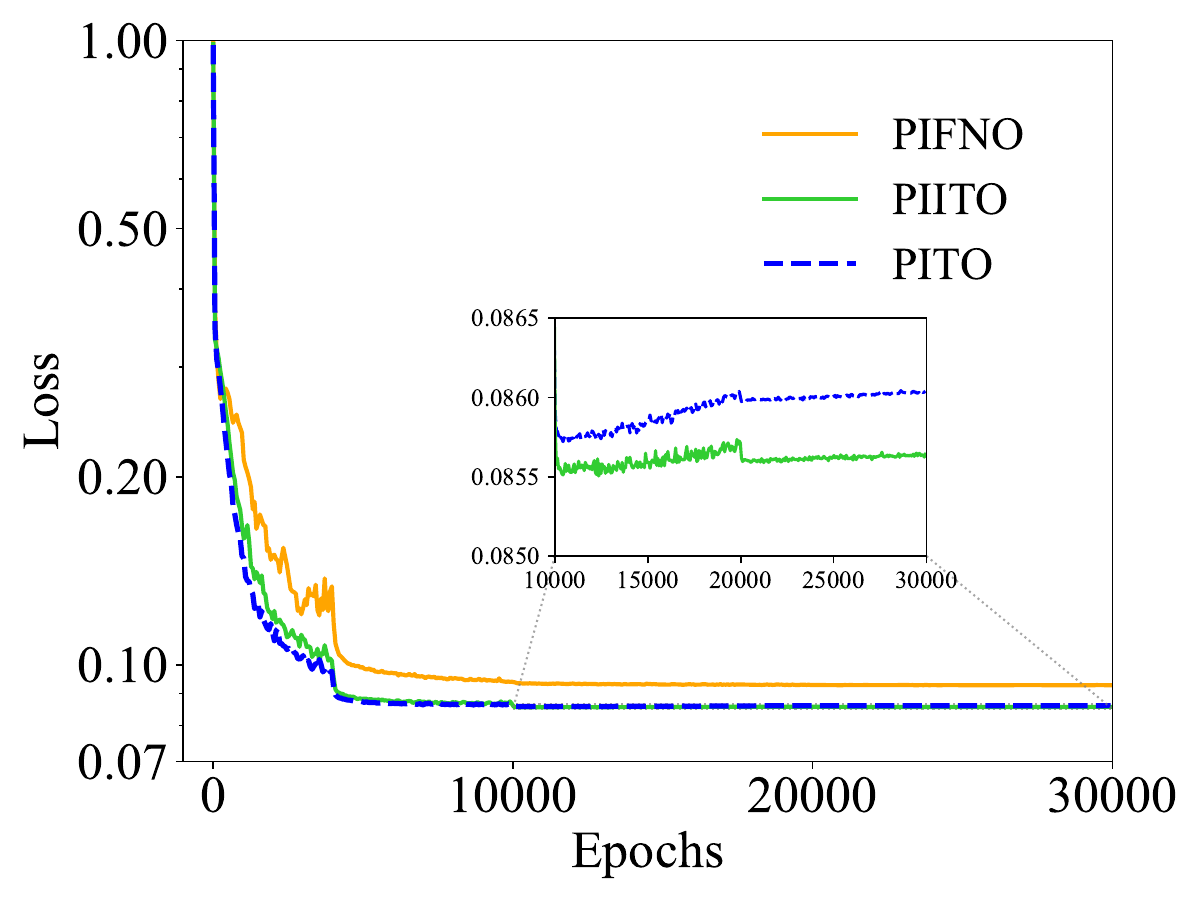} 
    \caption{Test losses of physics-informed models for decaying HIT at random initial condition.}
    \label{nDHIT.testloss}
\end{figure*}

\begin{table*}
    \centering
    \caption{Comparison of the minimum training and testing loss of different models from decaying HIT at random initial condition.}
    \label{nDHIT.loss}
   \begin{tabular}{ccc} % 列格式：左对齐，四个列居中 % 建议设定宽度，让表格不至于太窄
        \toprule
        Model & PDE Loss & Test Loss \\
        \midrule
        PIFNO   & 0.09727 & 0.09273 \\
        PITO  & 0.16671 & 0.08571 \\
        PIITO & 0.18613 & 0.08548 \\
        \bottomrule
    \end{tabular}
\end{table*}

We adopted the same parameter settings as those used for decaying isotropic turbulence at stationary initial condition. The initial flow field is generated with Gaussian random distribution at $t = 0$ in spectral space, which satisfies the Kolmogorov spectrum. The energy spectrum $E(k)$ of initial flow field is given by \cite{yuan2020deconvolutional}:
\begin{equation}
    E(k) = A_0 \left( \frac{k}{k_0} \right)^4 \exp \left[ -2 \left( \frac{k}{k_0} \right)^2 \right],
\end{equation}
where $A_0=2.7882$ and $k_0=4.5786$.

The evolutions of the test loss for the different models are illustrated in Fig.~\ref{nDHIT.testloss}, while Table \ref{nDHIT.loss} summarizes the corresponding minimum training and testing losses. It can be seen that the test losses of the physics-informed models are higher compared to those in the situation of statistically stationary initial conditions. Moreover, the converged test losses of the PITO and PIITO models are lower than those of PIFNO, implying that the neural operators with the Transformer framework have stronger fitting capabilities than that with FNO.

\begin{figure*}
    \centering
    \includegraphics[width=.38\textwidth]{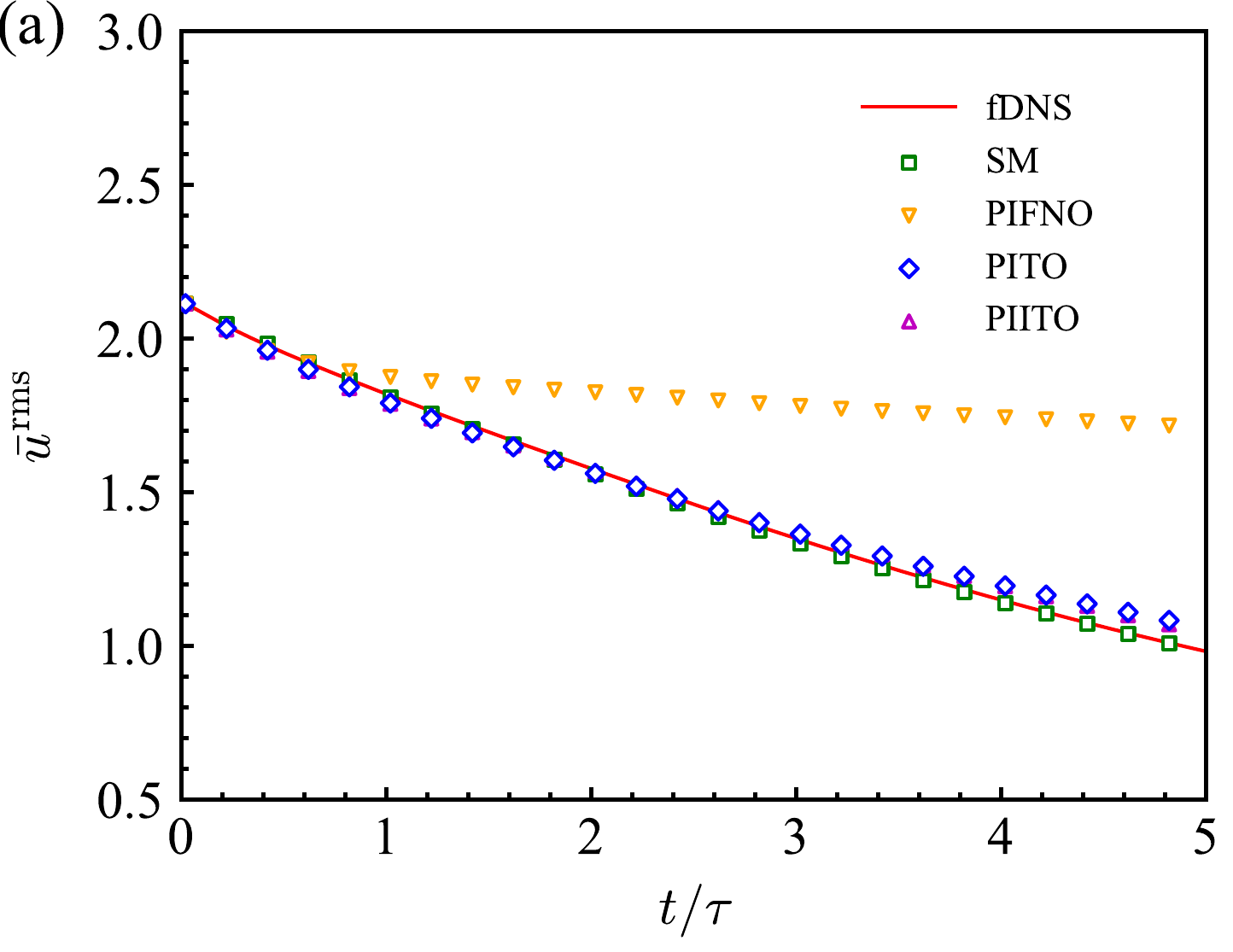} 
    \includegraphics[width=.38\textwidth]{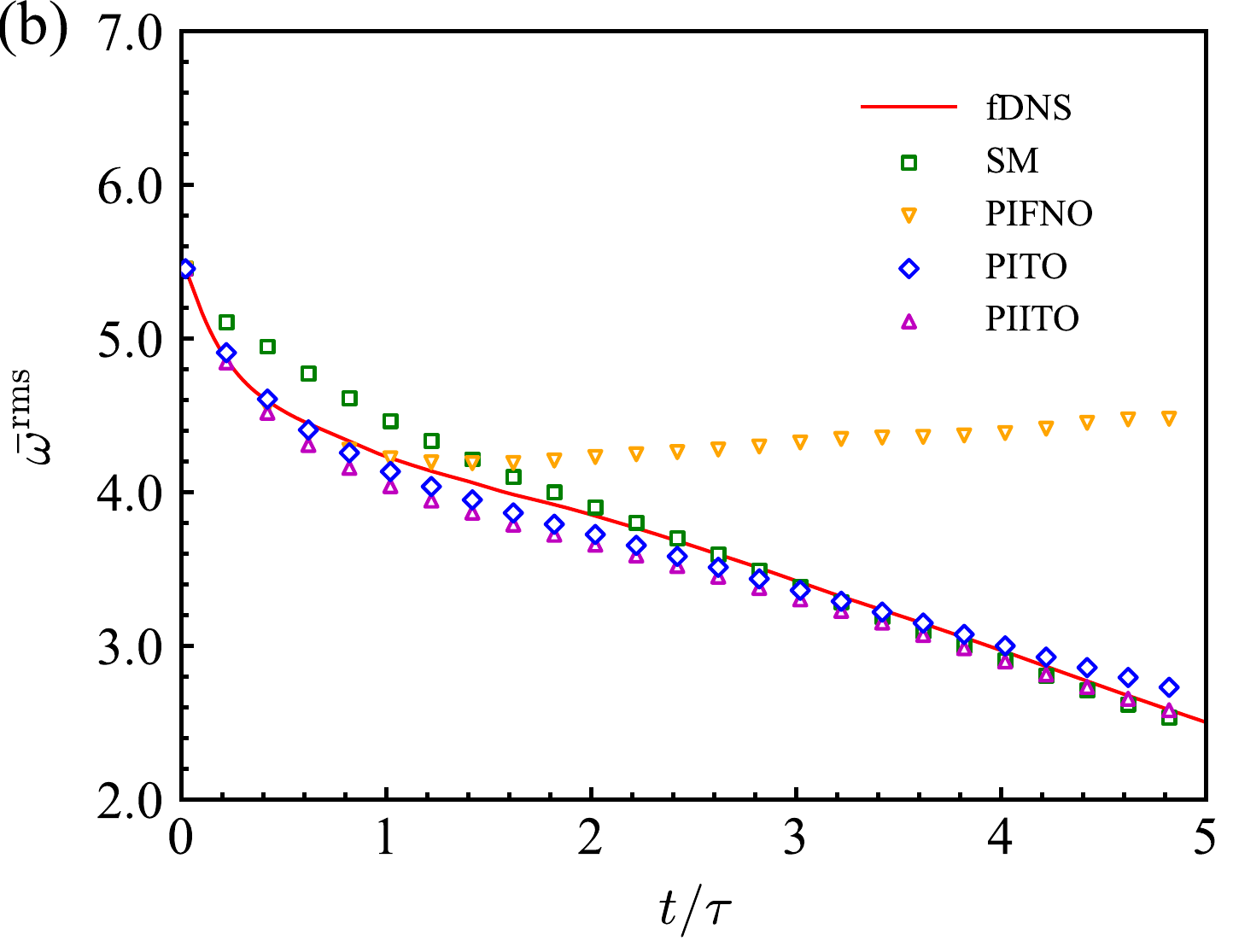}
    \caption{Temporal evolutions of (a) mean root-mean-square (rms) velocity and (b) the rms vorticity in decaying HIT at random initial condition.}
    \label{nDHIT.u w}
\end{figure*}

Here, five independently randomly generated velocity fields are used to perform the \textit{a posteriori} test. The ensemble average across these five cases is used to display statistical results. Fig.~\ref{nDHIT.u w} presents the root mean square (rms) values of velocity and vorticity. It can be seen that the PIFNO exhibits a significant error and becomes unstable at $t\approx \tau$. In contrast, both PITO and PIITO models show excellent stability and accuracy in the long-term predictions.

\begin{figure*}
    \centering
    % (a) FNO
    \includegraphics[width=.38\textwidth]{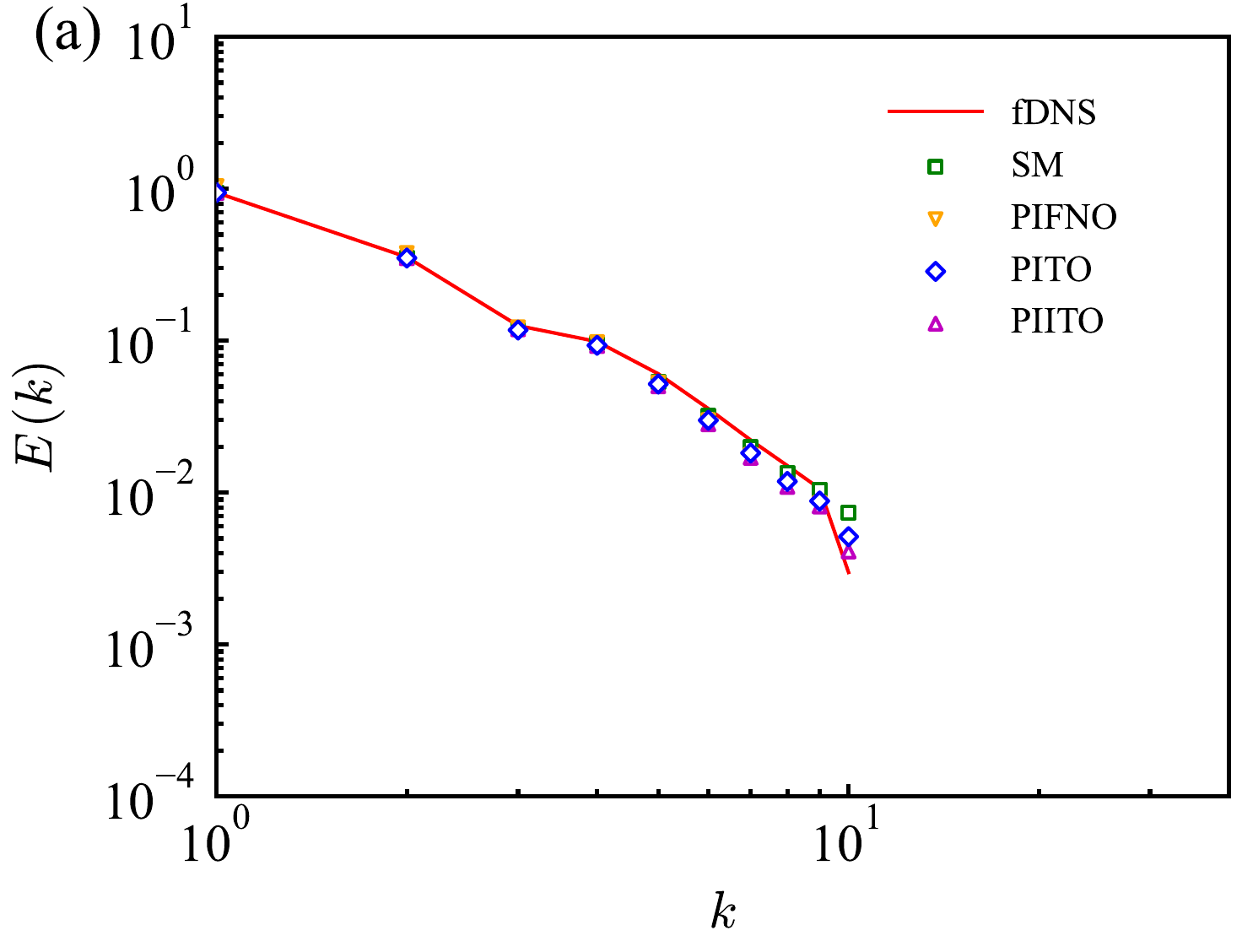} 
    \includegraphics[width=.38\textwidth]{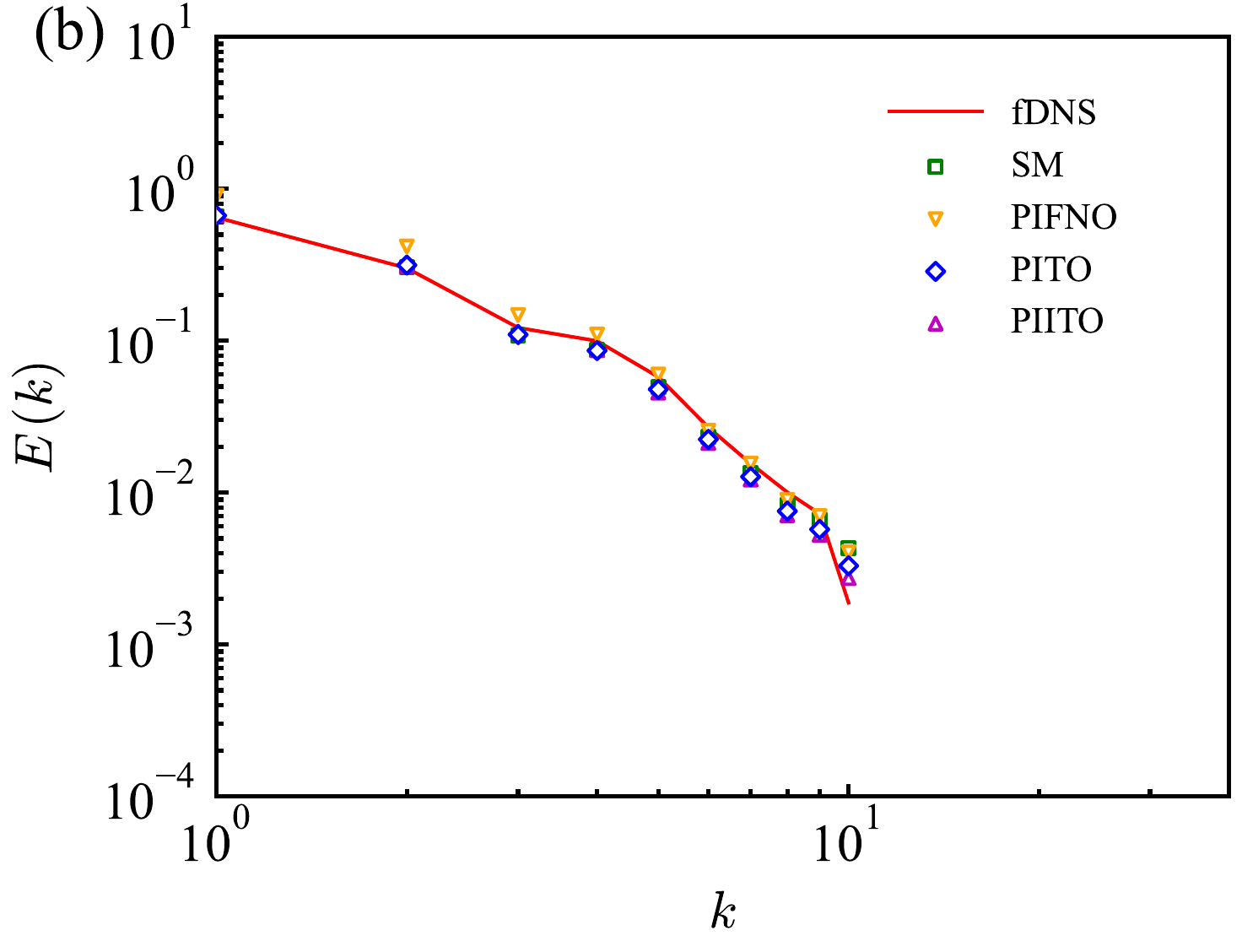}
    \includegraphics[width=.38\textwidth]{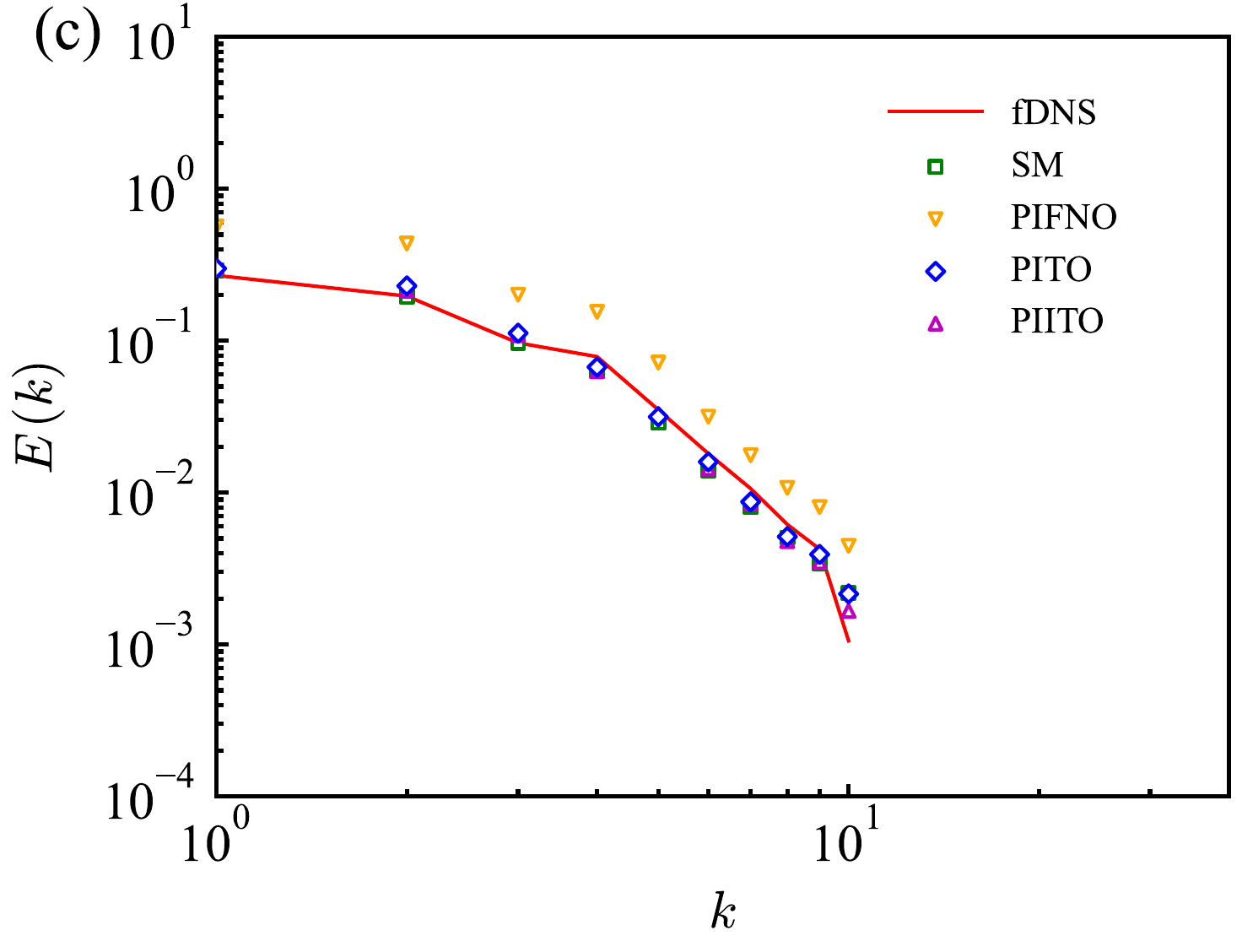} 
    \includegraphics[width=.38\textwidth]{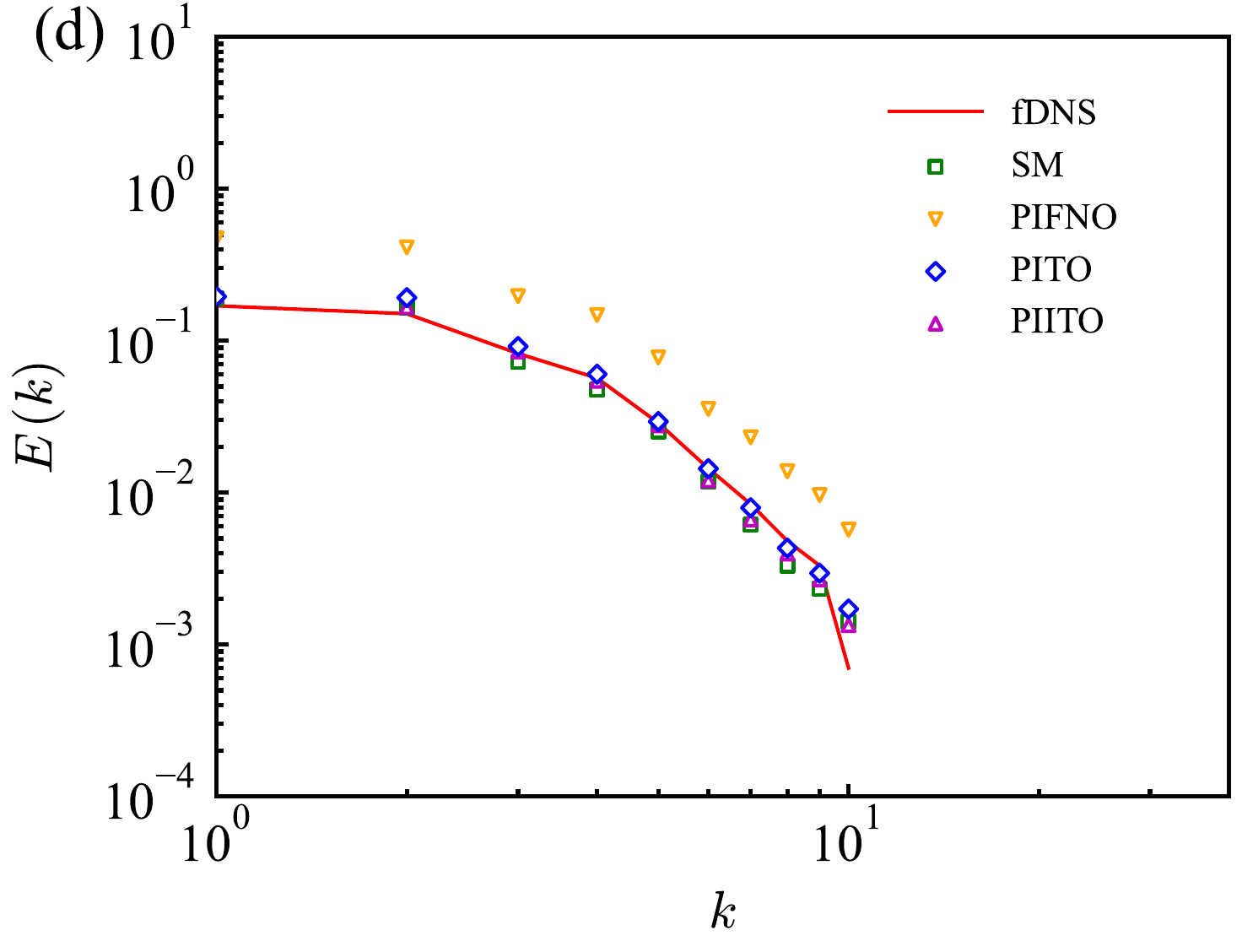}

    \caption{turbulent kinetic energy spectra in decaying HIT at random initial condition at different time instants: (a) $t \approx \tau$ (b) $t \approx2\tau$ (c) $t \approx4\tau$ (d) $t \approx5\tau$.}
    \label{nDHIT.energy}
\end{figure*}

Fig.~\ref{nDHIT.energy} illustrates predictions of kinetic energy spectrum by different models at different times. At $t \approx 4\tau$ and $ t \approx 5\tau $, the PIFNO predictions completely overestimates the energy spectrum, and the error continues to increase with time. In contrast, PITO and PIITO models are very close to both SM and fDNS results.

\begin{figure*}
    \centering
    \includegraphics[width=.38\textwidth]{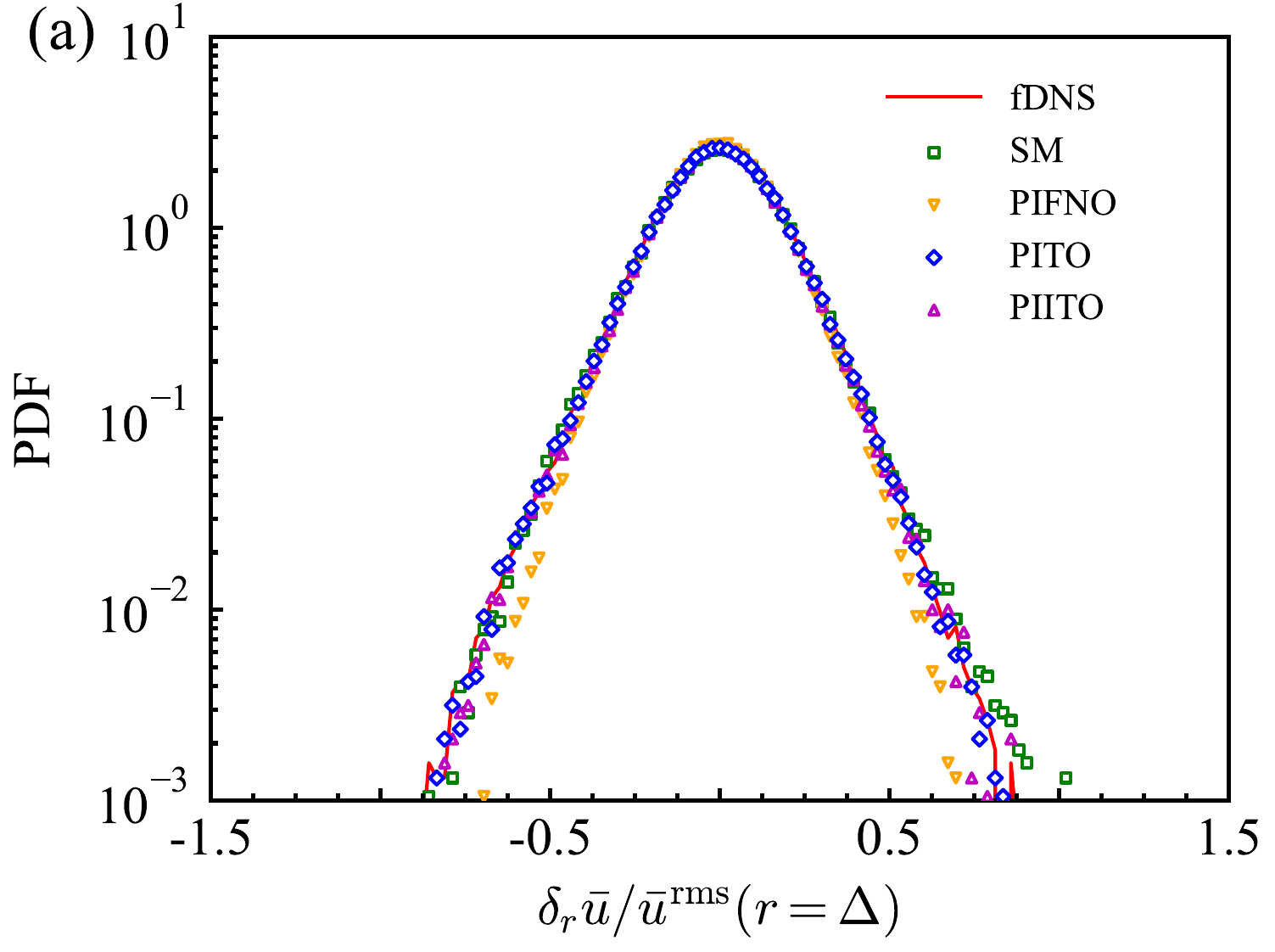} 
    \includegraphics[width=.38\textwidth]{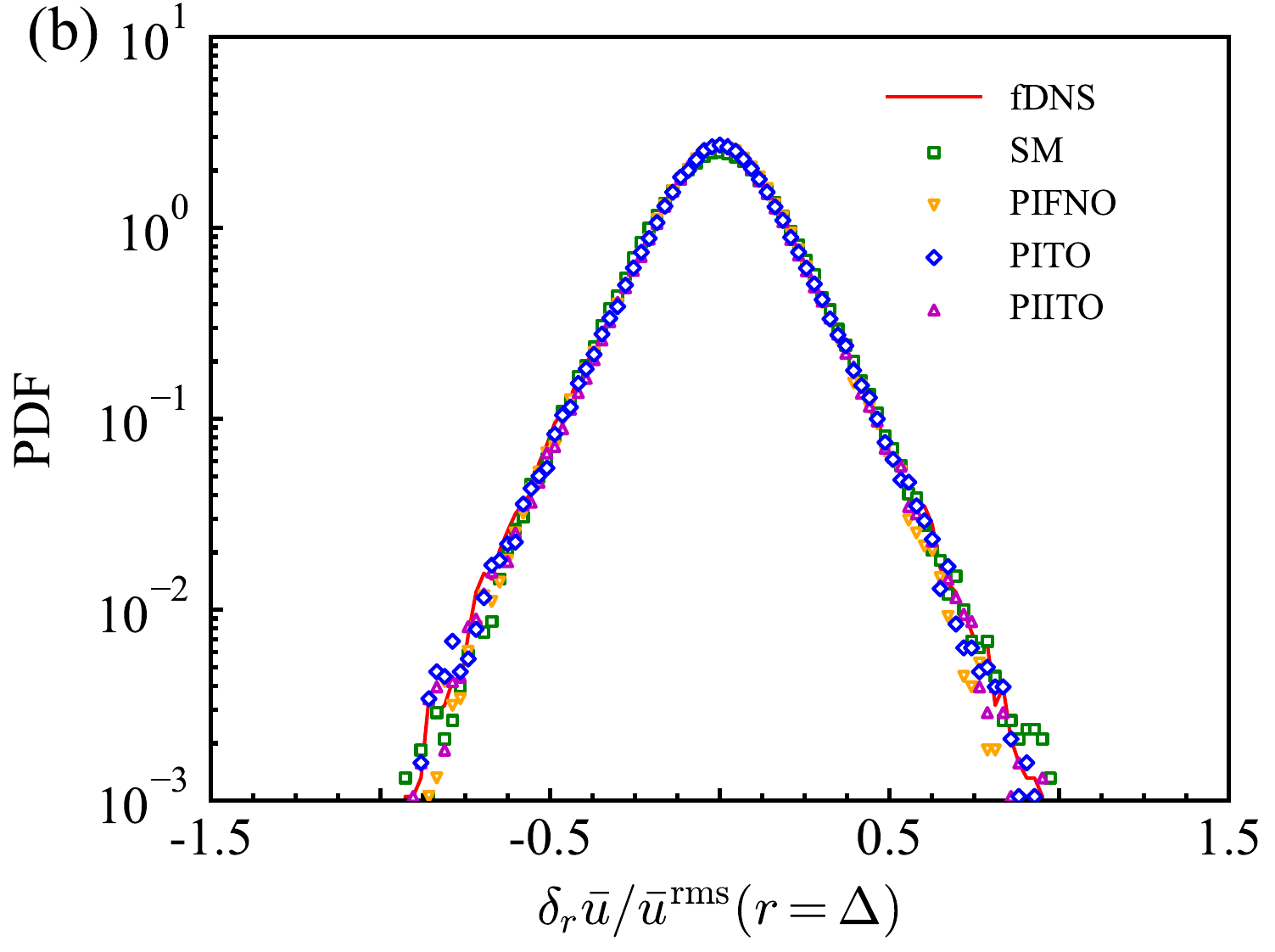}
    \includegraphics[width=.38\textwidth]{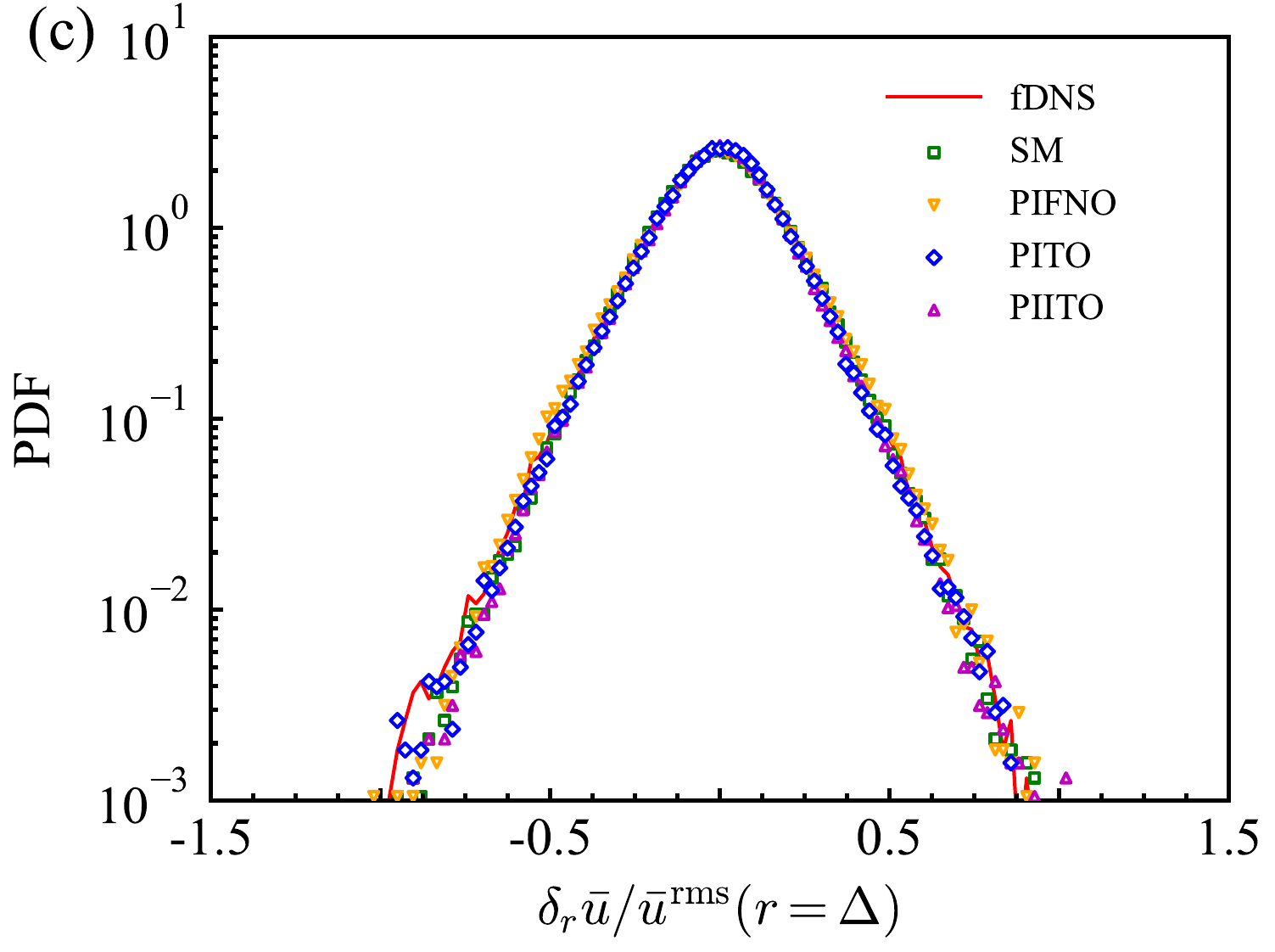} 
    \includegraphics[width=.38\textwidth]{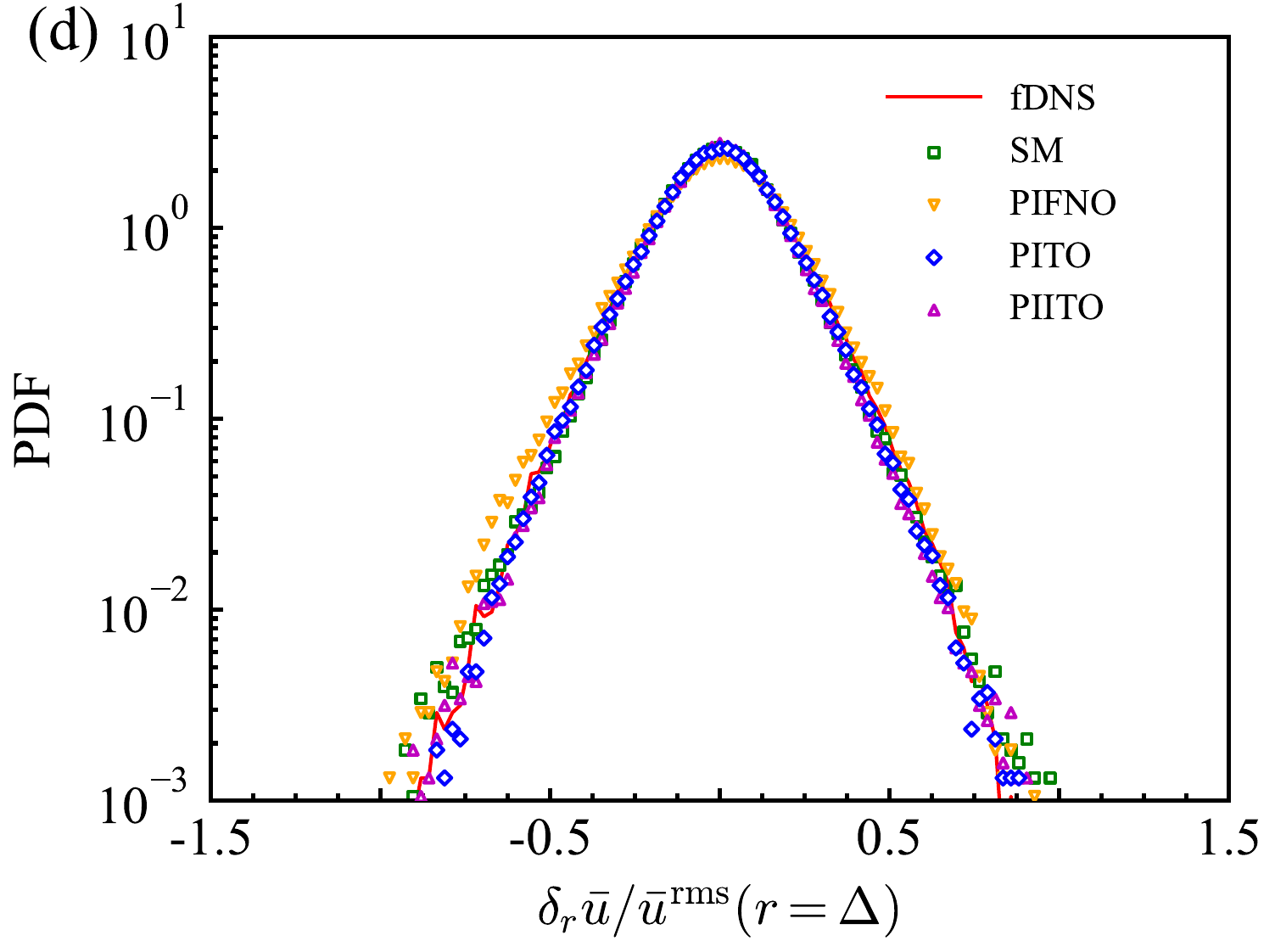} 
    \caption{Probability density functions (PDFs) of normalized velocity increments for different models in decaying HIT at random initial condition at different time instants: (a) $t \approx \tau$ (b) $t \approx2\tau$ (c) $t \approx4\tau$ (d) $t \approx5\tau$.}
    \label{nDHIT.deltau}
\end{figure*}

\begin{figure*}
    \centering
    % (a) FNO
    \includegraphics[width=.38\textwidth]{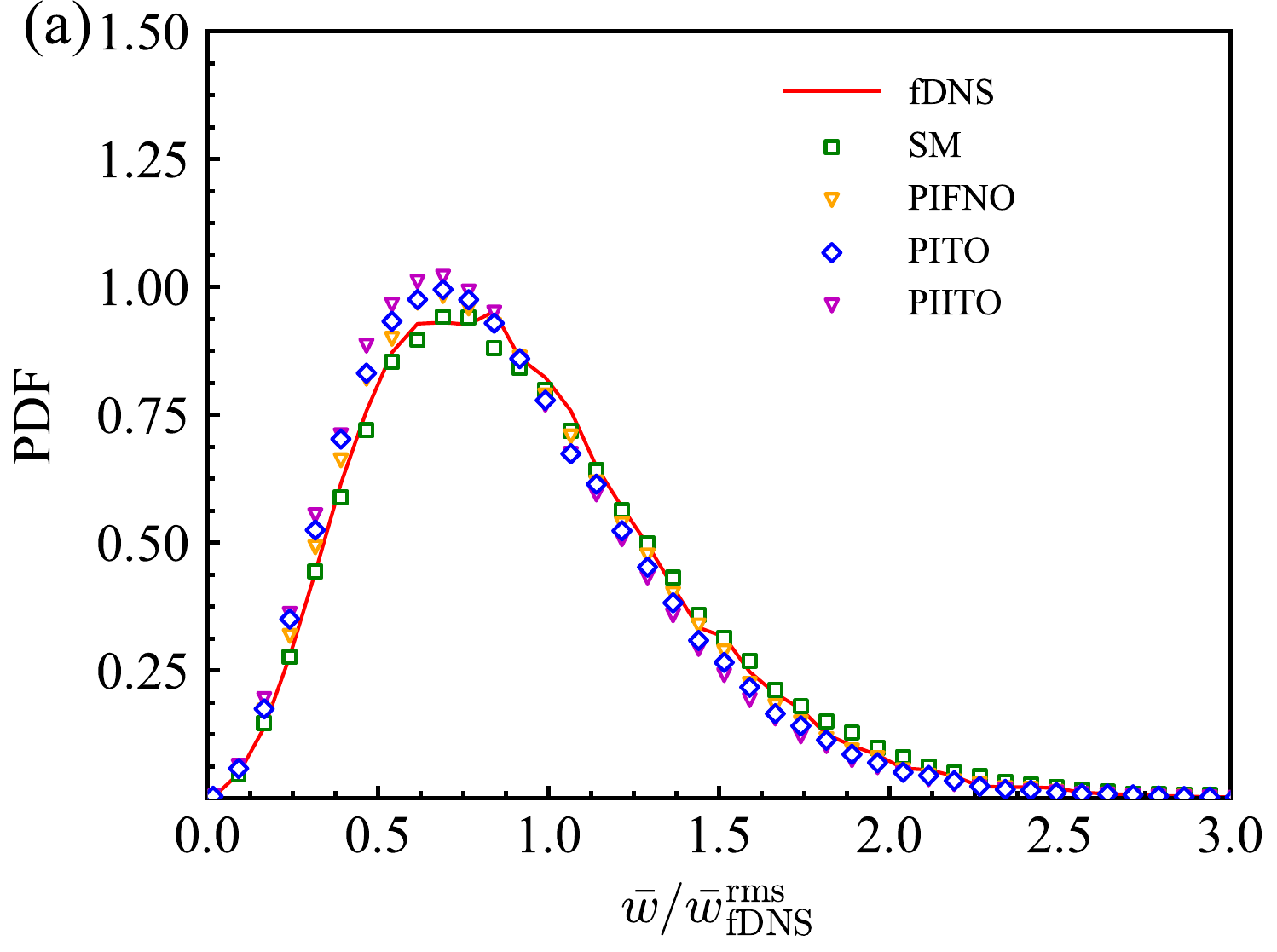} 
    \includegraphics[width=.38\textwidth]{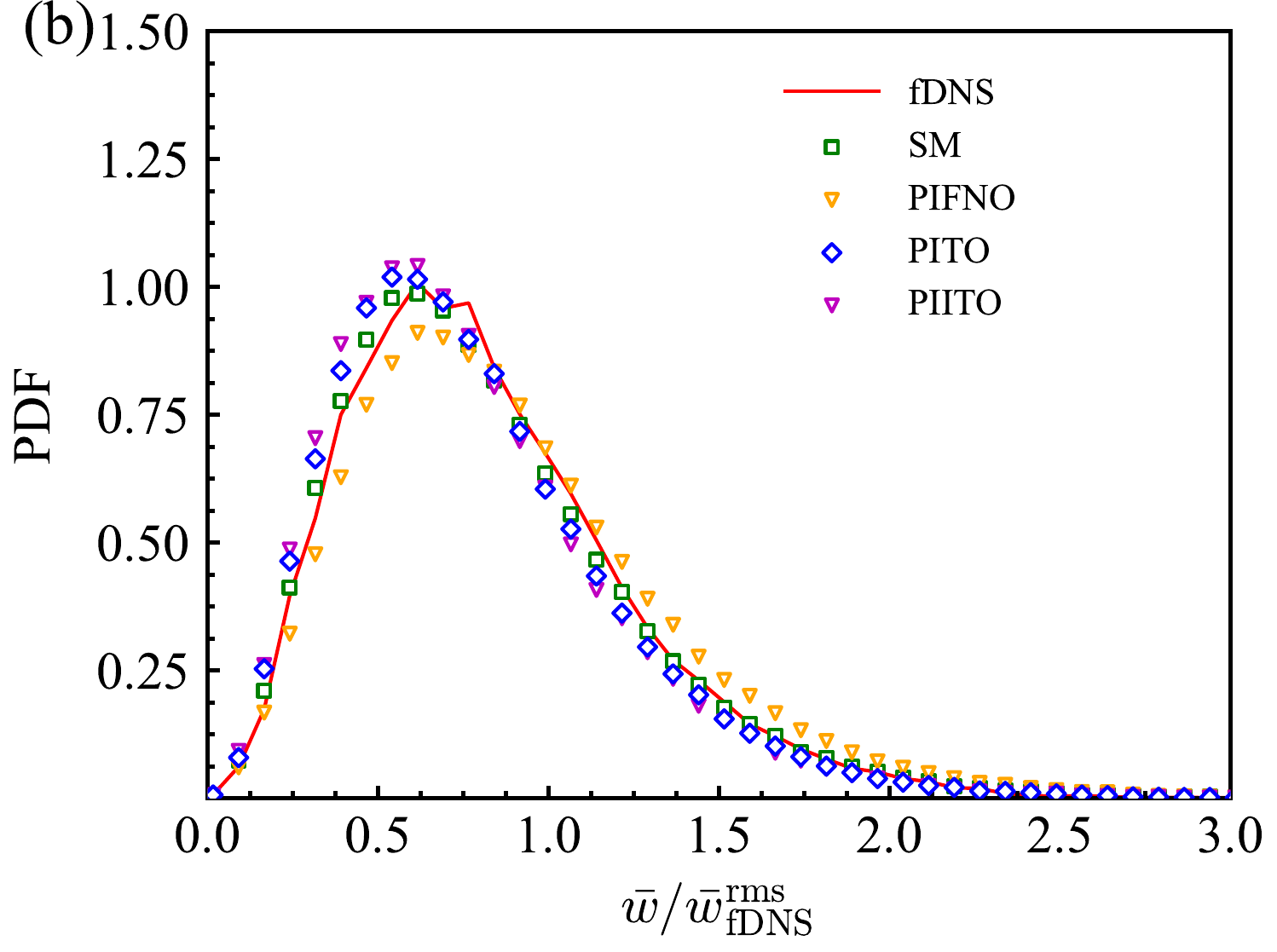}
    \includegraphics[width=.38\textwidth]{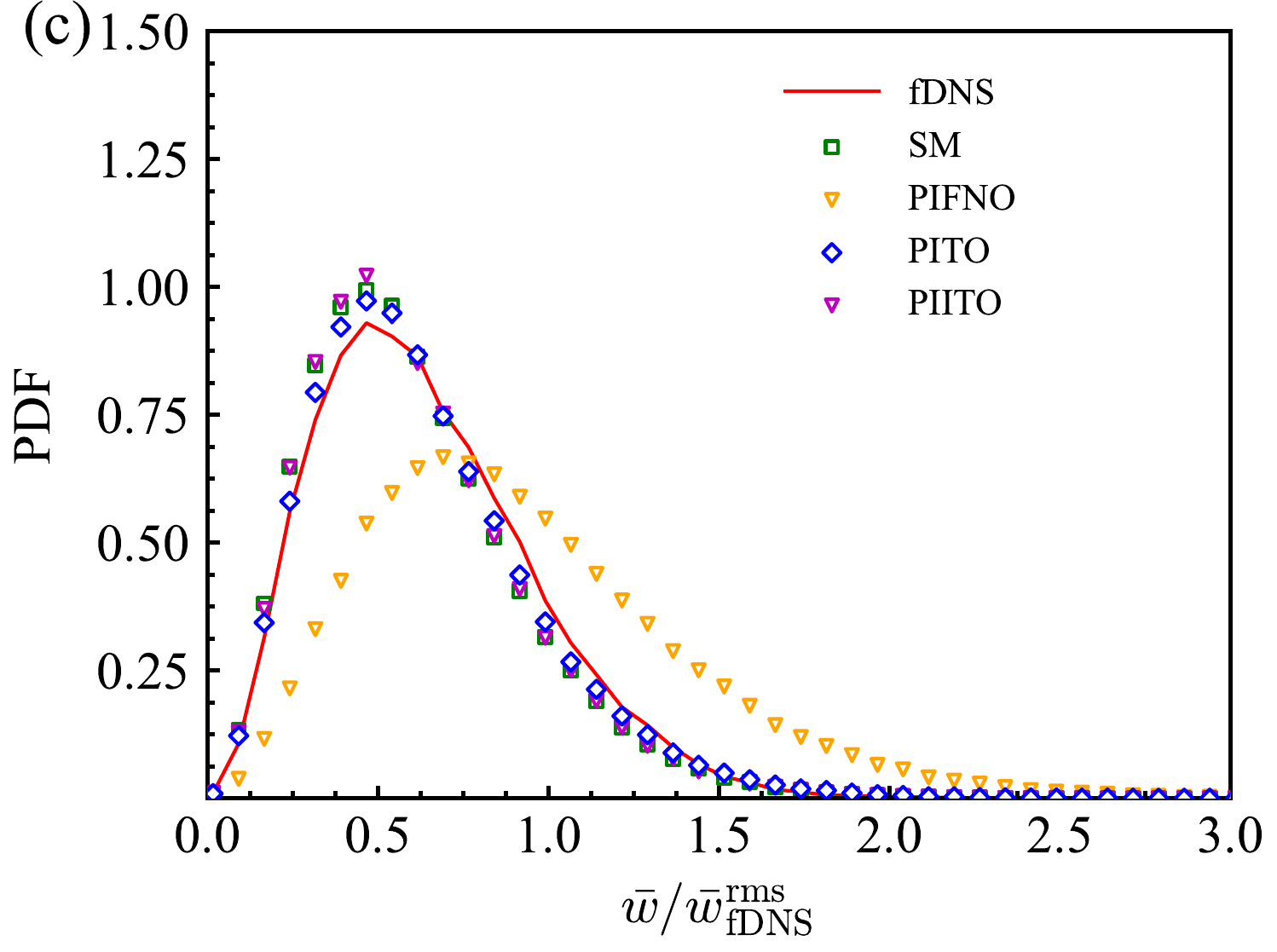} 
    \includegraphics[width=.38\textwidth]{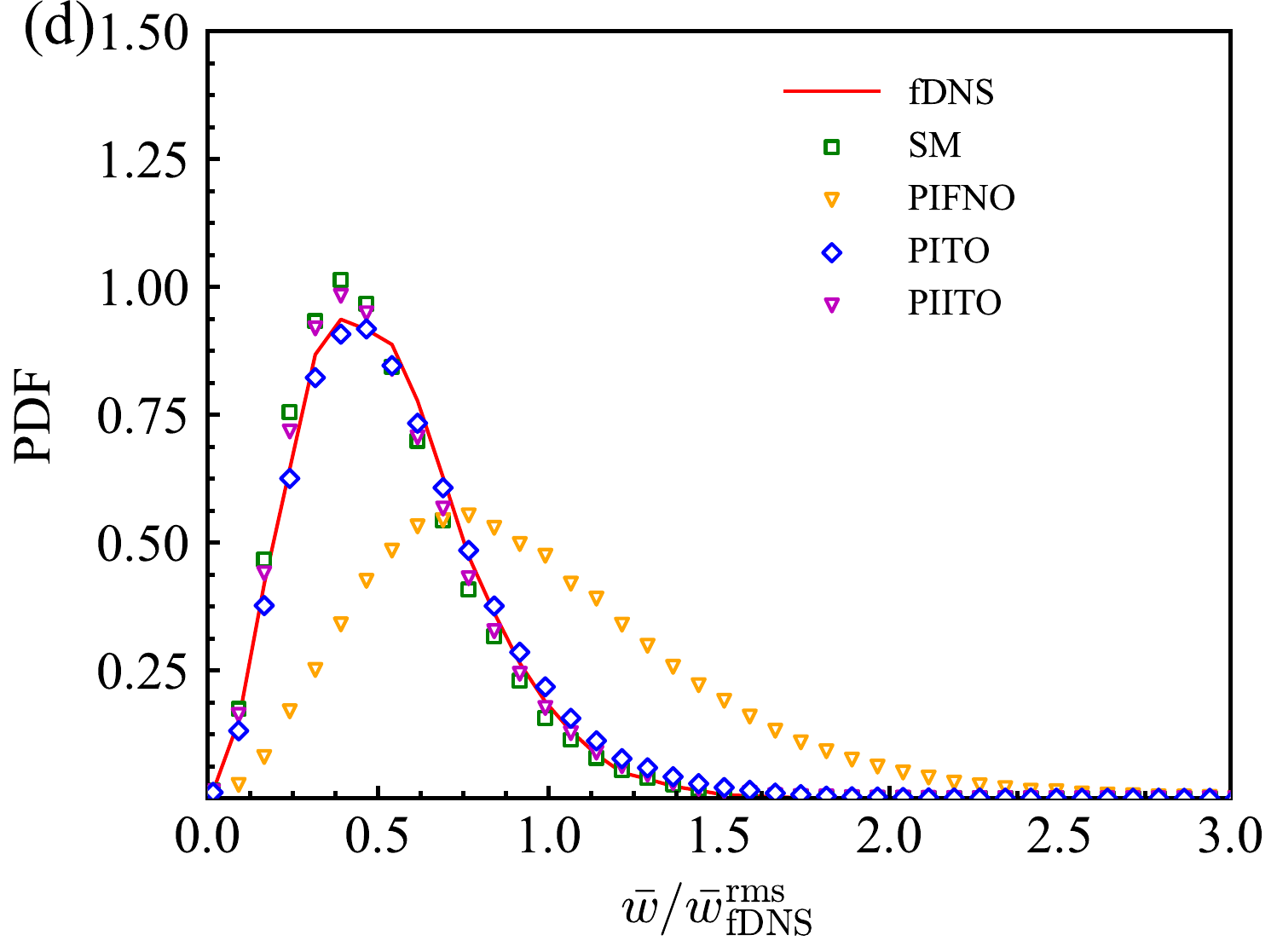} 
    \caption{PDFs of normalized vorticity magnitude for different models in decaying HIT at random initial condition at various time instants: (a) $t \approx \tau$ (b) $t \approx2\tau$ (c) $t \approx4\tau$ (d) $t \approx5\tau$.}
    \label{nDHIT.deltaw}
\end{figure*}

The probability density functions (PDFs) of the normalized velocity increments ${\delta_{r}\overline{u}}/{\overline{u}^{\mathrm{rms}}}$ with distance $(r = \Delta)$ are shown in Fig.~\ref{nDHIT.deltau}. Here, the longitudinal velocity increment is defined as $\delta_r \overline{u} = [\overline{\mathbf{u}}(\mathbf{x}+\mathbf{r}) - \overline{\mathbf{u}}(\mathbf{x})] \cdot \mathbf{\hat{r}}$. $\mathbf{\hat{r}} = \mathbf{r}/|\mathbf{r}|$ represents the unit vector in the separation direction. The PDFs predicted by the PIFNO significantly deviate from the fDNS results at $t\approx\tau$ and $t\approx5\tau$. However, the PDFs predicted by PITO and PIITO are in a good agreement with the fDNS data at different instants. 

The PDFs of normalized vorticity ${\overline{{w}}}/{\overline{{w}}_{\mathrm{fDNS}}^{\mathrm{rms}}}$ at different time instants are plotted in Fig.~\ref{nDHIT.deltaw}. The vorticity magnitude is normalized by the rms value of vorticity of fDNS. It can be seen that PIFNO fails to predict the PDF of vorticity at time $t \approx 4\tau$ and $t \approx 5\tau$. In contrast, the PDFs of vortcity predicted by PITO and PIITO exhibits a good agreement with fDNS and SM results at different time instants.

\begin{figure*}
    \centering
    \includegraphics[width=0.8\textwidth]{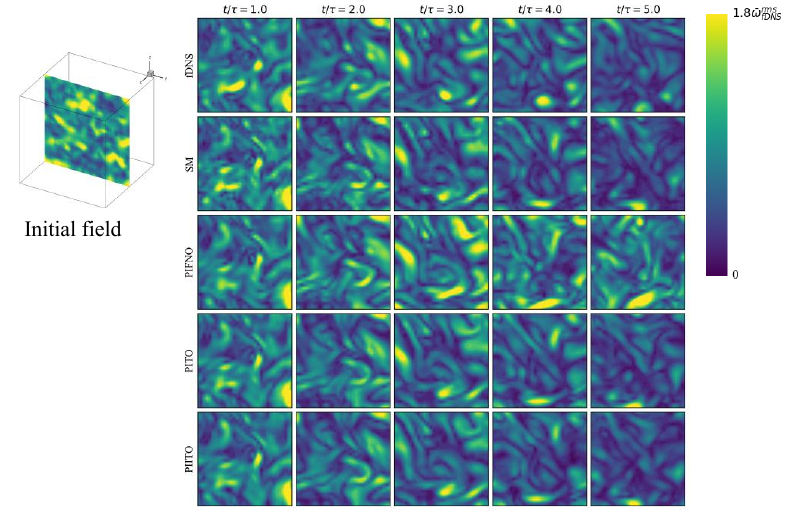} % 调整宽度比例为合适大小
    \caption{vorticity contours on the y-z plane predicted by different models for decaying HIT at random initial condition}
    \label{ndhit.2Dqiemian}
\end{figure*}

Vorticity contours on the y-z plane at the center are shown in Fig.~\ref{ndhit.2Dqiemian}.  The vorticity contours predicted by physics-informed models show a good agreement with the  SM and the fDNS results in short-term evolution $t/\tau \approx 1.0$ and $t/\tau \approx 2.0$. In long-term predictions, PIFNO consistently generated large-scale vortex structures, which are non-physical. In contrast, both PITO and PIITO models are very close to SM in long-term predictions.

\subsection{Forced homogeneous isotropic turbulence}
In this section, we also assess the performance of the physics-informed models in three-dimensional forced homogeneous isotropic turbulence. We employ the same parameters as those used in the decaying HIT simulations. The Taylor Reynolds number is $Re_\lambda \approx 60$ and the large-eddy turnover time is $\tau \approx 1.0$. The difference is that we do not remove the driving force after the turbulence reaches the statistically steady state. The SM model is implemented with Smagorinsky coefficient $C_\text{smag}= 0.1$.

In the \textit{a posteriori} analysis of forced HIT, the velocity field is extrapolated to a total time of $t \approx 5\tau$ through iterative inference. The learning time step of the model is $\Delta t_n = t_{n+1} - t_n = 10 \Delta t$, representing a 10-step forward predictions that covers a duration of $0.1\tau$.

\begin{figure*}
    \centering
    % (a) FNO
    \includegraphics[width=.45\textwidth]{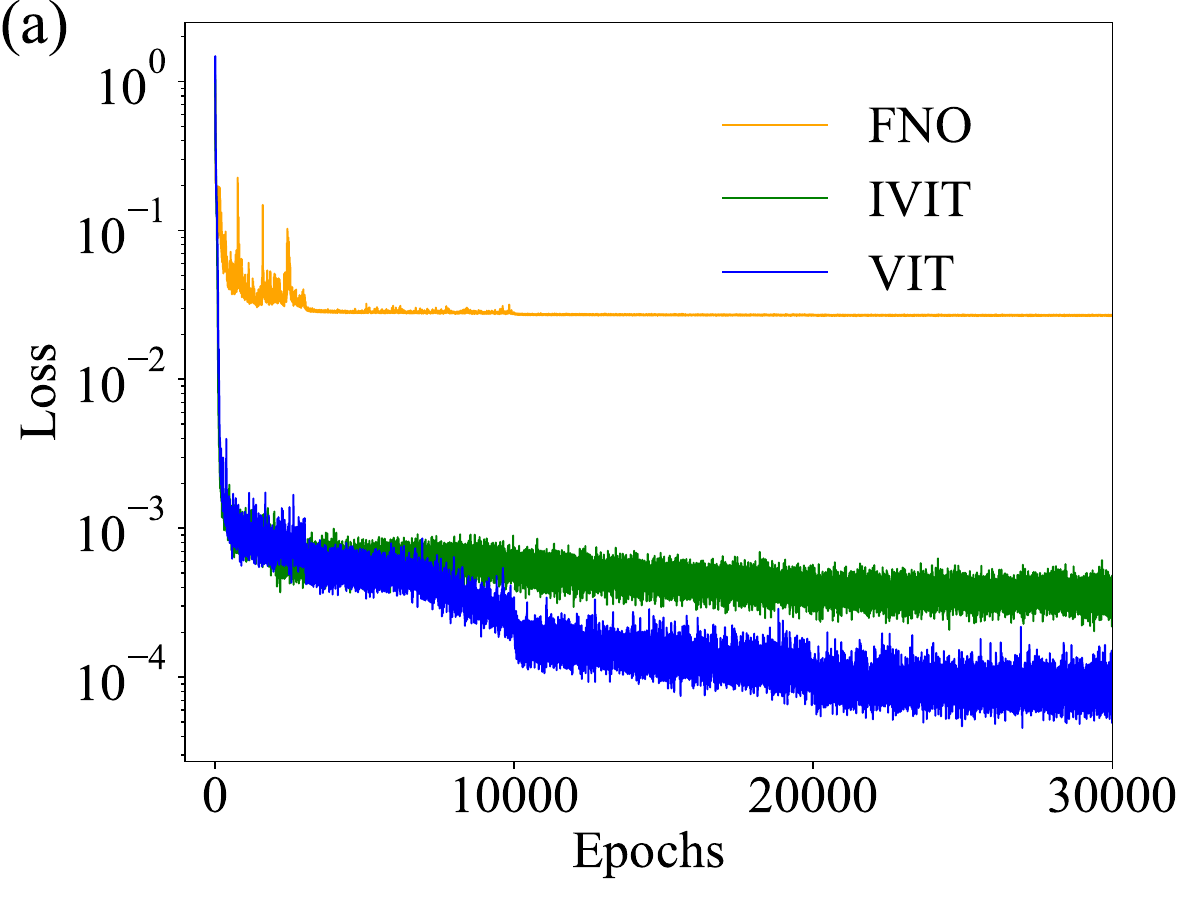} 
    \includegraphics[width=.45\textwidth]{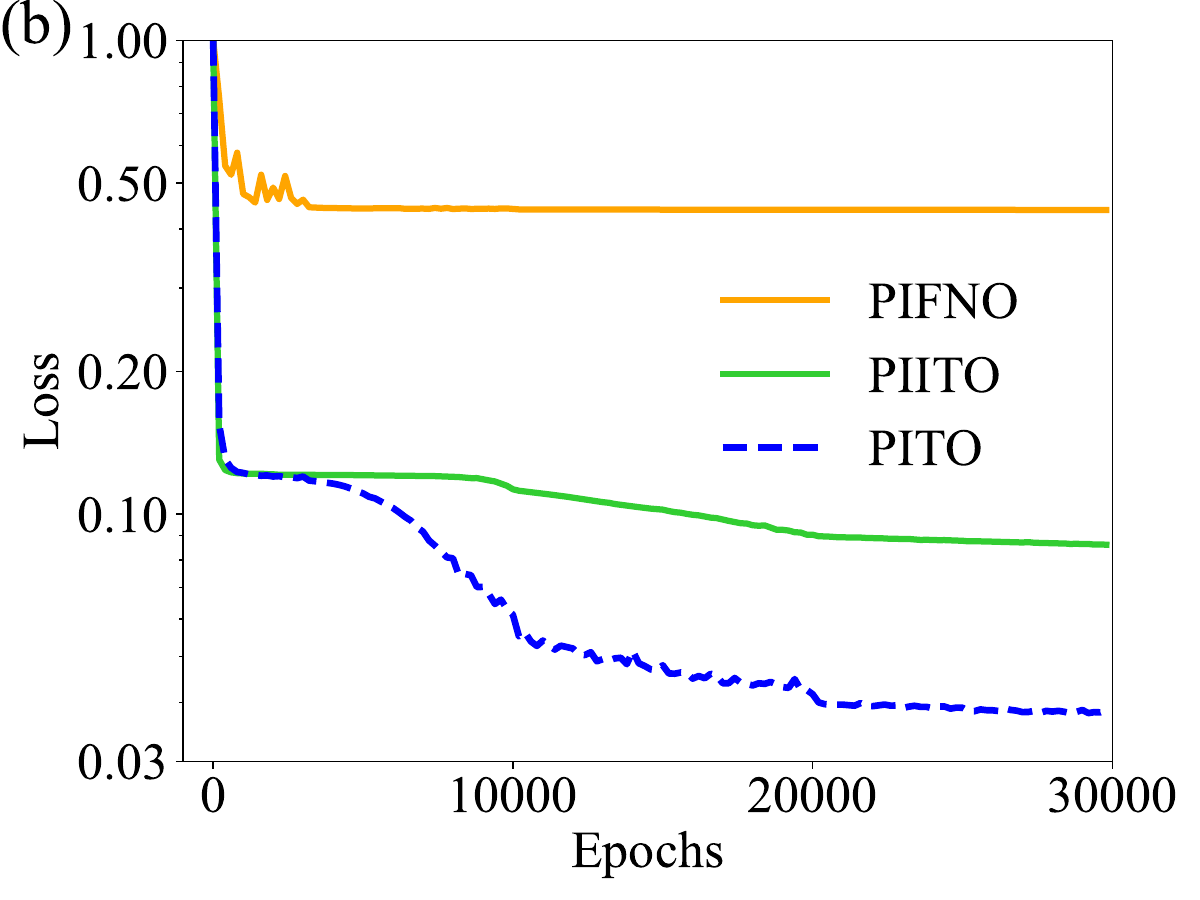}
    \caption{Evolutions of (a) PDE losses and (b) test losses curves for different physics-informed models in forced HIT.}
    \label{HIT.loss}
\end{figure*}

Fig.~\ref{HIT.loss} compares the train and test loss curves of the different models during operator learning. The training and testing losses of PIFNO converge to a high value and fail to decrease further. The PITO presents a stronger learning capability than PIFNO in predicting forced HIT with fewer parameters. Although the losses PIITO also converges to a relatively low level, but the converged value is higher compared to PITO. The results indicate that, the learning capability of the neural operators in the Transformer architecture is more effective than that of the Fourier architecture.

\begin{figure*}
    \centering
    % (a) FNO
    \includegraphics[width=.38\textwidth]{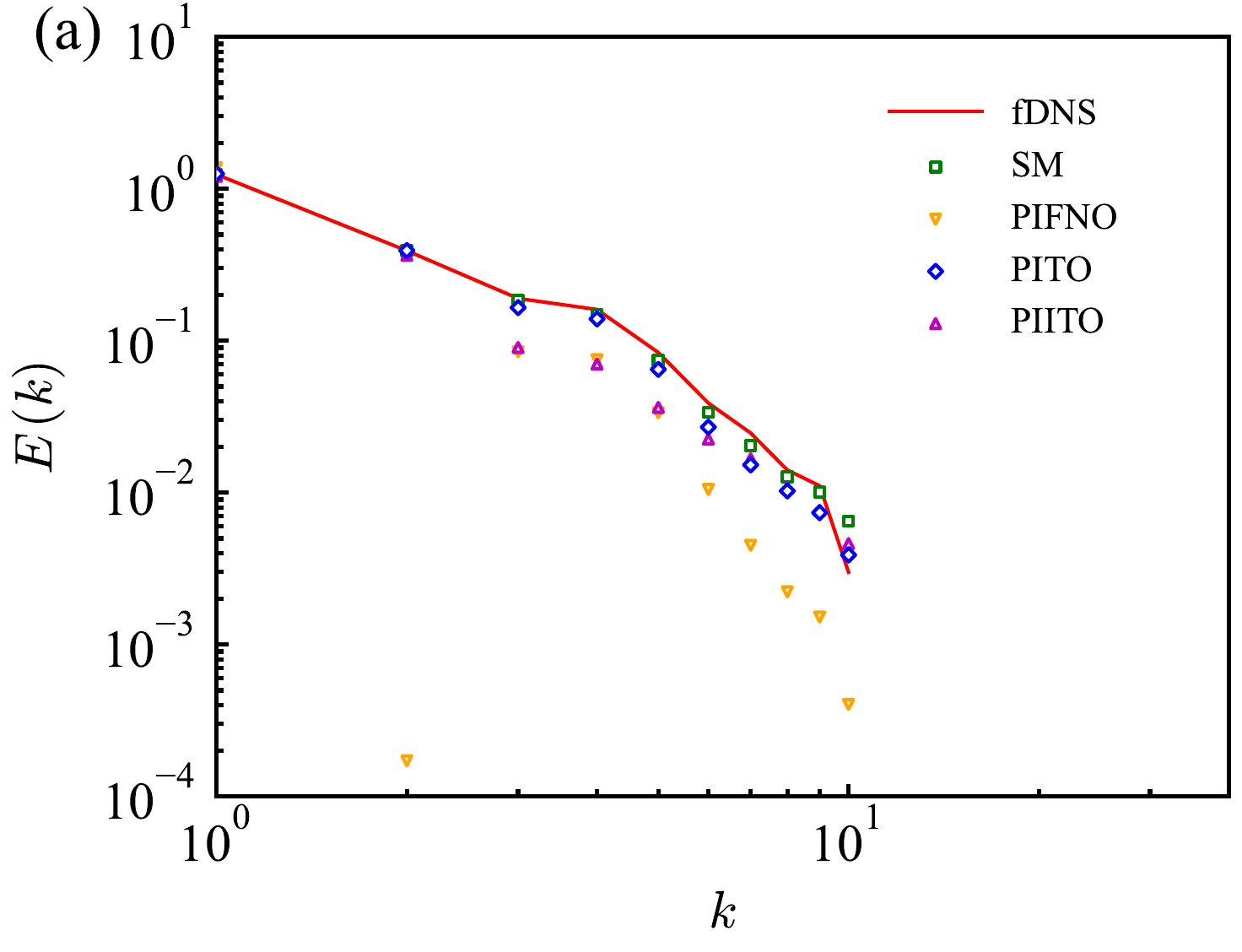} 
    \includegraphics[width=.38\textwidth]{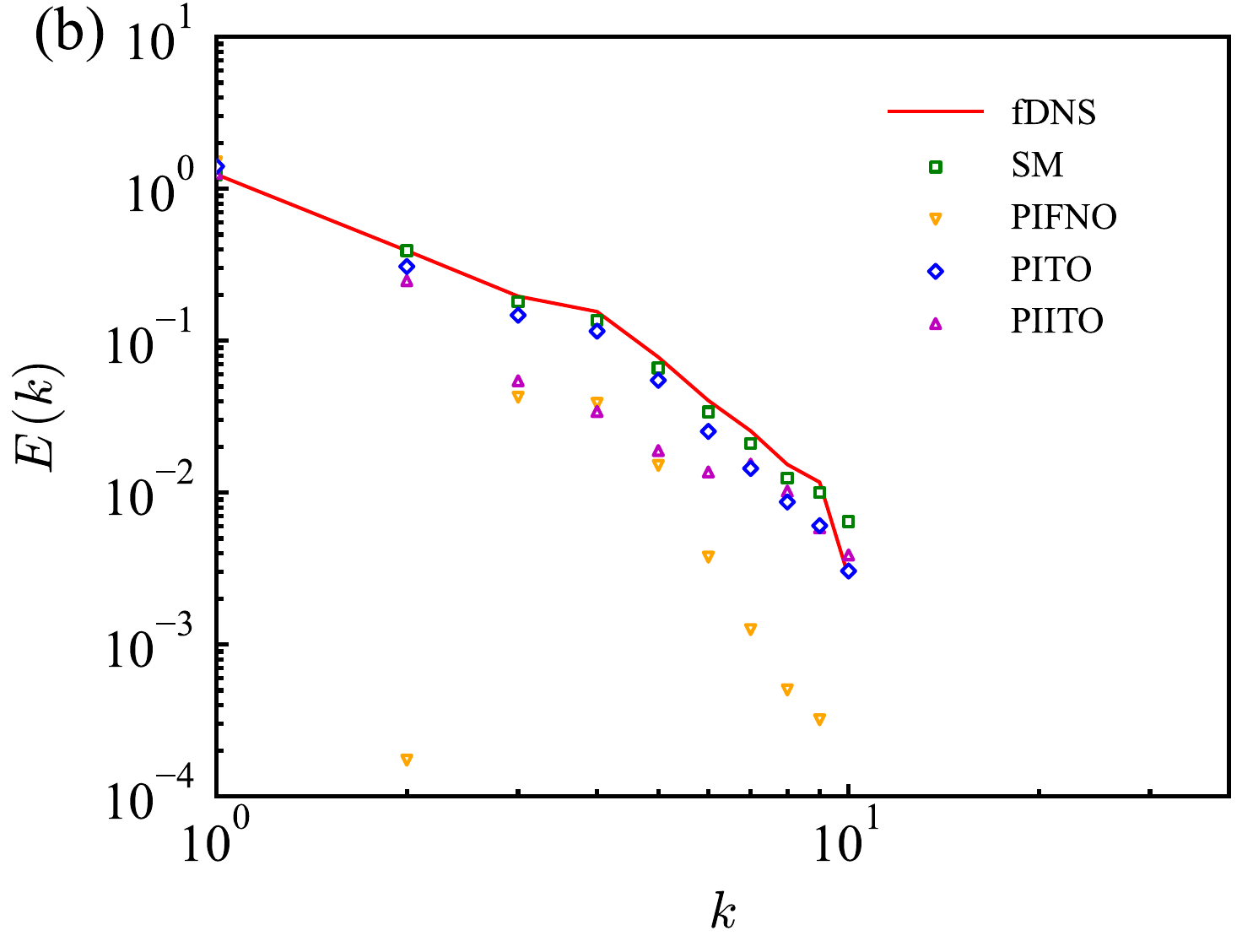}
    \includegraphics[width=.38\textwidth]{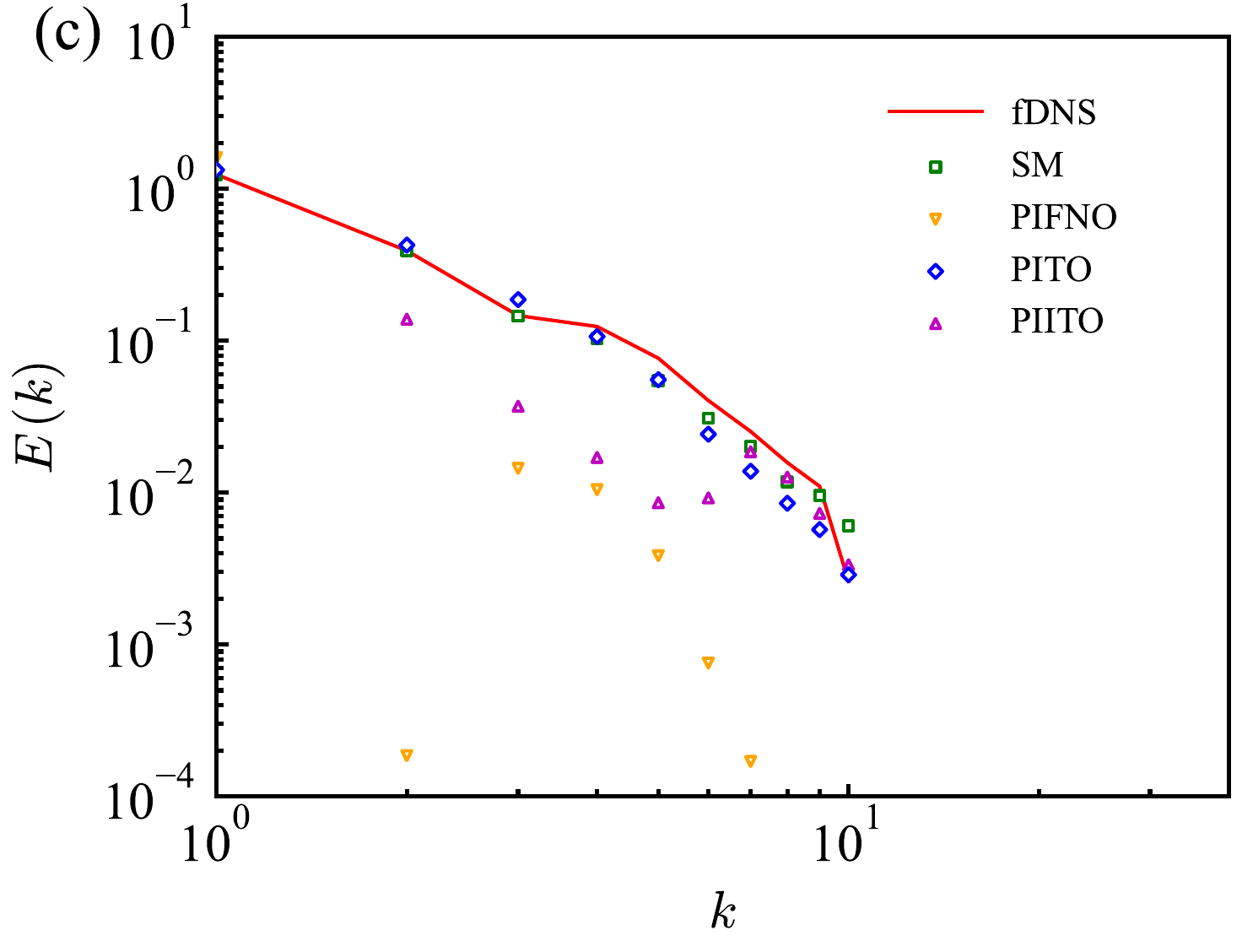} 
    \includegraphics[width=.38\textwidth]{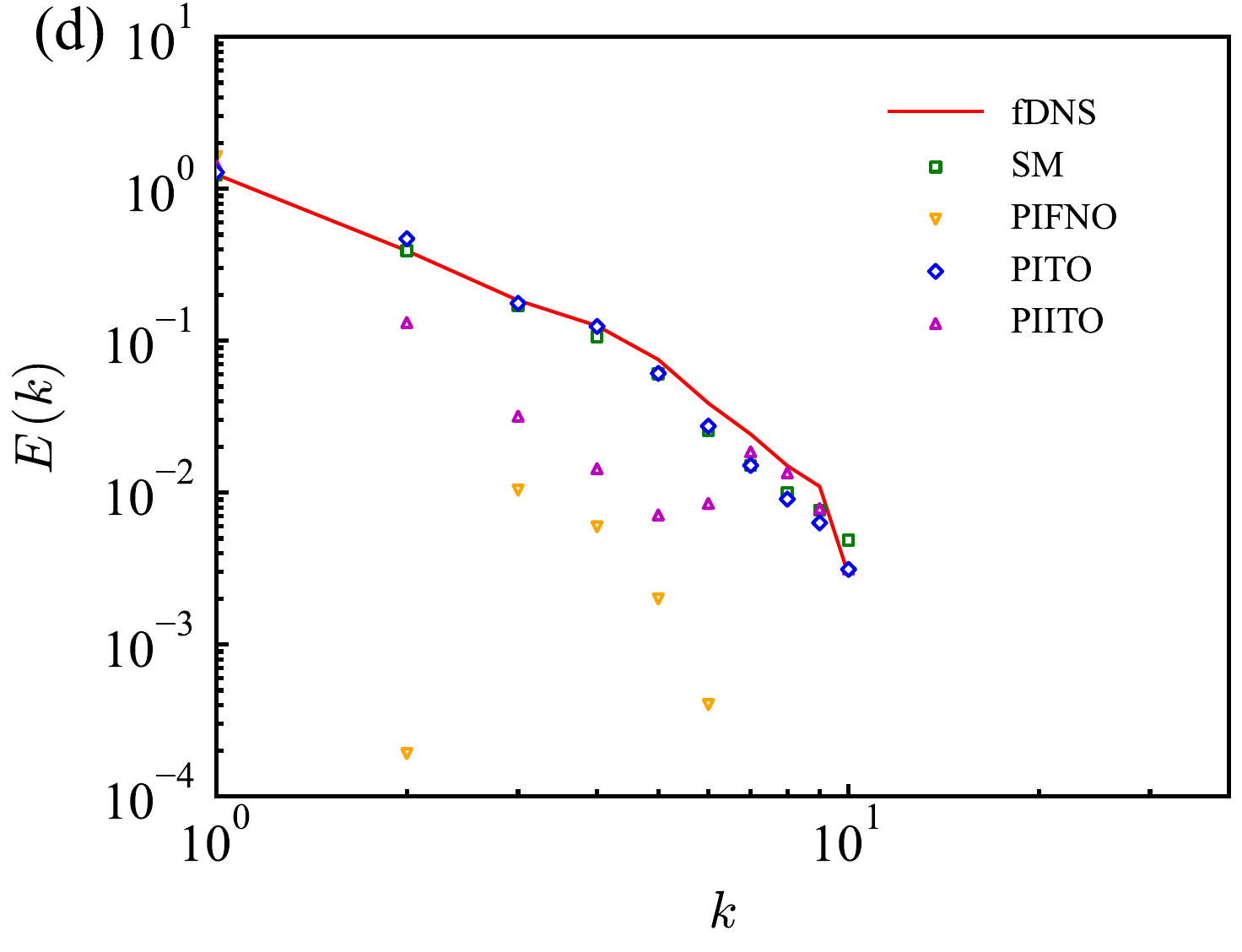} 
    \caption{Turbulent kinetic energy spectra predicted by different models in forced HIT at different time instants: (a) $t \approx \tau$ (b) $t \approx2\tau$ (c) $t \approx4\tau$ (d) $t \approx5\tau$.}
    \label{hit.energy}
\end{figure*}
The energy spectrum at four different instants are shown in Fig.~\ref{hit.energy}. SM results demonstrate excellent consistency with fDNS in long-term predictions. Both PIFNO and PIITO significantly underestimate the energy spectrum for $k \geq 2$. The observation indicates that PIFNO and PIITO fail to learn the multiscale flow structures in turbulence. The energy spectrum predicted by PITO is close to both SM and fDNS.

\begin{figure*}
    \centering
    % (a) FN
    \includegraphics[width=.38\textwidth]{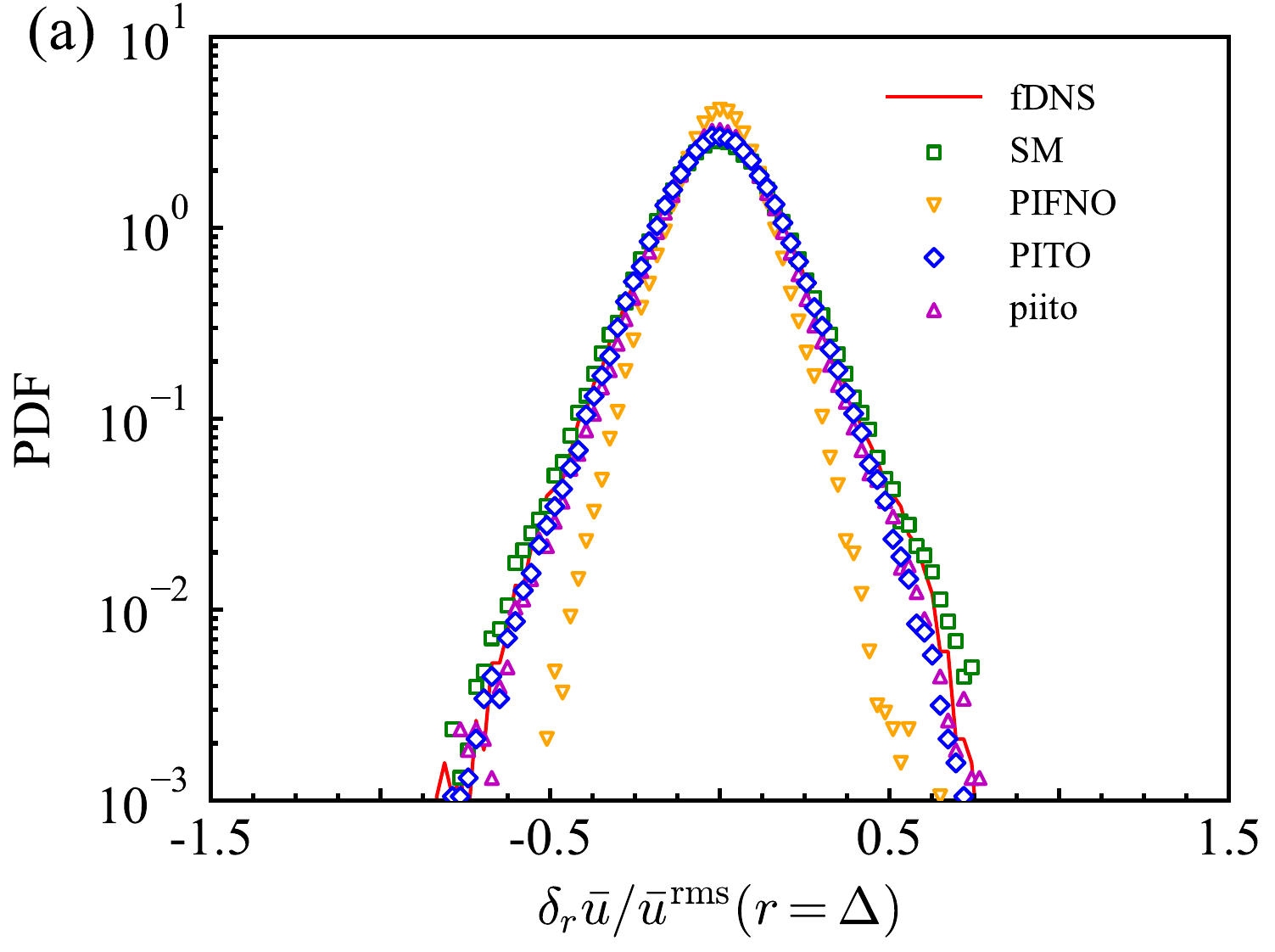} 
    \includegraphics[width=.38\textwidth]{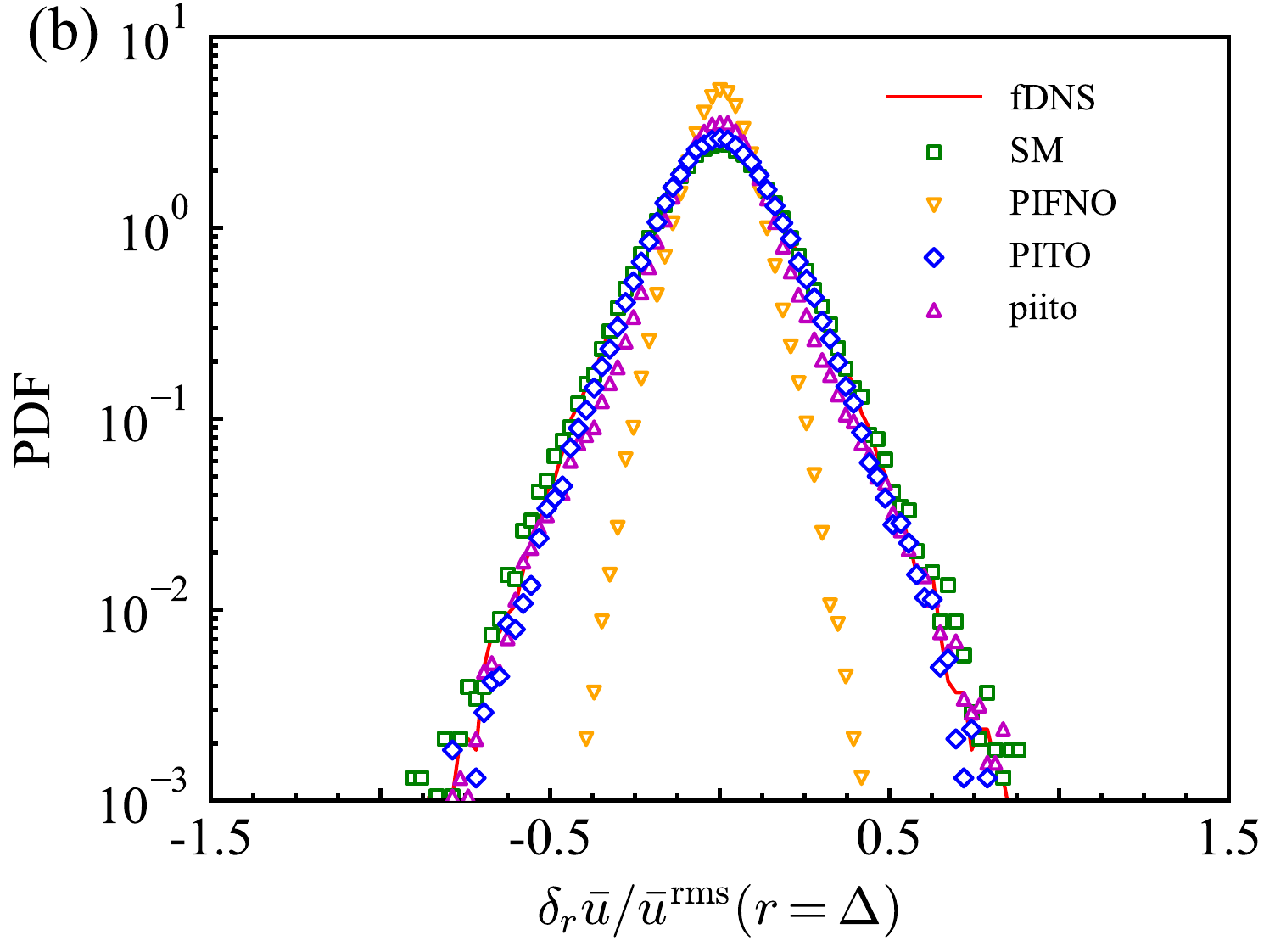}
    \includegraphics[width=.38\textwidth]{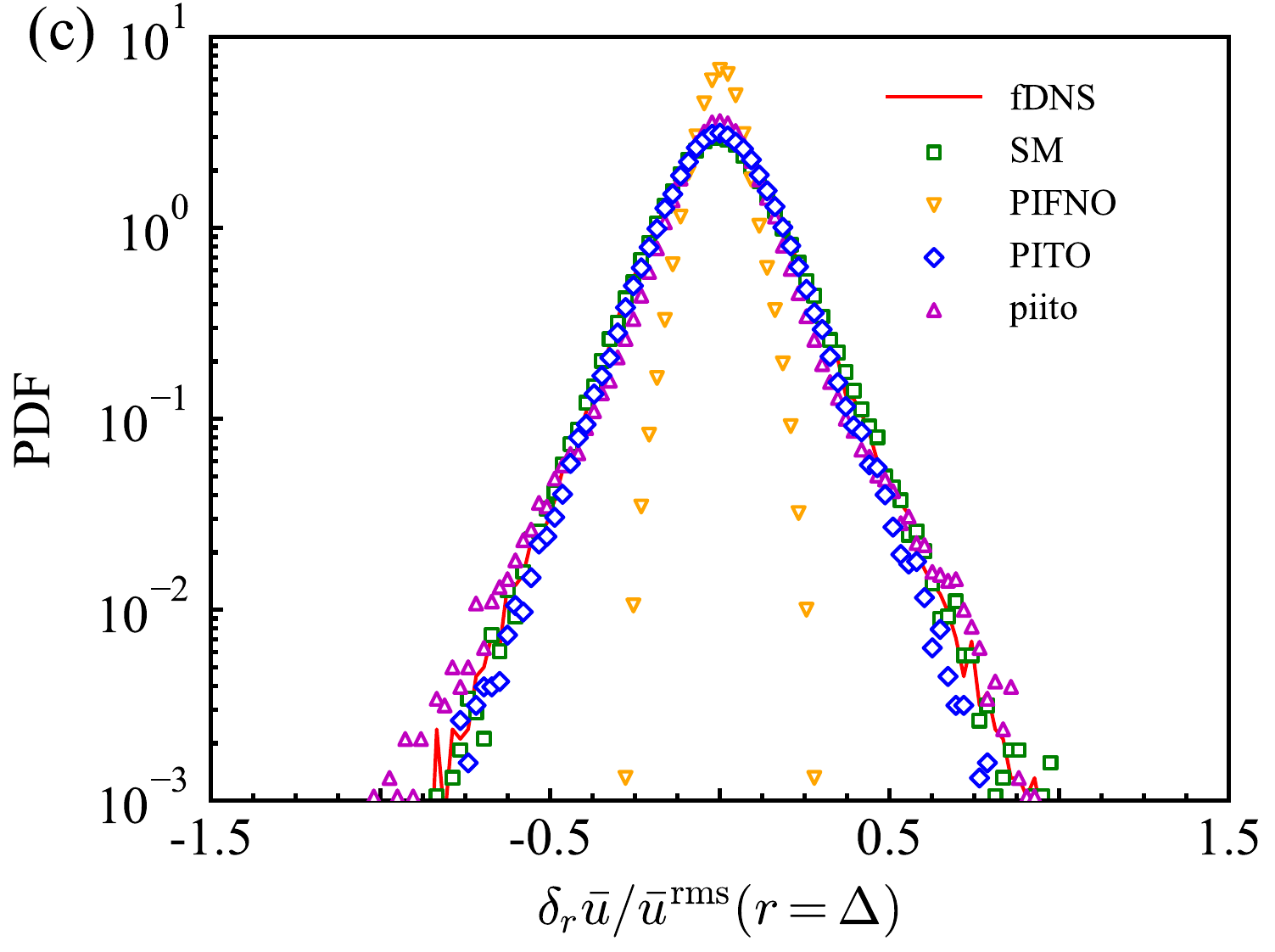} 
    \includegraphics[width=.38\textwidth]{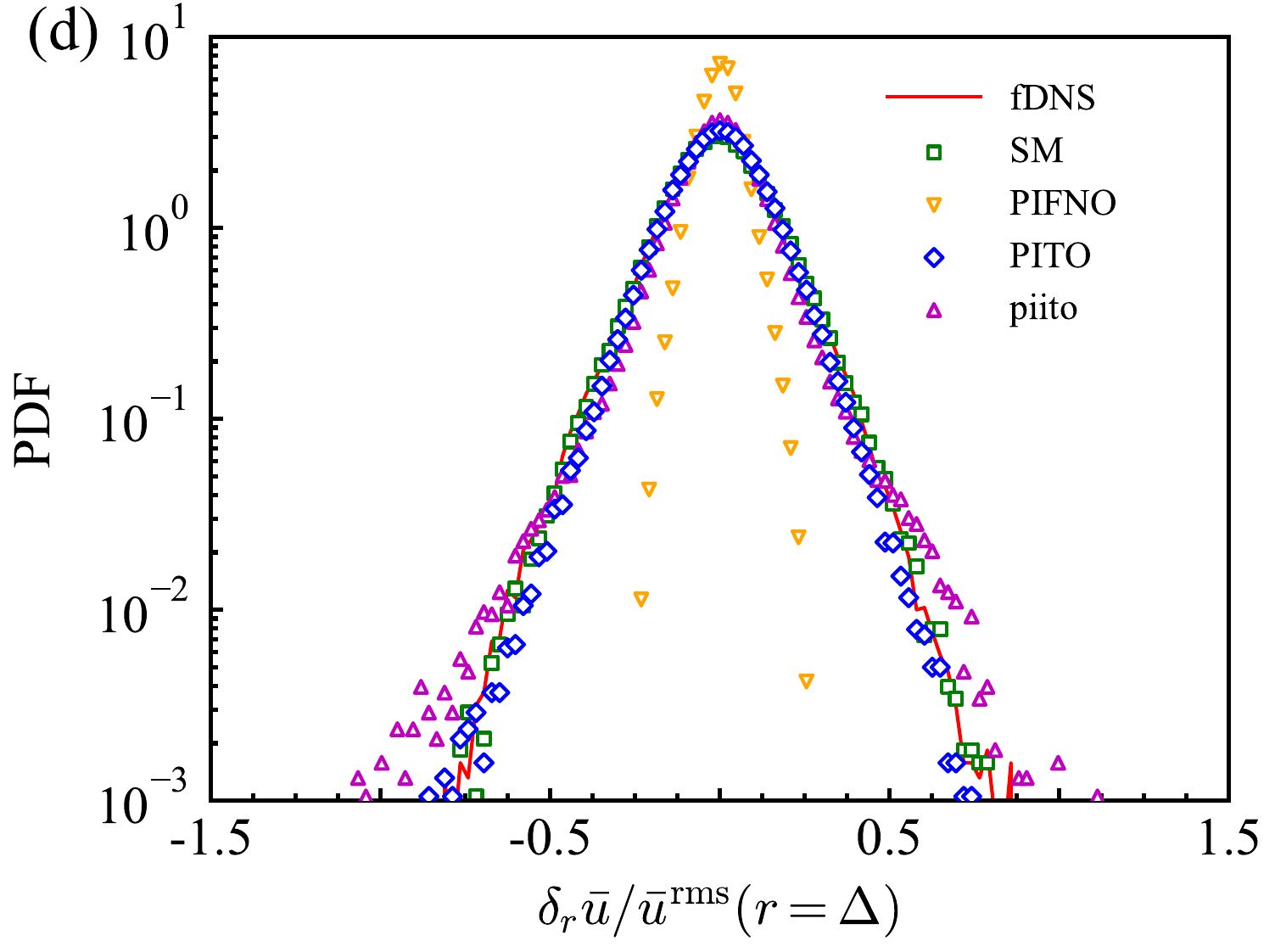} 
    \caption{PDFs of normalized velocity increments predicted by different models in forced HIT at various time instants: (a) $t \approx \tau$ (b) $t \approx2\tau$ (c) $t \approx4\tau$ (d) $t \approx5\tau$.}
    \label{HIT. DELTA U}
\end{figure*}

The PDFs of normalized velocity increments are shown in Fig.~\ref{HIT. DELTA U}. It is shown that PIFNO fails to predict the PDF of velocity increment. PIITO shows a good agreement with fDNS at time $t \approx \tau$, but exhibits some deviations from fDNS results later. However, PITO exhibits a high consistency with fDNS and SM across four distinct time steps, outperforming PIITO and PIFNO models in predicting the PDF of velocity increments.

It is noteworthy that PIITO exhibits a significant performance degradation in forced HIT compared to its performance in decaying HIT. The degrees of freedom of forced HIT is higher than decaying HIT due to its higher Reynolds number, and is more challenging for the prediction. While the parameter-sharing mechanism in PIITO significantly enhances model's efficiency, it restricts the model's total capacity. Within this implicit architecture, the model effectively operates with a parameter scale similar to that of a single layer. Consequently, PIITO fails to maintain a high level of predictive accuracy in the more complex forced HIT case.

\section{Effect of Hyperparameters on Model Performance}

\begin{figure*}
    \centering
    \includegraphics[width=0.6\textwidth]{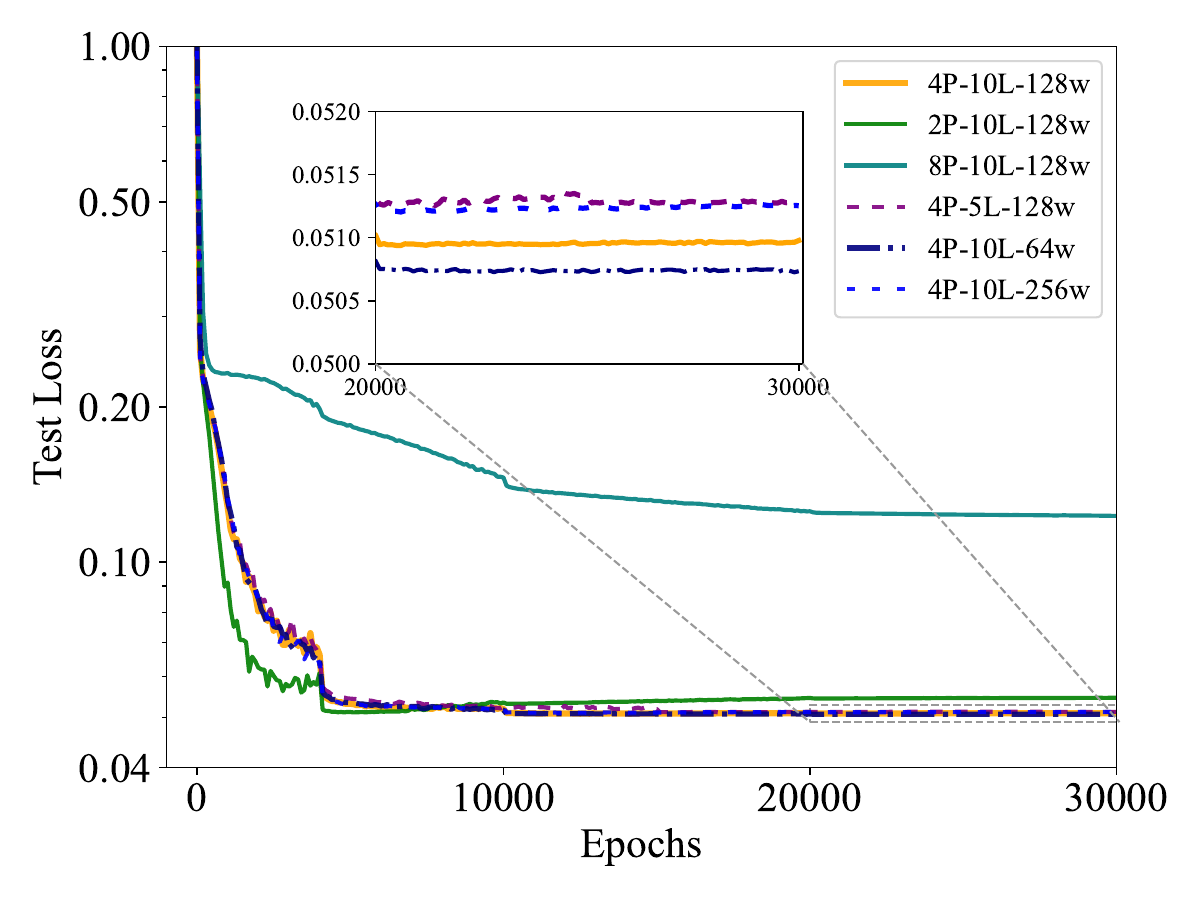} % 调整宽度比例为合适大小
    \caption{Impact of different hyperparameters (patch size, number of layers, and channel width) on the testing loss of the PITO model.}
    \label{ViT_discuss}
\end{figure*}

We further investigate effects of different hyperparameters on the overall performance of PITO model. Specifically, we examine the effects of the patch size $P$, layers $L$ and channel width $w$ on the accuracy of the model in decaying isotropic turbulence at stationary initial condition. As illustrated in Fig.~\ref{ViT_discuss}, the patch size $P$ emerges as the critical parameter governing the model's performance, while the testing loss exhibits relative insensitivity to variations in layers $L$ and channel width $w$.

From a neural operator standpoint, selecting the patch size $P$ is key to matching model complexity appropriately. When $P$ is small (e.g., $P=2$), the increase in the number of tokens leads to too high complexity and training difficulty of the model. Conversely, when $P$ is large (e.g., $P=8$), the significant reduction in the number of tokens results in a smaller parameter scale, giving rise to too weak expressive ability and low precision.

From the perspective of fluid mechanics, the patch size $P$ serves as a pivotal parameter that governs the trade-off between resolving localized flow fluctuations and capturing long-range global correlations across the entire domain. In PITO and PIITO, the model captures internal relationships within each patch through linear combinations, while global relations across patches are modeled via a self-attention mechanism. When $P$ is small (e.g., $P=2$), each patch becomes densely packed with fine-scale information. This causes the self-attention mechanism to prioritize small-scale local fluctuations, relatively preventing it from efficiently capturing the global flow patterns of large-scale structures. Conversely, when $P$ is large (e.g., $P=8$), each patch operates as a macro-scale unit encompassing a substantial segment of the flow field. While this configuration enables the model to effectively capture global governing laws, it concurrently renders the model insensitive to localized fluctuations arising within individual subregions.

\begin{figure*}
    \centering
    % (a) FNO
    \includegraphics[width=.45\textwidth]{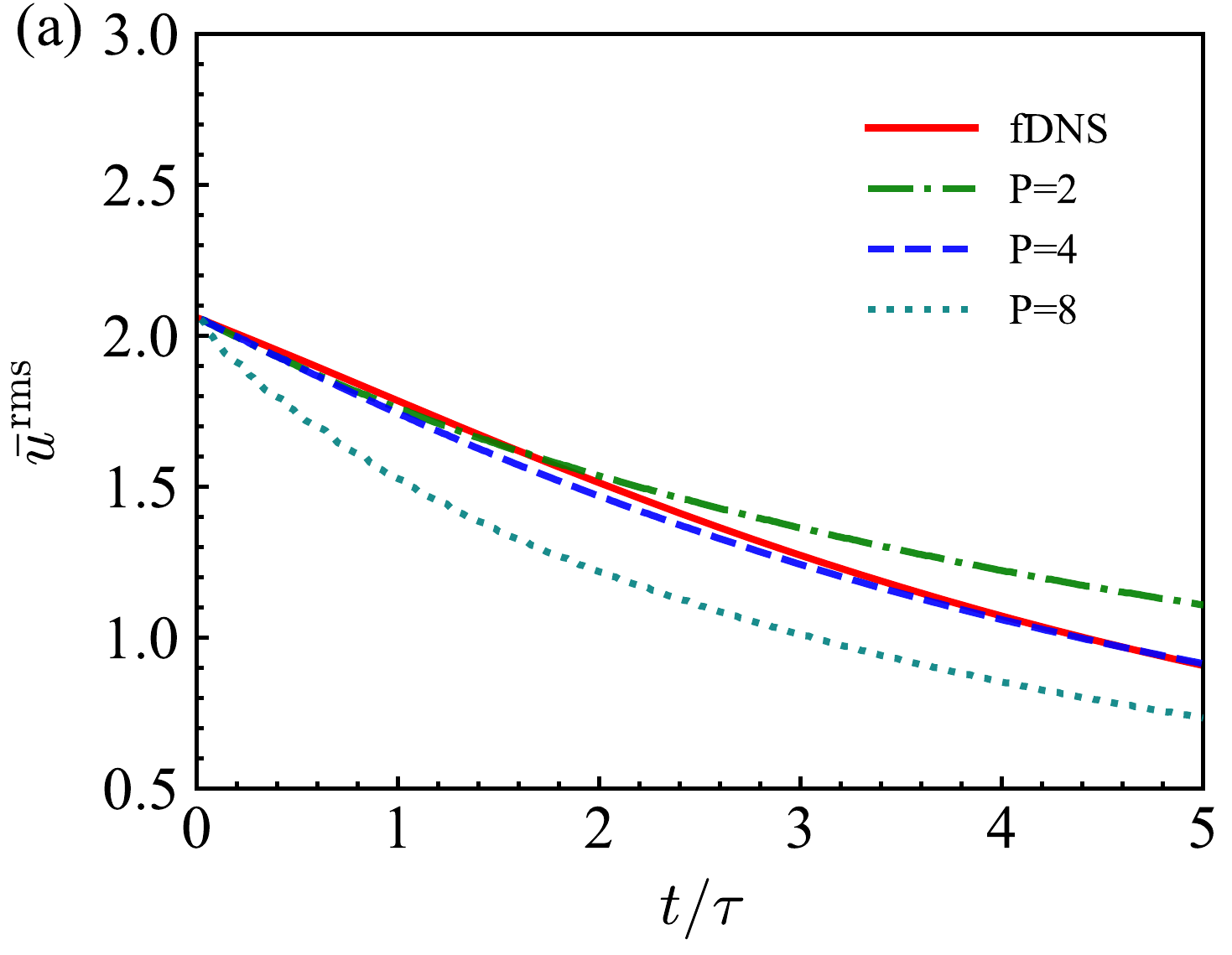} 
    \includegraphics[width=.45\textwidth]{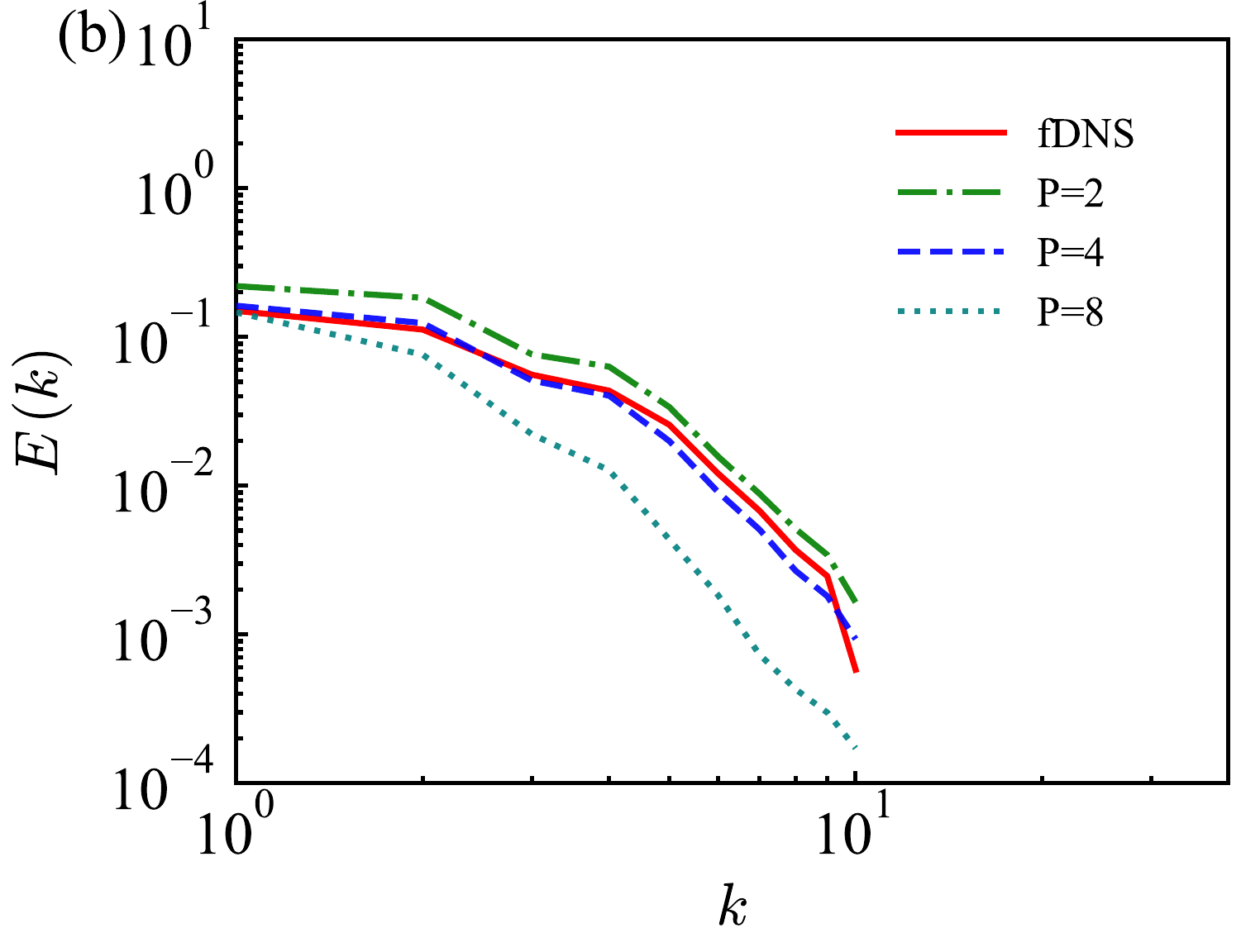}
    \caption{
Effect of patch size ($P$) on the predictions of (a) rms velocity and (b) kinetic energy spectra at $ t /\tau \approx5.0$ by the PITO model.}
\label{ViT_discuss tongji}
\end{figure*}

These theoretical insights are empirically validated through our \textit{a posteriori} performance tests. To systematically evaluate the model’s peak performance, we selected the model checkpoint corresponding to the 5,000th training epoch for downstream analysis. As demonstrated in Fig.~\ref{ViT_discuss tongji}, the configuration with $P=4$ achieves the optimal balance between local resolution and global stability, exhibiting a better agreement with fDNS results for both the root mean square velocity and the spectrum compared to the other two configurations. In contrast, the model with $P=8$ underestimates both the rms velocity and the high-wavenumber energy spectrum, confirming that oversized patch size fails to resolve fine-scale turbulent structures. The model with $P=2$ exhibits overestimation of energy at low wavenumbers and initially aligns with the reference but begins to deviate upwards after approximately $t/\tau > 2$. Consequently, a patch size of $P = 4$ emerges as the optimal resolution scale, achieving a robust balance between capturing small-scale local structures and preserving global physical consistency across the entire flow field. 

\section{Automatic Optimization Of Subgrid-Scale Coefficient}
In the Smagorinsky model for large eddy simulation (LES), the coefficient $C_\text{smag}$ is typically treated as an empirical constant. However, the optimal value of this coefficient is often unknown for different flow conditions. Leveraging the physics-informed architecture, we can embed $C_\text{smag}$ as a trainable parameter within the neural network, allowing the model to automatically identify the optimal coefficient value and minimize the PDE loss \cite{zhao2025lesnets}.

We generate datasets in decaying HIT at stationary initial condition for learning the $C_\text{smag}$ coefficient, which consists of a single set of either SM or fDNS dataset. These datasets are respectively derived from LES simulations employing the Smagorinsky model ($C_\text{smag}^2 = 0.01$) and DNS. It should be emphasized that these datasets are used only for updating the $C_\text{smag}$ coefficient \cite{zhao2025lesnets}. The SM and fDNS datasets are generated over the temporal interval from $t = 10,000\Delta t$ to $10,200\Delta t$ collected every $20\Delta t$. 

During training, the model takes original fDNS dataset and additional SM or fDNS datasets as inputs. Two independent Adam optimizers are employed: one for the neural operator and another for the Smagorinsky coefficient $C_\text{smag}$. The output velocity fields and $C_\text{smag}$ are then both used to compute $\mathcal{L}_{PDE}$. In this method, the Loss is constructed as follows \cite{zhao2025lesnets}:
\begin{equation}
    \mathcal{L} = \mathcal{L}_{PDE} + \gamma \mathcal{L}_{cs},
\end{equation}
where $\gamma$ is a weighting coefficient. The coefficient loss, $\mathcal{L}_{cs}$ is given by the $L^2$ norm of the residual:
\begin{equation}
    \mathcal{L}_{cs} = \|\partial_t \mathbf{\bar{u}} + \mathbf{\bar{u}} \cdot \nabla \mathbf{\bar{u}} + \nabla \bar{p} - \nu \nabla^2 \mathbf{\bar{u}} - \nabla \cdot \boldsymbol{\tau} \|_{L^2(T;D)}^2.
\end{equation}

\begin{figure*}
    \centering
    % (a) FNO
    \includegraphics[width=.38\textwidth]{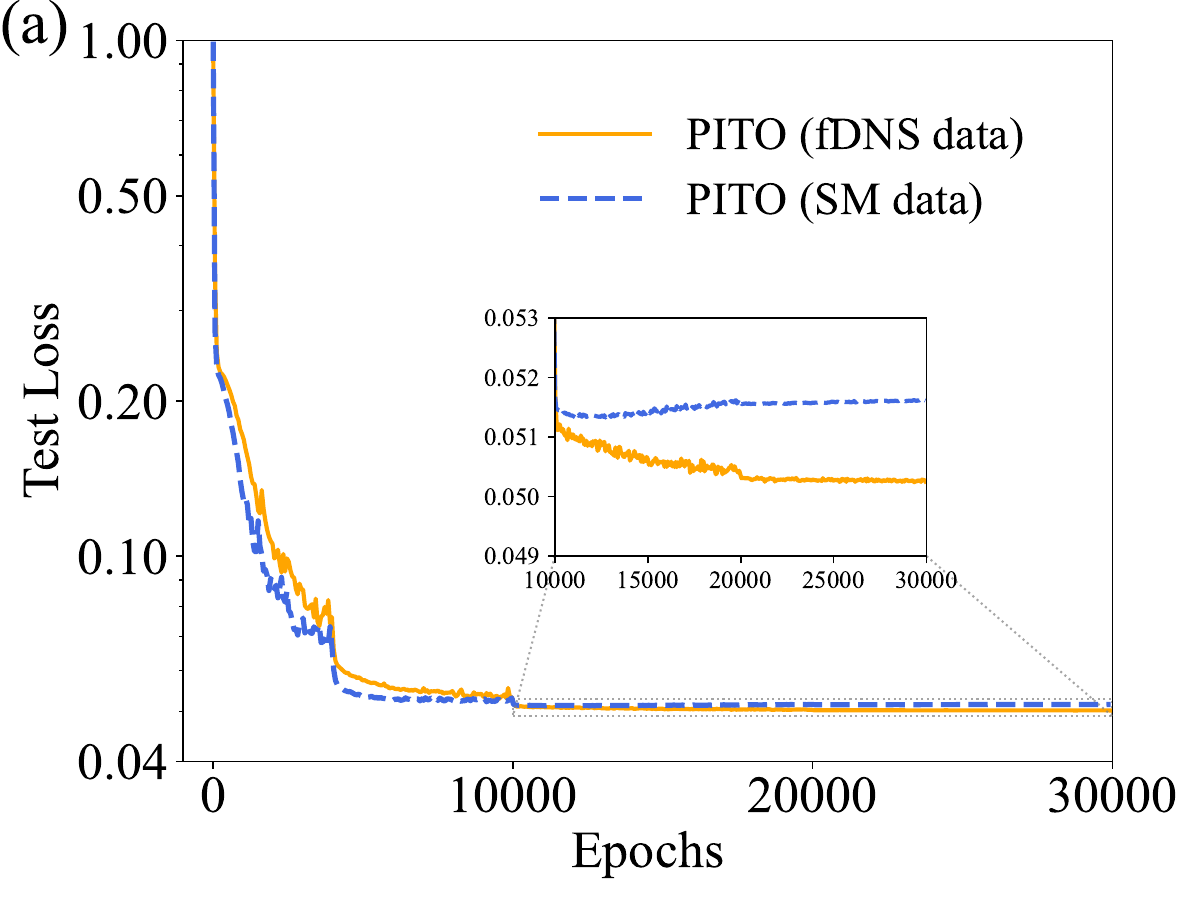} 
    \includegraphics[width=.38\textwidth]{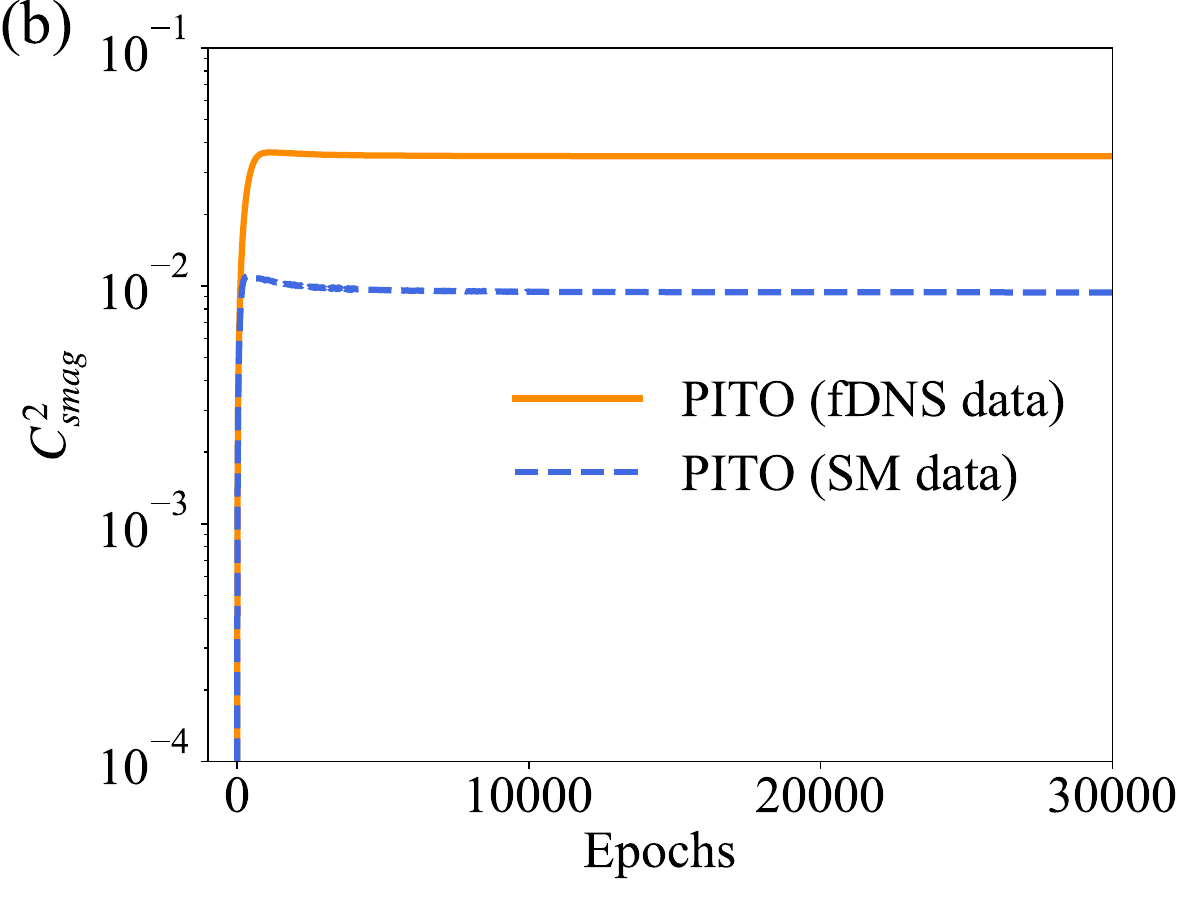}
    \includegraphics[width=.38\textwidth]{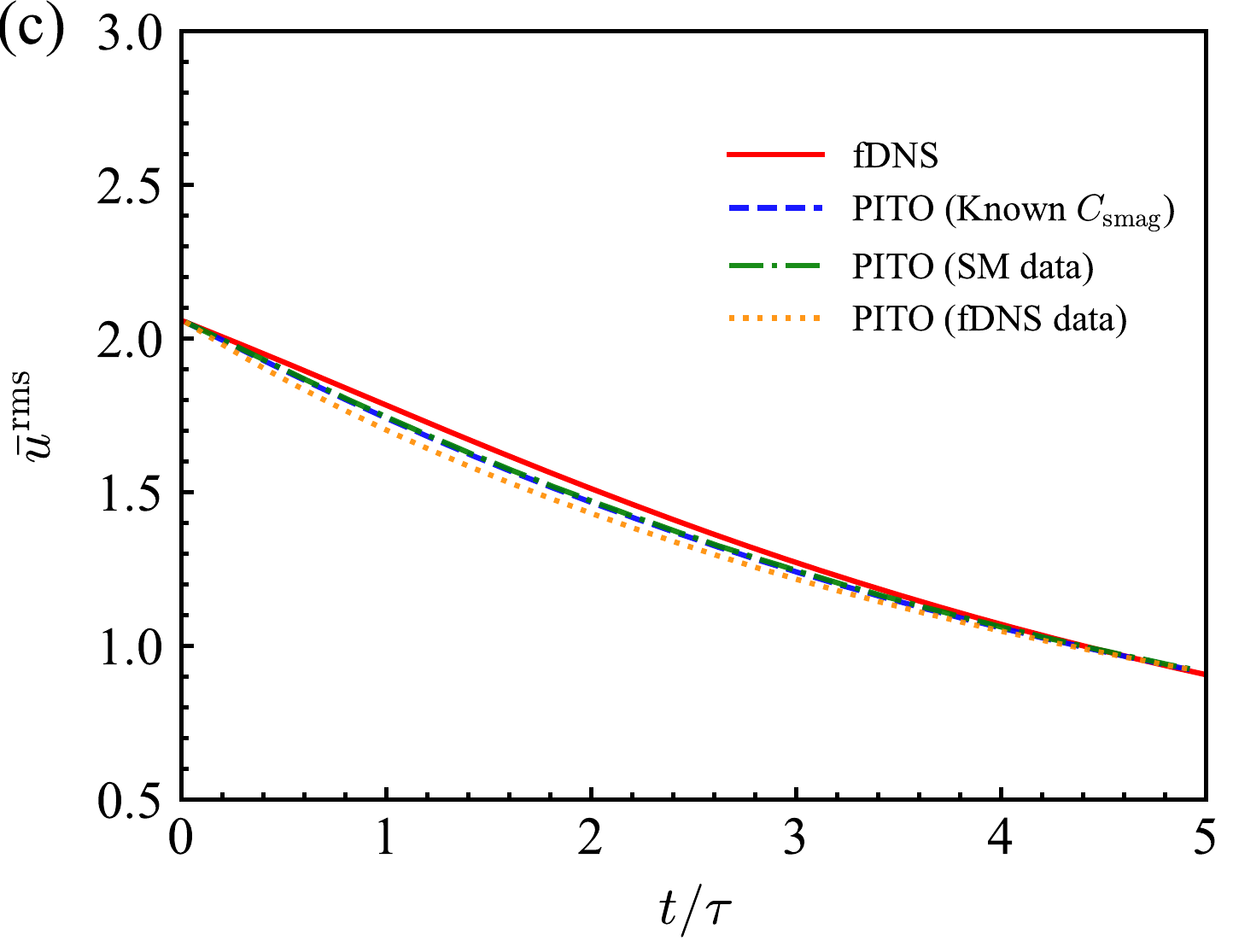} 
    \includegraphics[width=.38\textwidth]{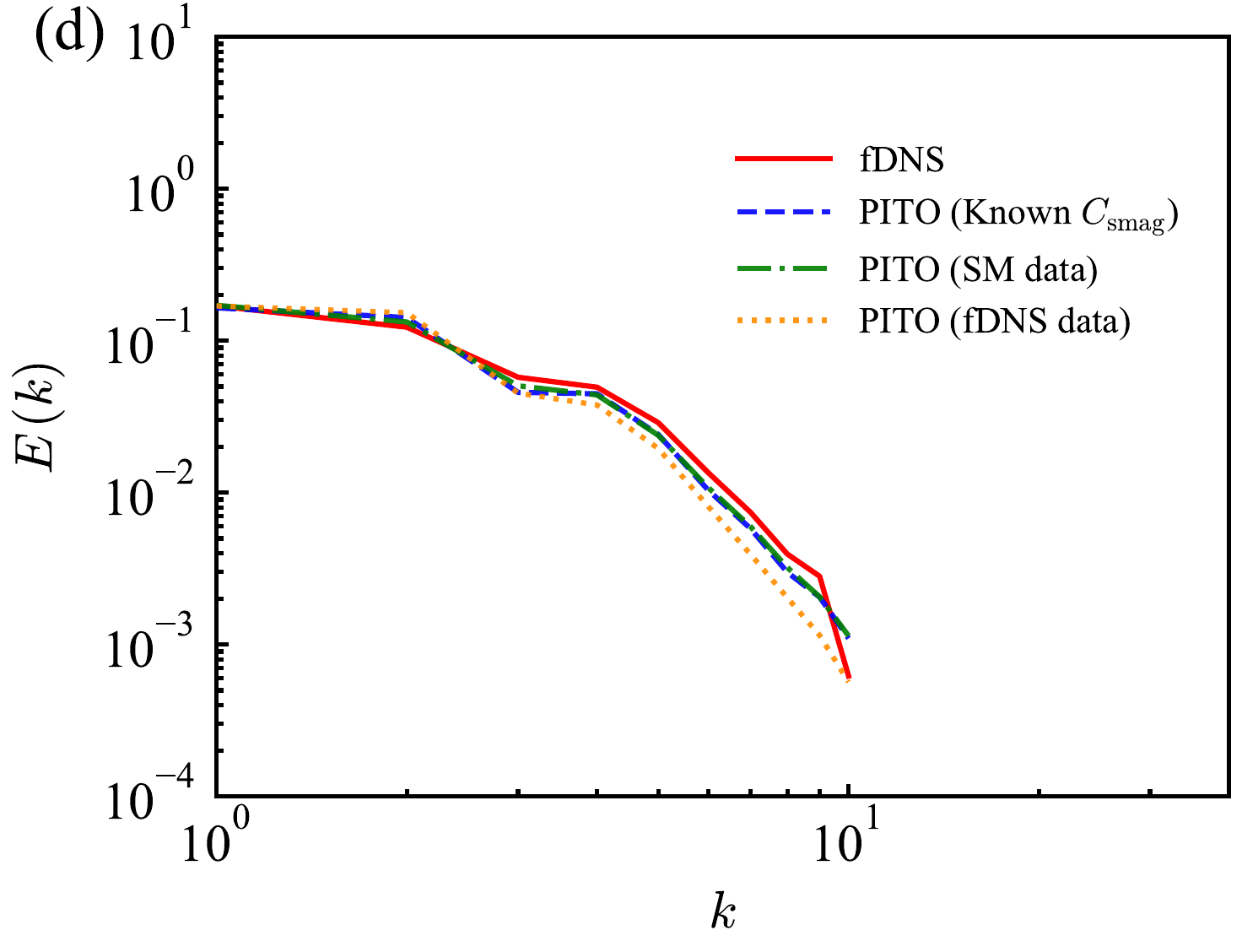}
    \caption{
Performance of the automatic optimization procedure in decaying HIT at at stationary initial condition: (a) evolutions of test loss curves; (b) evolutions of $C^2_\text{smag}$; (c) evolutions of the rms velocity and (d) energy spectra at $t \approx 5\tau$.}
\label{auto}
\end{figure*}

With the parameter set to $\gamma = 50$ \cite{zhao2025lesnets}, we evaluate the automatic learning procedure of the Smagorinsky coefficient using PITO model in the decaying HIT at stationary initial condition. The evolutions of the loss curve and the automatic learning of the coefficient $C_\text{smag}$ are shown in Figs.~\ref{auto}(a) and (b). The Smagorinsky coefficient $C_\text{smag}$ reaches final converged values of 0.0969 and 0.1871 for the SM and fDNS datasets, respectively. Furthermore, in the \textit{a posteriori} tests in Figs.~\ref{auto}(c) and (d), the model demonstrates an excellent agreement with the fDNS results, performing as effectively as the original model.

It is observed that the converged $C_\text{smag}$ values trained from the SM and fDNS datasets exhibit a significant discrepancy. Particularly, the model trained on the SM dataset demonstrates a relatively better performance. fDNS data gives more realistic flow dynamics at coarse grids, but is harder for operator learning. In contrast, LES data using SM is directly generated at coarse grids, so it is easier for operator learning at the same coarse grids. It is worth noting that the prediction of the neural operator model trained under the constraint of the large eddy simulation equation is not necessarily the same as that of the original large eddy simulation equation, due to the training error.

\begin{figure*}
    \centering
    % (a) FNO
    \includegraphics[width=.38\textwidth]{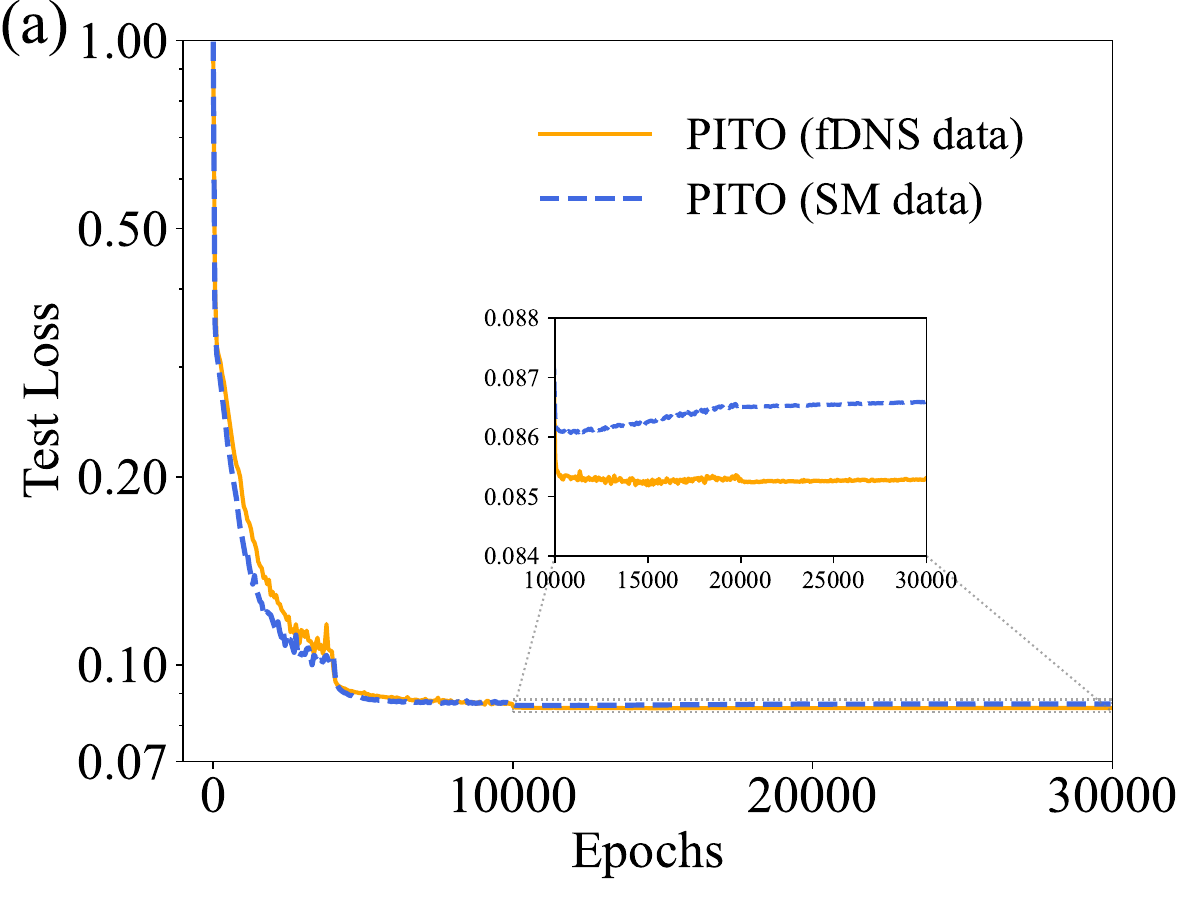} 
    \includegraphics[width=.38\textwidth]{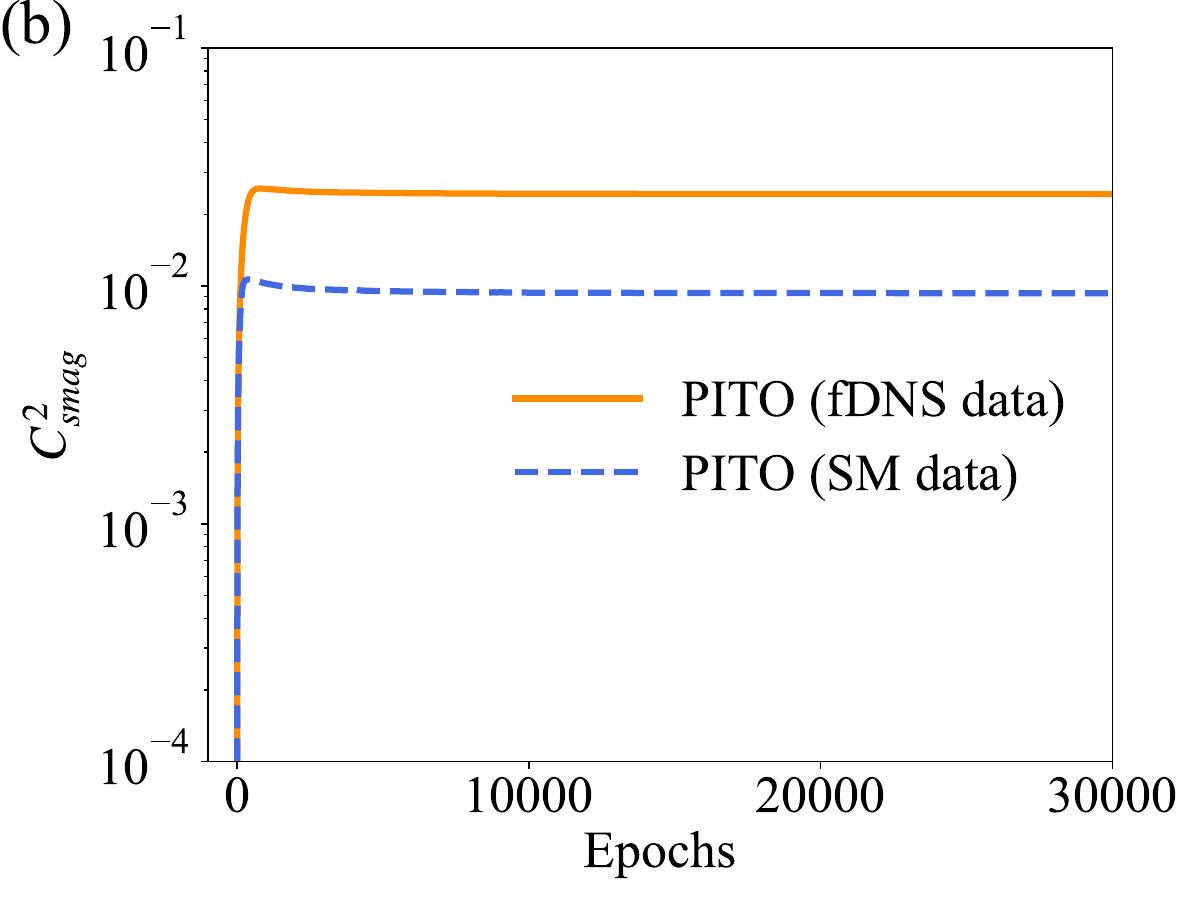}
    \includegraphics[width=.38\textwidth]{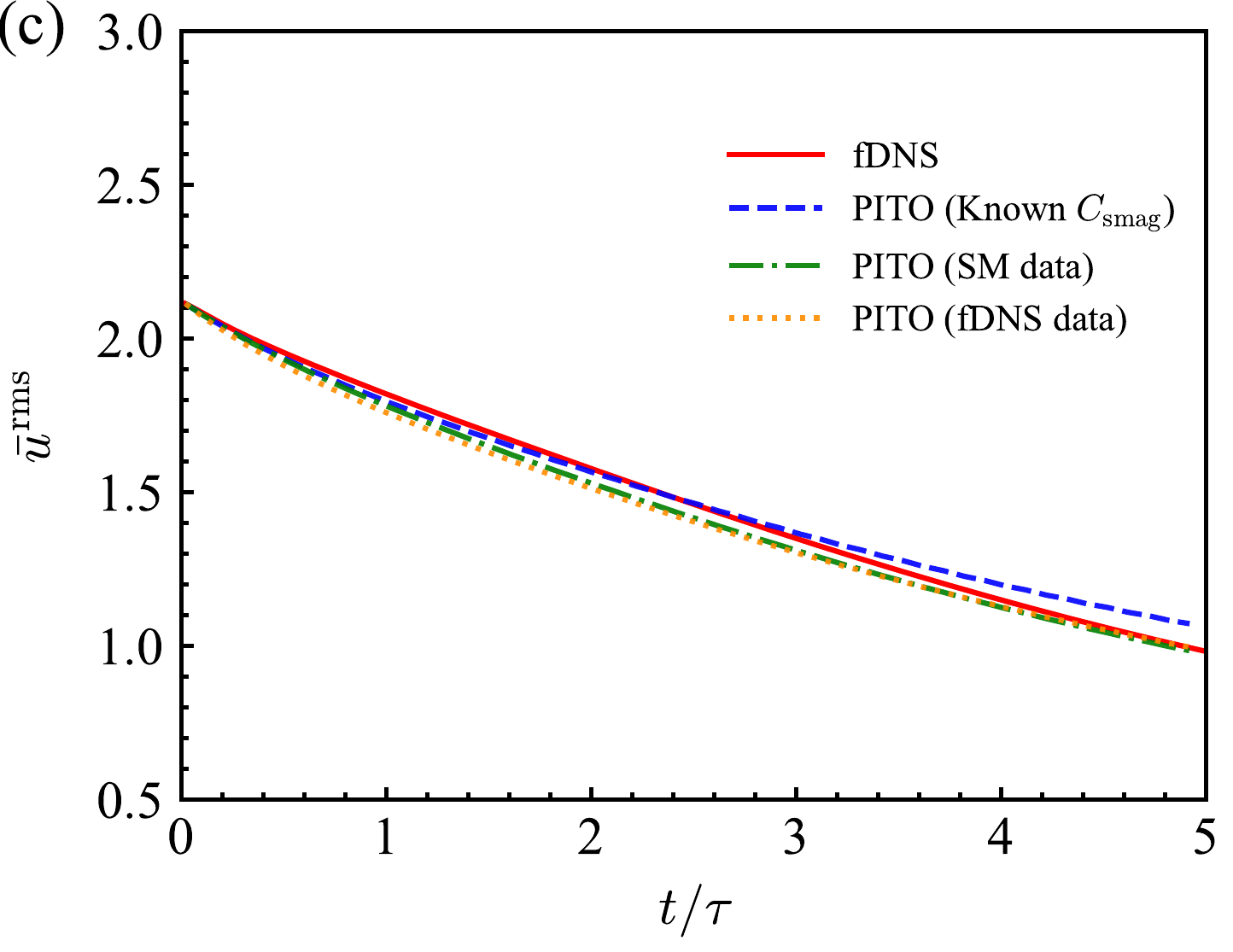} 
    \includegraphics[width=.38\textwidth]{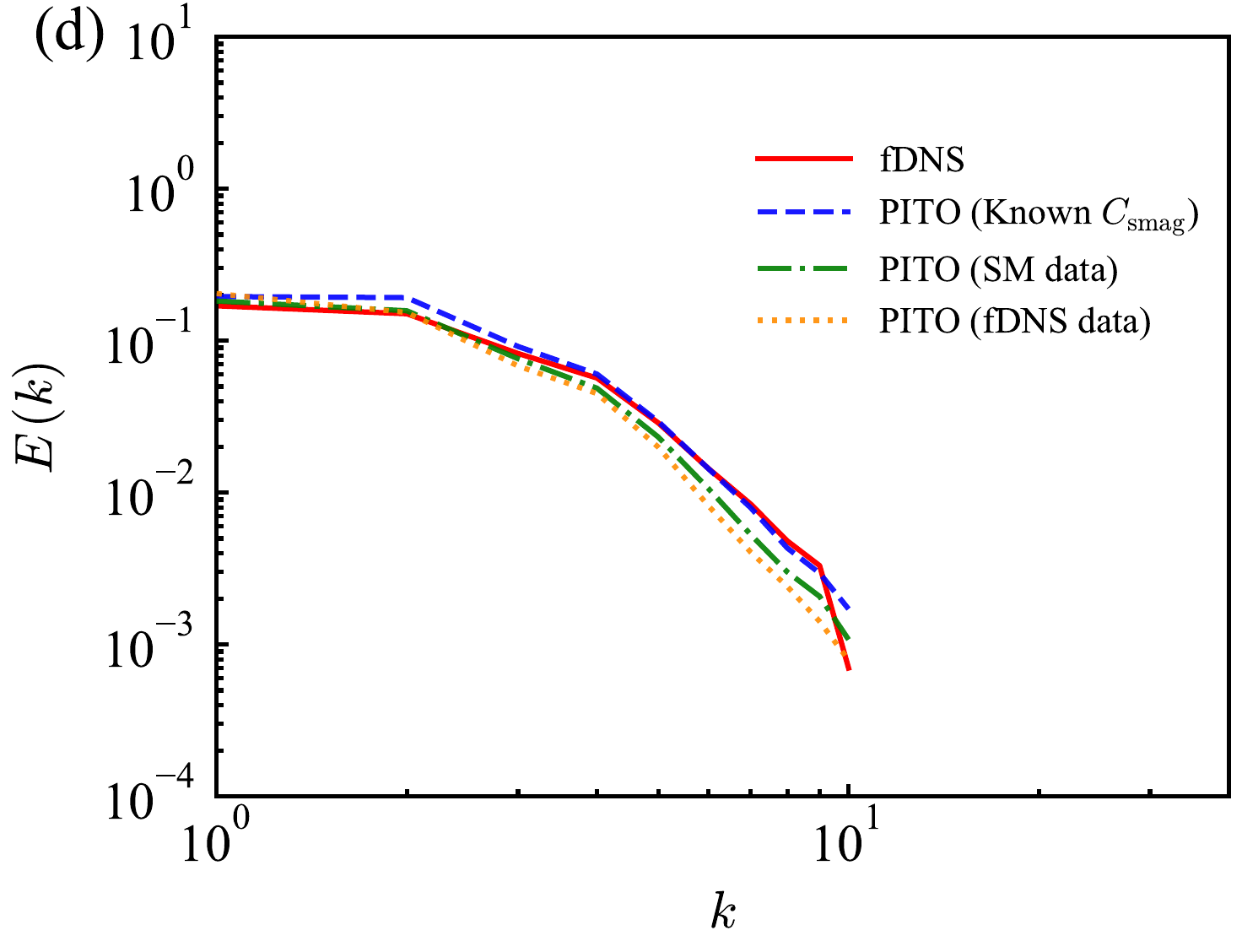}
    \caption{
Performance of the automatic optimization procedure in decaying HIT at at random initial condition: (a) evolutions of test loss curves; (b) evolutions of $C^2_\text{smag}$; (c) evolutions of the rms velocity and (d) energy spectra at $t \approx 5\tau$.}
\label{auto_tongji}
\end{figure*}

Additionally, to further validate the applicability of the method under different initial conditions, we test the automatic optimization methods in decaying HIT at random initial condition, employing a single set of either SM or fDNS dataset at this initial condition. The evolutions of the loss curves and the automatic learning of the coefficient $C_\text{smag}$ are presented in Figs.~\ref{auto_tongji}(a) and (b). The coefficient $C_\text{smag}$ converges to 0.0965 for the SM dataset and 0.1559 for the fDNS dataset. Notably, the learned $C_\text{smag}$ values from the SM dataset under both stationary ($0.0969$) and random ($0.0965$) initial conditions are remarkably close to the ground truth value ($C_\text{smag} = 0.1$). Figs.~\ref{auto_tongji}(c) and (d) show the temporal evolutions of rms velocity and energy spectrum $E(k)$ at $t \approx 5.0\tau$. It can be seen that, our model is capable of automatically learning the $C_\text{smag}$ coefficient from minimal data. Particularly, the model can learn $C_\text{smag}$ directly from fDNS data, meaning that it can be trained without prior knowledge of the SGS model parameters.

\section{Conclusions}
In this study, we proposed a physics-informed Transformer operator (PITO) and its implicit variant (PIITO) for 3D turbulence. By embedding the LES equations with Smagorinsky SGS model directly into the loss function, Transformer neural operator models can be trained without labeled data to predict velocity at the next time step based solely on the current state. In decaying HIT at the statistically steady state initial condition, compared to PIFNO, both PITO and PIITO achieve a higher accuracy while significantly reducing memory requirements. Additionally, in the situation of random initial condition, PITO and PIITO can still maintain high-precision predictions, whereas PIFNO gradually diverges with iterations. In forced HIT, where PIFNO fails to predict the correct energy spectrum, PITO can give quite accurate predictions for the statistical properties of flow fields.

In terms of computational efficiency, PITO and PIITO achieve a 40-fold speedup over the SM model. Remarkably, both PITO and PIITO reduce the parameter count by approximately 68.5\% and 96.9\%, and GPU memory footprint by 79.5\% and 91.3\% in decaying HIT compared to PIFNO, demonstrating outstanding resource efficiency without compromising precision.

The proposed PITO and PIITO models, although achieving satisfactory results in simple 3D turbulence, exhibit certain limitations. Current models are only applicable to regular grids, and subsequent improvements are needed to adapt the model for irregular grids. The current implementation employs the most basic constant-coefficient eddy viscosity model as a physical constraint, which faces challenges in accuracy when applied to complex turbulent flows, necessitating the adoption of more advanced subgrid-scale models in future work. Future work will focus on extending PITO framework to non-uniform and irregular grids and complex geometries using more advanced SGS models. We will explore more diverse network structures to enhance the performance of neural operators, such as diffusion models, which have recently demonstrated great potential in flow prediction problem.

\section*{Code availability}
The code and dataset in this study are available on GitHub at \url{https://github.com/Gzh-sustech/PITO}.
\section*{Acknowledgments}
		This work was supported by the National Natural Science Foundation of China (NSFC) (Grant Nos. 12588301, 12302283, and 12172161); the NSFC Excellence Research Group Program for `Multiscale Problems in Nonlinear Mechanics' (No. 12588201); the Shenzhen Science and Technology Program (Grant Nos. SYSPG20241211173725008, and KQTD20180411143441009); and the Department of Science and Technology of Guangdong Province (Grant Nos. 2019B21203001, 2020B1212030001, and 2023B1212060001). Additional support was provided by the Innovation Capability Support Program of Shaanxi (Program No. 2023-CX-TD-30) and the Center for Computational Science and Engineering of Southern University of Science and Technology.

% \nocite{*}

\appendix
\section{Fourier Neural Operator (FNO)}
\begin{figure*}
    \centering
    \includegraphics[width=0.8\textwidth]{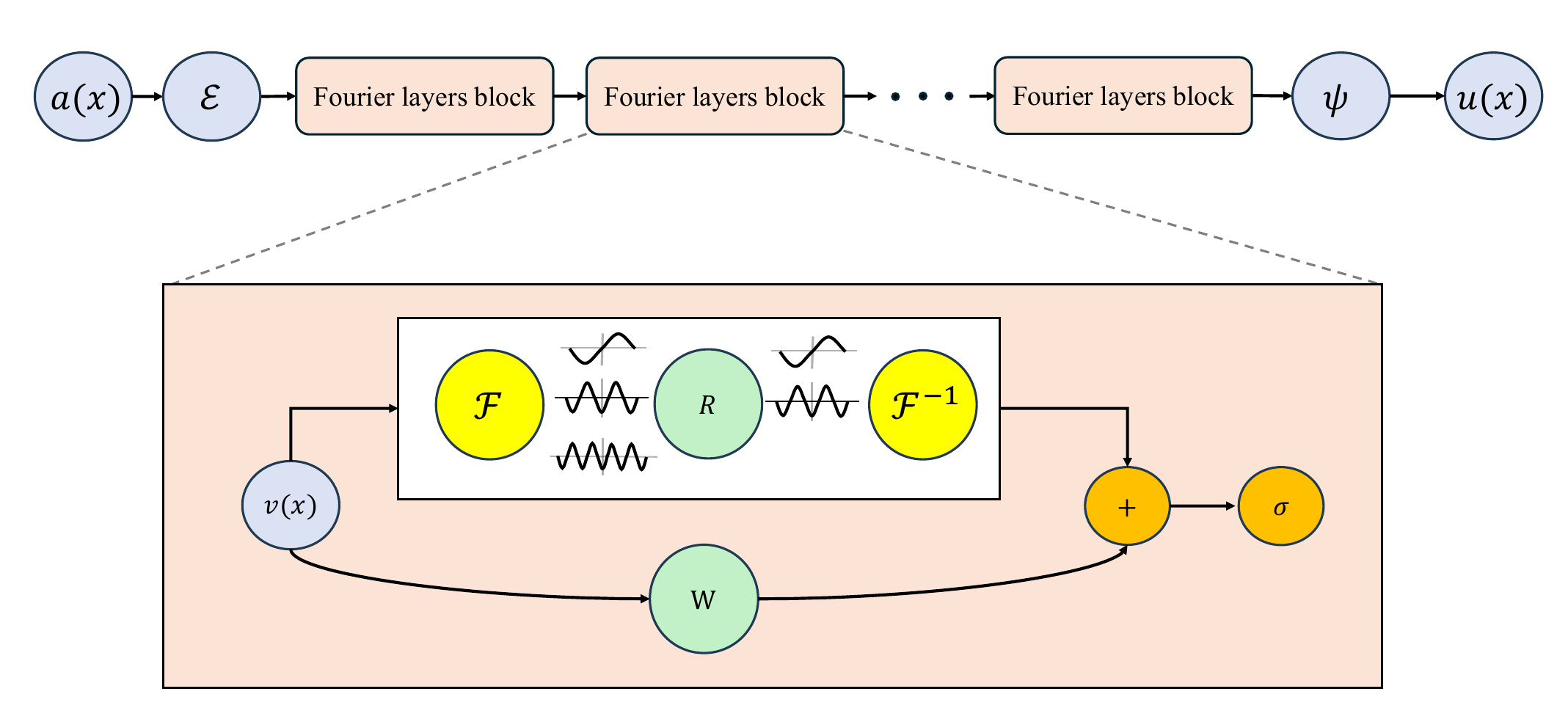} % 调整宽度比例为合适大小
    \caption{The architecture of FNO.}
    \label{FNO}
\end{figure*}
The Fourier neural operator (FNO) is a neural operator that parameterizes the integral kernel in the spectral domain \cite{li2020fourier}. The FNO architecture is shown in Fig.~\ref{FNO}. In the FNO architecture, each Fourier layer updates the latent feature $v_l$ by combining a local linear transformation with a spectral convolution:
\begin{equation}
v_{l+1}(x) := \sigma \big( W v_l(x) + \left( \mathcal{K}(a; \phi) v_l \right)(x) \big), \quad \forall x \in D,
\end{equation}
where $\mathcal{K}(a; \phi)$ denotes a kernel integral operator determined by the input field $a$ and parameterized by $\phi$. $W:\mathbb{R}^{d_v} \to \mathbb{R}^{d_v}$ is a linear transformation.
In FNO, the kernel integral operator $\mathcal{K}$ is given by
\begin{equation}
(\mathcal{K}(\phi)v_l)(x) = \mathcal{F}^{-1} \left( R_{\phi} \cdot (\mathcal{F} v_l) \right)(x), \quad \forall x \in D,
\end{equation}
where $\mathcal{F}$ and $\mathcal{F}^{-1}$ denote the forward and inverse Fast Fourier Transform (FFT), respectively. The weight tensor $R_{\phi}$ functions as a linear mapping in the frequency domain and truncates high-frequency components beyond a cutoff wavenumber $k_{\max}$. The following equation can be derived by multiplying $R_{\phi}$ and$ \mathcal{F} v_l$:
\begin{equation}
(R_{\phi} \cdot (\mathcal{F} v_l))_{k,i} = \sum_{j=1}^{d_v} R_{\phi k,i,j}(\mathcal{F} v_l)_{k,j}, \quad k = 1, \dots, k_{\max}.
\end{equation}

\end{multicols}


\begin{thebibliography}{100}
\providecommand{\url}[1]{{#1}}
\providecommand{\urlprefix}{URL }
\expandafter\ifx\csname urlstyle\endcsname\relax
  \providecommand{\doi}[1]{DOI~\discretionary{}{}{}#1}\else
  \providecommand{\doi}{DOI~\discretionary{}{}{}\begingroup \urlstyle{rm}\Url}\fi

\bibitem{pope}
Pope, S.B.: Turbulent Flows.
\newblock Cambridge University Press (2000)

\bibitem{moin1998direct}
Moin, P., Mahesh, K.: Direct numerical simulation: a tool in turbulence research.
\newblock Annual review of fluid mechanics \textbf{30}(1), 539--578 (1998)

\bibitem{sagaut2006large}
Sagaut, P.: Large eddy simulation for incompressible flows: an introduction.
\newblock Springer (2006)

\bibitem{smagorinsky1963general}
Smagorinsky, J.: General circulation experiments with the primitive equations: I. the basic experiment.
\newblock Monthly weather review \textbf{91}(3), 99--164 (1963)

\bibitem{brunton2020machine}
Brunton, S.L., Noack, B.R., Koumoutsakos, P.: Machine learning for fluid mechanics.
\newblock Annual review of fluid mechanics \textbf{52}(1), 477--508 (2020)

\bibitem{karniadakis2021physics}
Karniadakis, G.E., Kevrekidis, I.G., Lu, L., Perdikaris, P., Wang, S., Yang, L.: Physics-informed machine learning.
\newblock Nature Reviews Physics \textbf{3}(6), 422--440 (2021)

\bibitem{beck2019deep}
Beck, A., Flad, D., Munz, C.D.: Deep neural networks for data-driven {LES} closure models.
\newblock Journal of Computational Physics \textbf{398}, 108910 (2019)

\bibitem{xie2020modeling}
Xie, C., Wang, J., E, W.: Modeling subgrid-scale forces by spatial artificial neural networks in large eddy simulation of turbulence.
\newblock Physical Review Fluids \textbf{5}(5), 054606 (2020)

\bibitem{kurz2023deep}
Kurz, M., Offenh{\"a}user, P., Beck, A.: Deep reinforcement learning for turbulence modeling in large eddy simulations.
\newblock International journal of heat and fluid flow \textbf{99}, 109094 (2023)

\bibitem{maulik2018data}
Maulik, R., San, O., Rasheed, A., Vedula, P.: Data-driven deconvolution for large eddy simulations of kraichnan turbulence.
\newblock Physics of Fluids \textbf{30}(12) (2018)

\bibitem{zhu2019machine}
Zhu, L., Zhang, W., Kou, J., Liu, Y.: Machine learning methods for turbulence modeling in subsonic flows around airfoils.
\newblock Physics of Fluids \textbf{31}(1) (2019)

\bibitem{ling2016reynolds}
Ling, J., Kurzawski, A., Templeton, J.: {Reynolds} averaged turbulence modelling using deep neural networks with embedded invariance.
\newblock Journal of Fluid Mechanics \textbf{807}, 155--166 (2016)

\bibitem{wang2018investigations}
Wang, Z., Luo, K., Li, D., Tan, J., Fan, J.: Investigations of data-driven closure for subgrid-scale stress in large-eddy simulation.
\newblock Physics of Fluids \textbf{30}(12) (2018)

\bibitem{zhou2019subgrid}
Zhou, Z., He, G., Wang, S., Jin, G.: Subgrid-scale model for large-eddy simulation of isotropic turbulent flows using an artificial neural network.
\newblock Computers \& Fluids \textbf{195}, 104319 (2019)

\bibitem{novati2021automating}
Novati, G., de~Laroussilhe, H.L., Koumoutsakos, P.: Automating turbulence modelling by multi-agent reinforcement learning.
\newblock Nature Machine Intelligence \textbf{3}(1), 87--96 (2021)

\bibitem{yang2019predictive}
Yang, X., Zafar, S., Wang, J.X., Xiao, H.: Predictive large-eddy-simulation wall modeling via physics-informed neural networks.
\newblock Physical Review Fluids \textbf{4}(3), 034602 (2019)

\bibitem{srinivasan2019predictions}
Srinivasan, P.A., Guastoni, L., Azizpour, H., Schlatter, P., Vinuesa, R.: Predictions of turbulent shear flows using deep neural networks.
\newblock Physical Review Fluids \textbf{4}(5), 054603 (2019)

\bibitem{ruhling2023dyffusion}
R{\"u}hling~Cachay, S., Zhao, B., Joren, H., Yu, R.: Dyffusion: A dynamics-informed diffusion model for spatiotemporal forecasting.
\newblock Advances in neural information processing systems \textbf{36}, 45259--45287 (2023)

\bibitem{nakamura2021convolutional}
Nakamura, T., Fukami, K., Hasegawa, K., Nabae, Y., Fukagata, K.: Convolutional neural network and long short-term memory based reduced order surrogate for minimal turbulent channel flow.
\newblock Physics of Fluids \textbf{33}(2) (2021)

\bibitem{bukka2021assessment}
Bukka, S.R., Gupta, R., Magee, A.R., Jaiman, R.K.: Assessment of unsteady flow predictions using hybrid deep learning based reduced-order models.
\newblock Physics of Fluids \textbf{33}(1) (2021)

\bibitem{raissi2017physics}
Raissi, M., Perdikaris, P., Karniadakis, G.E.: Physics informed deep learning (part i): Data-driven solutions of nonlinear partial differential equations.
\newblock arXiv preprint arXiv:1711.10561  (2017)

\bibitem{cai2021physics}
Cai, S., Mao, Z., Wang, Z., Yin, M., Karniadakis, G.E.: Physics-informed neural networks (pinns) for fluid mechanics: A review.
\newblock Acta Mechanica Sinica \textbf{37}(12), 1727--1738 (2021)

\bibitem{sun2026reconstruction}
Sun, B., Cai, S., Du, Q., Wang, Z., Chen, Y., Tian, Z., Zhu, J.: Reconstruction of fields based on physics-informed neural networks with sensor placement optimization.
\newblock Acta Mechanica Sinica \textbf{42}(7), 725195 (2026)

\bibitem{zhu2026physics}
Zhu, Y., Chen, W., Deng, J., Bian, X.: Physics-informed neural networks for hidden boundary detection and flow field reconstruction.
\newblock Acta Mechanica Sinica \textbf{42}(7), 725273 (2026)

\bibitem{Wu2024high}
Wu, H., Zhang, K., Zhou, D., Chen, W.L., Han, Z., Cao, Y.: High-flexibility reconstruction of small-scale motions in wall turbulence using a generalized zero-shot learning.
\newblock Journal of Fluid Mechanics \textbf{990}, R1 (2024)

\bibitem{Wu2025_general}
Wu, H., Cao, Y., Chen, Y., Laima, S., Chen, W.L., Zhou, D., Li, H.: Generalizable super-resolution turbulence reconstruction from minimal training data.
\newblock Journal of Fluid Mechanics \textbf{1024}, A27 (2025)

\bibitem{du2024conditional}
Du, P., Parikh, M.H., Fan, X., Liu, X.Y., Wang, J.X.: Conditional neural field latent diffusion model for generating spatiotemporal turbulence.
\newblock Nature Communications \textbf{15}(1), 10416 (2024)

\bibitem{jiang2025integrating}
Jiang, Y., Wang, Y., Yang, H., Wang, J.: Integrating fourier neural operator with diffusion model for autoregressive predictions of three-dimensional turbulence.
\newblock arXiv preprint arXiv:2512.12628  (2025)

\bibitem{oommen2026learning}
Oommen, V., Khodakarami, S., Bora, A., Wang, Z., Karniadakis, G.E.: Learning turbulent flows with generative models for super resolution and sparse flow reconstruction.
\newblock Nature Communications  (2026)

\bibitem{raissi2019physics}
Raissi, M., Perdikaris, P., Karniadakis, G.E.: Physics-informed neural networks: A deep learning framework for solving forward and inverse problems involving nonlinear partial differential equations.
\newblock Journal of Computational physics \textbf{378}, 686--707 (2019)

\bibitem{wang2020towards}
Wang, R., Kashinath, K., Mustafa, M., Albert, A., Yu, R.: Towards physics-informed deep learning for turbulent flow prediction.
\newblock In: Proceedings of the 26th ACM SIGKDD international conference on knowledge discovery \& data mining, pp. 1457--1466 (2020)

\bibitem{jin2021nsfnets}
Jin, X., Cai, S., Li, H., Karniadakis, G.E.: Nsfnets ({Navier-Stokes} flow nets): Physics-informed neural networks for the incompressible {Navier-Stokes} equations.
\newblock Journal of Computational Physics \textbf{426}, 109951 (2021)

\bibitem{zhang2026physics}
Zhang, W., Suo, W., Song, J., Cao, W.: Physics-informed neural networks ({PINNs}) as intelligent computing technique for solving partial differential equations: Limitation and future prospects.
\newblock SCIENCE CHINA Physics, Mechanics \& Astronomy \textbf{69}(1), 214602 (2026)

\bibitem{wang2022and}
Wang, S., Yu, X., Perdikaris, P.: When and why {PINNs} fail to train: A neural tangent kernel perspective.
\newblock Journal of Computational Physics \textbf{449}, 110768 (2022)

\bibitem{wang2021understanding}
Wang, S., Teng, Y., Perdikaris, P.: Understanding and mitigating gradient flow pathologies in physics-informed neural networks.
\newblock SIAM Journal on Scientific Computing \textbf{43}(5), A3055--A3081 (2021)

\bibitem{jagtap2020adaptive}
Jagtap, A.D., Kawaguchi, K., Karniadakis, G.E.: Adaptive activation functions accelerate convergence in deep and physics-informed neural networks.
\newblock Journal of Computational Physics \textbf{404}, 109136 (2020)

\bibitem{yang2021b}
Yang, L., Meng, X., Karniadakis, G.E.: B-pinns: Bayesian physics-informed neural networks for forward and inverse {PDE} problems with noisy data.
\newblock Journal of Computational Physics \textbf{425}, 109913 (2021)

\bibitem{wang2025simulating}
Wang, S., Sankaran, S., Fan, X., Stinis, P., Perdikaris, P.: Simulating three-dimensional turbulence with physics-informed neural networks.
\newblock arXiv preprint arXiv:2507.08972  (2025)

\bibitem{wang2025gradient}
Wang, S., Bhartari, A.K., Li, B., Perdikaris, P.: Gradient alignment in physics-informed neural networks: A second-order optimization perspective.
\newblock arXiv preprint arXiv:2502.00604  (2025)

\bibitem{zhang2025physics}
Zhang, Z., Su, X., Yuan, X.: Physics-informed neural networks with coordinate transformation to solve high {Reynolds} number boundary layer flows.
\newblock Journal of Computational Physics p. 114338 (2025)

\bibitem{cao2024solver}
Cao, W., Song, J., Zhang, W.: A solver for subsonic flow around airfoils based on physics-informed neural networks and mesh transformation.
\newblock Physics of Fluids \textbf{36}(2) (2024)

\bibitem{cao2025analysis}
Cao, W., Zhang, W.: An analysis and solution of ill-conditioning in physics-informed neural networks.
\newblock Journal of Computational Physics \textbf{520}, 113494 (2025)

\bibitem{zhao2023pinnsformer}
Zhao, Z., Ding, X., Prakash, B.A.: Pinnsformer: A transformer-based framework for physics-informed neural networks.
\newblock arXiv preprint arXiv:2307.11833  (2023)

\bibitem{lu2021learning}
Lu, L., Jin, P., Pang, G., Zhang, Z., Karniadakis, G.E.: Learning nonlinear operators via {DeepONet} based on the universal approximation theorem of operators.
\newblock Nature machine intelligence \textbf{3}(3), 218--229 (2021)

\bibitem{li2023geometry}
Li, Z., Kovachki, N., Choy, C., Li, B., Kossaifi, J., Otta, S., Nabian, M.A., Stadler, M., Hundt, C., Azizzadenesheli, K., et~al.: Geometry-informed neural operator for large-scale {3D} {PDE}s.
\newblock Advances in Neural Information Processing Systems \textbf{36}, 35836--35854 (2023)

\bibitem{li2024Transformer}
Li, Z., Liu, T., Peng, W., Yuan, Z., Wang, J.: A transformer-based neural operator for large-eddy simulation of turbulence.
\newblock Physics of Fluids \textbf{36}(6) (2024)

\bibitem{luo2024fourier}
Luo, T., Li, Z., Yuan, Z., Peng, W., Liu, T., Wang, L.L., Wang, J.: Fourier neural operator for large eddy simulation of compressible {Rayleigh--Taylor} turbulence.
\newblock Physics of Fluids \textbf{36}(7) (2024)

\bibitem{peng2023linear}
Peng, W., Yuan, Z., Li, Z., Wang, J.: Linear attention coupled {Fourier} neural operator for simulation of three-dimensional turbulence.
\newblock Physics of Fluids \textbf{35}(1) (2023)

\bibitem{hao2023gnot}
Hao, Z., Wang, Z., Su, H., Ying, C., Dong, Y., Liu, S., Cheng, Z., Song, J., Zhu, J.: Gnot: A general neural operator transformer for operator learning.
\newblock In: International Conference on Machine Learning, pp. 12556--12569. PMLR (2023)

\bibitem{li2020fourier}
Li, Z., Kovachki, N., Azizzadenesheli, K., Liu, B., Bhattacharya, K., Stuart, A., Anandkumar, A.: Fourier neural operator for parametric partial differential equations.
\newblock arXiv preprint arXiv:2010.08895  (2020)

\bibitem{you2022learning}
You, H., Zhang, Q., Ross, C.J., Lee, C.H., Yu, Y.: Learning deep implicit {Fourier} neural operators ({IFNO}s) with applications to heterogeneous material modeling.
\newblock Computer Methods in Applied Mechanics and Engineering \textbf{398}, 115296 (2022)

\bibitem{tran2021factorized}
Tran, A., Mathews, A., Xie, L., Ong, C.S.: Factorized {Fourier} neural operators.
\newblock arXiv preprint arXiv:2111.13802  (2021)

\bibitem{peng2022attention}
Peng, W., Yuan, Z., Wang, J.: Attention-enhanced neural network models for turbulence simulation.
\newblock Physics of Fluids \textbf{34}(2) (2022)

\bibitem{li2022fourier}
Li, Z., Peng, W., Yuan, Z., Wang, J.: Fourier neural operator approach to large eddy simulation of three-dimensional turbulence.
\newblock Theoretical and Applied Mechanics Letters \textbf{12}(6), 100389 (2022)

\bibitem{li2023long}
Li, Z., Peng, W., Yuan, Z., Wang, J.: Long-term predictions of turbulence by implicit {U-Net} enhanced {Fourier} neural operator.
\newblock Physics of Fluids \textbf{35}(7) (2023)

\bibitem{wang2024prediction}
Wang, Y., Li, Z., Yuan, Z., Peng, W., Liu, T., Wang, J.: Prediction of turbulent channel flow using {Fourier} neural operator-based machine-learning strategy.
\newblock Physical Review Fluids \textbf{9}(8), 084604 (2024)

\bibitem{li2023fourier}
Li, Z., Huang, D.Z., Liu, B., Anandkumar, A.: Fourier neural operator with learned deformations for {PDE}s on general geometries.
\newblock Journal of Machine Learning Research \textbf{24}(388), 1--26 (2023)

\bibitem{vaswani2017attention}
Vaswani, A., Shazeer, N., Parmar, N., Uszkoreit, J., Jones, L., Gomez, A.N., Kaiser, {\L}., Polosukhin, I.: Attention is all you need.
\newblock Advances in neural information processing systems \textbf{30} (2017)

\bibitem{khan2022Transformers}
Khan, S., Naseer, M., Hayat, M., Zamir, S.W., Khan, F.S., Shah, M.: Transformers in vision: A survey.
\newblock ACM computing surveys (CSUR) \textbf{54}(10s), 1--41 (2022)

\bibitem{liu2021swin}
Liu, Z., Lin, Y., Cao, Y., Hu, H., Wei, Y., Zhang, Z., Lin, S., Guo, B.: Swin transformer: Hierarchical vision transformer using shifted windows.
\newblock In: Proceedings of the IEEE/CVF international conference on computer vision, pp. 10012--10022 (2021)

\bibitem{dosovitskiy2020image}
Dosovitskiy, A.: An image is worth 16x16 words: Transformers for image recognition at scale.
\newblock arXiv preprint arXiv:2010.11929  (2020)

\bibitem{wu2024transolver}
Wu, H., Luo, H., Wang, H., Wang, J., Long, M.: Transolver: A fast transformer solver for {PDEs} on general geometries.
\newblock arXiv preprint arXiv:2402.02366  (2024)

\bibitem{hao2024dpot}
Hao, Z., Su, C., Liu, S., Berner, J., Ying, C., Su, H., Anandkumar, A., Song, J., Zhu, J.: Dpot: Auto-regressive denoising operator transformer for large-scale pde pre-training.
\newblock arXiv preprint arXiv:2403.03542  (2024)

\bibitem{chen2024omniarch}
Chen, T., Zhou, H., Li, Y., Wang, H., Gao, C., Shi, R., Zhang, S., Li, J.: Omniarch: Building foundation model for scientific computing.
\newblock arXiv preprint arXiv:2402.16014  (2024)

\bibitem{wang2026reconstruction}
Wang, D., Song, Y., Chen, M., Cao, Y., Xia, W., Yu, C.: Reconstruction of time-resolved three-dimensional flow fields based on a multi-domain fusion transformer.
\newblock Physics of Fluids \textbf{38}(1) (2026)

\bibitem{wang2025parametric}
Wang, Y., Shelyag, S., Schluter, J.: Parametric super-resolution of turbulent channel flows from spatially sparse data using a transformer-based neural operator.
\newblock Physics of Fluids \textbf{37}(12) (2025)

\bibitem{zheng2024aerodit}
Zheng, H., Dai, Z., Pan, B., Wang, C., Zhang, B., Xiang, H., Fan, D.: Aerodit: Diffusion transformers for {Reynolds}-averaged navier-stokes simulations of airfoil flows.
\newblock arXiv preprint arXiv:2412.17394  (2024)

\bibitem{fan2025cascaded}
Fan, Y., Wei, C., Wong, J.C., Ooi, C.C., Wang, H., Chiu, P.H.: Cascaded phyformer-srnet: A physics-informed transformer-based framework with cascaded refinement for spatiotemporal super-resolution of turbulent flows.
\newblock Physics of Fluids \textbf{37}(7) (2025)

\bibitem{lei2025efficient}
Lei, F., Min, K., Zhang, J., Zhu, C.: An efficient diffusion transformer-based method for reconstructing three-dimensional turbulent flows from two-dimensional observations.
\newblock Physics of Fluids \textbf{37}(3) (2025)

\bibitem{cao2021choose}
Cao, S.: Choose a transformer: {Fourier} or {Galerkin}.
\newblock Advances in neural information processing systems \textbf{34}, 24924--24940 (2021)

\bibitem{patil2023autoregressive}
Patil, A., Viquerat, J., Hachem, E.: Autoregressive transformers for data-driven spatiotemporal learning of turbulent flows.
\newblock APL Machine Learning \textbf{1}(4) (2023)

\bibitem{xu2023super}
Xu, Q., Zhuang, Z., Pan, Y., Wen, B.: Super-resolution reconstruction of turbulent flows with a transformer-based deep learning framework.
\newblock Physics of Fluids \textbf{35}(5) (2023)

\bibitem{niu2025spatiotemporal}
Niu, X., Chen, G., Zhigang, C., Li, X., Wang, M., Shuai, L.: Spatiotemporal intelligent inference model of flows over complex {3D} configurations via multiscale grid fusion.
\newblock Aerospace Science and Technology p. 110886 (2025)

\bibitem{li2023scalable}
Li, Z., Shu, D., Barati~Farimani, A.: Scalable transformer for {PDE} surrogate modeling.
\newblock Advances in Neural Information Processing Systems \textbf{36}, 28010--28039 (2023)

\bibitem{yang2024implicit}
Yang, H., Li, Z., Wang, X., Wang, J.: An implicit factorized transformer with applications to fast prediction of three-dimensional turbulence.
\newblock Theoretical and Applied Mechanics Letters \textbf{14}(6), 100527 (2024)

\bibitem{yang2026implicit}
Yang, H., Wang, Y., Wang, J.: Implicit factorized transformer approach to fast prediction of turbulent channel flows.
\newblock SCIENCE CHINA Physics, Mechanics \& Astronomy \textbf{69}(1), 214606 (2026)

\bibitem{miotto2023flow}
Miotto, R.F., Wolf, W.R.: Flow imaging as an alternative to non-intrusive measurements and surrogate models through vision transformers and convolutional neural networks.
\newblock Physics of Fluids \textbf{35}(4) (2023)

\bibitem{jinhua2025general}
Jinhua, L., Rongqian, C., Zelun, L., Jiaqi, L., Yue, B., Hao, W., Yancheng, Y.: A general framework for airfoil flow field reconstruction based on transformer-guided diffusion models.
\newblock Chinese Journal of Aeronautics p. 103624 (2025)

\bibitem{sun2024improving}
Sun, Y., Pang, S., Qiu, Z., Zhang, Y.: Improving fluid identification in well logging using continuous wavelet transform and vision transformers: An innovative approach.
\newblock Physics of Fluids \textbf{36}(10), 106622 (2024)

\bibitem{cui2024Transformer}
Cui, Q., Zhang, M., Xiao, M., Ni, G.: Transformer based deep learning accelerated numerical simulation for incompressible flow.
\newblock Physics of Fluids \textbf{36}(12) (2024)

\bibitem{liu2025efficient}
Liu, W., Guan, Y., Yan, X., He, S., Jiang, M., Gong, J., Chang, C., Shen, X., Liu, Y., Zhang, G.: An efficient prediction and historical matching method for reservoir dynamics using advanced attention-based vision transformer.
\newblock Physics of Fluids \textbf{37}(1) (2025)

\bibitem{zeng2025swin}
Zeng, H., Xu, Q., Qin, Z., Wen, B.: A swin-transformer-based model for super-resolution reconstruction of turbulent flows.
\newblock Physics of Fluids \textbf{37}(7) (2025)

\bibitem{huo2025aero}
Huo, T., Li, B., Wang, Z., Zhang, D., Zhao, Z., Gao, J.: Aero-engine combustion flame segmentation via vision transformer.
\newblock Physics of Fluids \textbf{37}(12) (2025)

\bibitem{wu2024fast}
Wu, Y., Ba, D., Du, J., Zhang, M., Fan, Z., Xu, X.: Fast prediction of compressor flow field based on a deep attention symmetrical neural network.
\newblock Physics of Fluids \textbf{36}(11) (2024)

\bibitem{deng2023prediction}
Deng, Z., Wang, J., Liu, H., Xie, H., Li, B., Zhang, M., Jia, T., Zhang, Y., Wang, Z., Dong, B.: Prediction of transonic flow over supercritical airfoils using geometric-encoding and deep-learning strategies.
\newblock Physics of Fluids \textbf{35}(7) (2023)

\bibitem{ovadia2024vito}
Ovadia, O., Kahana, A., Stinis, P., Turkel, E., Givoli, D., Karniadakis, G.E.: Vito: Vision transformer-operator.
\newblock Computer Methods in Applied Mechanics and Engineering \textbf{428}, 117109 (2024)

\bibitem{wang2024cvit}
Wang, S., Seidman, J.H., Sankaran, S., Wang, H., Pappas, G.J., Perdikaris, P.: Cvit: Continuous vision transformer for operator learning.
\newblock arXiv preprint arXiv:2405.13998  (2024)

\bibitem{wang2021learning}
Wang, S., Wang, H., Perdikaris, P.: Learning the solution operator of parametric partial differential equations with physics-informed {DeepONet}s.
\newblock Science advances \textbf{7}(40), eabi8605 (2021)

\bibitem{li2024physics}
Li, Z., Zheng, H., Kovachki, N., Jin, D., Chen, H., Liu, B., Azizzadenesheli, K., Anandkumar, A.: Physics-informed neural operator for learning partial differential equations.
\newblock ACM/IMS Journal of Data Science \textbf{1}(3), 1--27 (2024)

\bibitem{goswami2023physics}
Goswami, S., Bora, A., Yu, Y., Karniadakis, G.E.: Physics-informed deep neural operator networks.
\newblock In: Machine learning in modeling and simulation: methods and applications, pp. 219--254. Springer (2023)

\bibitem{lin2023operator}
Lin, B., Mao, Z., Wang, Z., Karniadakis, G.E.: Operator learning enhanced physics-informed neural networks for solving partial differential equations characterized by sharp solutions.
\newblock arXiv preprint arXiv:2310.19590  (2023)

\bibitem{jiao2024solving}
Jiao, A., Yan, Q., Harlim, J., Lu, L.: {Solving forward and inverse PDE problems on unknown manifolds via physics-informed neural operators}.
\newblock arXiv preprint arXiv:2407.05477  (2024)

\bibitem{chen2024physics}
Chen, S., Givi, P., Zheng, C., Jia, X.: {Physics-enhanced Neural Operator for Simulating Turbulent Transport}.
\newblock arXiv preprint arXiv:2406.04367  (2024)

\bibitem{wang2024beyond}
Wang, C., Berner, J., Li, Z., Zhou, D., Wang, J., Bae, J., Anandkumar, A.: {Beyond Closure Models: Learning Chaotic-Systems via Physics-Informed Neural Operators}.
\newblock arXiv preprint arXiv:2408.05177  (2024)

\bibitem{ehlers2025bridging}
Ehlers, S., Stender, M., Hoffmann, N.: Bridging ocean wave physics and deep learning: Physics-informed neural operators for nonlinear wavefield reconstruction in real-time.
\newblock Physics of Fluids \textbf{37}(10) (2025)

\bibitem{VINO}
Eshaghi, M.S., Anitescu, C., Thombre, M., Wang, Y., Zhuang, X., Rabczuk, T.: Variational physics-informed neural operator {(VINO)} for solving partial differential equations.
\newblock Computer Methods in Applied Mechanics and Engineering \textbf{437}, 117785 (2025)

\bibitem{zhang2025omnifluids}
Zhang, R., Meng, Q., Wan, H., Liu, Y., Ma, Z.M., Sun, H.: Omnifluids: Unified physics pre-trained modeling of fluid dynamics.
\newblock arXiv preprint arXiv:2506.10862  (2025)

\bibitem{zhao2025lesnets}
Zhao, S., Li, Z., Fan, B., Wang, Y., Yang, H., Wang, J.: {LESnets} (large-eddy simulation nets): Physics-informed neural operator for large-eddy simulation of turbulence.
\newblock Journal of Computational Physics p. 114125 (2025)

\bibitem{lorsung2024physics}
Lorsung, C., Li, Z., Farimani, A.B.: Physics informed token transformer for solving partial differential equations.
\newblock Machine Learning: Science and Technology \textbf{5}(1), 015032 (2024)

\bibitem{germano1991dynamic}
Germano, M., Piomelli, U., Moin, P., Cabot, W.H.: A dynamic subgrid-scale eddy viscosity model.
\newblock Physics of fluids a: Fluid dynamics \textbf{3}(7), 1760--1765 (1991)

\bibitem{el2021implicit}
El~Ghaoui, L., Gu, F., Travacca, B., Askari, A., Tsai, A.: Implicit deep learning.
\newblock SIAM Journal on Mathematics of Data Science \textbf{3}(3), 930--958 (2021)

\bibitem{bai2019deep}
Bai, S., Kolter, J.Z., Koltun, V.: Deep equilibrium models.
\newblock Advances in neural information processing systems \textbf{32} (2019)

\bibitem{eswaran1988examination}
Eswaran, V., Pope, S.B.: An examination of forcing in direct numerical simulations of turbulence.
\newblock Computers \& Fluids \textbf{16}(3), 257--278 (1988)

\bibitem{canuto2007spectral}
Canuto, C., Quarteroni, A., Hussaini, M.Y., Zang~Jr, T.A.: Spectral methods: evolution to complex geometries and applications to fluid dynamics.
\newblock Springer (2007)

\bibitem{peyret2002spectral}
Peyret, R.: Spectral methods for incompressible viscous flow, vol. 148.
\newblock Springer (2002)

\bibitem{kingma2014adam}
Kingma, D.P.: Adam: A method for stochastic optimization.
\newblock arXiv preprint arXiv:1412.6980  (2014)

\bibitem{hendrycks2016gaussian}
Hendrycks, D.: Gaussian error linear units {(Gelus)}.
\newblock arXiv preprint arXiv:1606.08415  (2016)

\bibitem{yuan2020deconvolutional}
Yuan, Z., Xie, C., Wang, J.: Deconvolutional artificial neural network models for large eddy simulation of turbulence.
\newblock Physics of Fluids \textbf{32}(11) (2020)

\end{thebibliography}
\end{document}